\documentstyle[12pt,psfig]{article}
\titlepage
\begin{document}
\setcounter{page}{1}
\title{
  Flavor Production  in Pb(160AGeV) on Pb Collisions:
   Effect of Color Ropes and Hadronic Rescattering
      }
\author{H.\ Sorge \thanks{ E-mail: sorge@th.physik.uni-frankfurt.de }
\\ Institut f. theoretische Physik,  Universit\"at Frankfurt
\\ Gesellschaft f.\  Schwerionenforschung, Darmstadt
     }
\date{}
\maketitle

\vspace{1.7cm}
\begin{abstract}
  Collective interactions in the
   preequilibrium quark matter and
   hadronic resonance gas stage of ultrarelativistic
   nucleus-nucleus collisions
   are studied in the framework of the
   the transport theoretical
  approach RQMD.
  The paper reviews
  string fusion into color ropes and
   hadronic rescattering which
  serve as models for
   these interactions.
  Hadron production  in
  central Pb(160AGeV) on Pb collisions
   has been calculated. The  changes
    of the final flavor composition are more
   pronounced than in
   previous RQMD studies of
    light ion induced reactions at
   200AGeV.
  The ratio
   of created  quark pairs
  $s\bar{s}$/($u\bar{u}$+$d\bar{d}$)
   is enhanced by  a factor of 2.4
   in comparison to $pp$ results.
  Color rope formation  increases
   the initially produced  antibaryons
    to 3 times the value in the `NN mode',
  but only one quarter of the produced antibaryons
  survives because of
  subsequent strong
  absorption.
  The differences in the final particle composition for
  Pb on Pb collisions compared to   S induced reactions
   are attributed to the
    hadronic resonance gas stage which is
   baryon-richer and lasts longer.
\end{abstract}
\vspace{-21.0cm}

\newpage

\section{Introduction}

  The ground state of quantum chromodynamics
    shows peculiar
   properties, confinement of  colored
   degrees of freedom
   and spontaneous breaking of chiral symmetry.
   It is expected that
 at high temeratures and large baryon densities
    chiral symmetry gets restored and quarks are no longer
  confined.
  The central goal  of modern heavy ion physics
  is to explore  these properties of strongly
  interacting matter \cite{QM95,QM93}.
   Several experiments on nuclear targets
 are  performed nowadays using
    the Au(11.6AGeV) beam at BNL or the
    Pb(160AGeV) beam at CERN.
    The first series of experiments with light
   ion beams (p,O,Si,S) has just been
     finished in these laboratories.

  The work on which this paper is based has started some years ago
   \cite{SOR89}.
  The goal
    was set to develop a complete transport theoretical scenario of
   nucleus-nucleus reactions,  from the initial state
    of two nuclei before overlap to the final state
     after the strong interactions have ceased (freeze-out).
  The developed model has been dubbed relativistic quantum molecular
  dynamics (RQMD).
   RQMD is a semi-classical  microscopic approach which combines classical
    propagation with  stochastic interactions.
    Strings and resonances can be  excited in elementary collisions.
    Their fragmentation and decay leads to production of particles.
  Overlapping  strings
   do  not  fragment  independently  from  each other
  but  form
  `ropes',  chromoelectric flux-tubes
  whose sources are
   charge states
  in higher dimensional representations
   of color $SU_3$
  \cite{BIR84}.
   The nature of the active degrees of freedom
   in  RQMD  depends on the relevant
   length and time scales of the processes considered.
   In low energy collisions
    (around 1 AGeV)
   RQMD reduces  to
    solving transport equations for a system of
    nucleons, other hadrons and eventually resonances
    interacting in binary collisions and via mean fields
    (similarly to BUU \cite{BDK84},
     QMD \cite{AIC87}, and so on).
     At large beam energies ($> $ 10 AGeV)
    the description of a projectile hadron interacting in a
      medium
    (in the simplest case a cold nucleus)
    as a sequence of separated hadron or resonance collisions breaks
   down \cite{KOP75}.
   A multiple collision series can be  formulated on the
    subhadronic (quark) level.
    Following the paths of the ingoing constituent quarks,
      a Glauber-type
     multiple collision
     series  is generated in RQMD,
   with cross sections taken from
    the additive  quark model \cite{BIA76}.
   The  secondaries  which  emerge from  the fragmenting
   strings, ropes and  resonances may interact
   with  each other and the  original
   ingoing  hadrons (rescattering and mean-field interaction).

  The RQMD model
  has been successfully applied to nuclear reactions  at
    ultrarelativistic energies
   (see e.g.\ \cite{NAG94}-\cite{SOR95PLB}).
   The use of transport models like RQMD is not restricted to the
   study of the generated final state.
   It is of equal importance that such models can be used
     to study the influence of
    various types of interactions and medium effects on
      final state observables and
    to achieve a better insight into the {\it transient }
   stages of heavy ion collisions.
    Using the information from calculated RQMD events
    these questions have been  addressed already in the
     literature,
     at 200AGeV mostly for S on A collisions
   \cite{SOR92}-\cite{SOR95PLB}.
   Here and in  a  follow-up paper \cite{SOR95PRCB}
   I wish to expand on those results and  study
     the amount of collectivity for
     central Pb(160AGeV) on Pb collisions.
    This is
    the  heaviest
    projectile-target combination for which experiments
    are currently undertaken at CERN-SPS.
    Any kind of collectivity
    -- physics of dense matter beyond mere superposition of
      independent nucleon-nucleon interactions --
    is expected to be strongest  in a system with minimized
      surface-over-volume ratio.
   The paper presented here will focus on the influence
    of some important building blocks of RQMD --
    color ropes and hadronic rescattering -- on the
     particle chemistry and  distributions in
   phasespace.
    In particular, I shall consider the
    yields and momentum distributions of strange hadrons and
   antibaryons.
    Their production is dynamically suppressed in
     elementary hadronic interactions which makes them
      a useful probe for collectivity in
    the transient stage of $AA$ collisions \cite{KOC86}.
    The follow-up paper  mainly discusses the
     dynamical evolution of the created source
    in space-time
    until freeze-out.

  The achieved degree of  collectivity
  in the  first --  the prehadronic
   or quark matter -- stage of nucleus-nucleus collisions
   is a central topic of heavy ion research.
  Soft multiparticle production
   is commonly described by fragmentation
    of  excited color strings
   (see for reviews on this subject
   \cite{LUND83,WER93,DPM94}).
  It is expected that
   independent string fragmentation,
    which means no collectivity at all,
    breaks down
   in central nucleus-nucleus collisions,
   because the string density gets too large.
   Interactions of overlapping strings
   is  modeled as color rope formation
   in RQMD.
 Ropes can be viewed
 as a model
  of locally deconfined
 quark matter,
 which is dominated by longitudinal
 excitations and therefore of relevance for
 the preequilibrium stage in
 nucleus-nucleus collisions.
   The  evolution in a rope is entirely governed
 by nonperturbative dynamics (a distance-independent
  rope tension).
  Other `second generation' transport approaches
   also  go  beyond
   independent string fragmentation,
    for instance
     string fusion in QGSM \cite{AME93},
   the Spanish version of the dual-parton model \cite{ABFP95},
   or quark matter droplet formation in VENUS \cite{AIC93}.
  A common result of these approaches is that
   strangeness and baryon pair  production
    which is dynamically suppressed in elementary reactions
   becomes more favorable.
  On the other side,
  arguments have been put forward to include
    diquarks and an unsuppressed  strange quark component into the
    nucleon sea
   `from the beginning'
    which can be transformed into real particles
    by multiple collision effects\cite{CAP95}.

 Collectivity
  may also emerge
  in the hadronic stage of nuclear reactions,
    resonance matter formation
  -- as studied in the models RQMD \cite{SOR91,HOF95}
  and ARC \cite{ARC92} -- or
   mean fields \cite{SCH91,KO92}.
   In recent years
    the interactions of  resonances in a dense medium
     have found a lot of attention.
    For instance,
    resonances may act
     as `energy storage'
    in multi-step collision processes which are of importance
    for heavy particle production \cite{SOR91,WOL93,HJA94}.
  Strictly spoken, there is no scattering ($S$) matrix
   for resonances, because they are unstable. However,
    introducing them as quasi-particles makes sense under
    some limiting conditions  ($\Gamma /M \ll 1$).
    Note that
     resonance interactions can get important only in a
    system at sufficient density.
     The  collision frequency
     has to be larger than the inverse of
      resonance lifetime (typically 1-2 fm/c).

   The RQMD approach has evolved over the last years,
   because various  interaction pieces have been put in
    step by step.
    After the important role of
    resonances
    had been recognized,
    a model was constructed for  annihilation of
     mesons and baryons, including
     resonances,
     into nonstrange baryon resonances \cite{SOR91}
      and into a `string continuum' \cite{SOR93ZFP}.
    Independent string fragmentation was replaced by
     color rope formation in case of overlapping strings
     \cite{SOR92}.
     Low energy
     hadronic  interactions -- $s$ channel resonance
      formation, $t$ channel meson and Reggeon exchange --
     have been modeled in the meson-meson \cite{BER94} and the
    meson-baryon sector with net strangeness \cite{SOR95PLB}.
    In this paper we are going
    to  review  these interactions in some detail.
   Some recent work
    to further develop the RQMD model
    for applications up to collider energies
    (multi-string excitations
     according to the AGK cutting rules \cite{AGK74})
   will be described
    elsewhere \cite{SOR95AGK}.
 \footnote{
    For the topics of interest here
   multi-string excitations in $N$$N$ collisions
   are of minor importance.
    On the one side, they
    tend to enhance the string densities.
    However, with respect to
     rope formation
    most of this effect
     is canceled
    at the comparably small
    CERN-SPS energy, because
     the average string length
    decreases in comparison to the case with 2-string
    excitations only.
   }
    These interactions  have
     been implemented into the  computer code RQMD
    (most recent version 2.1)
    which was developed by the author.

\section{Strings and ropes in $hh$ and $AA$ collisions}

  The  so-called recombination approaches
 like RQMD assume that in elementary $h$$h$ interactions
 the hadron  wave function is
decomposed into  spectator  and
interacting quarks.
  The quark spectators
  neutralize their  color
  while keeping
  their gluon  cloud
    coherently bound
    (constituent quark picture).
 See Ref.~\cite{SOR93ZFP} for a more detailed
  description.
    The interactions are dominantly
     initiated by slow quanta.
    The interacting constituent quark looses
   part of its gluon cloud,
     because the interacting components
   get out of phase with the
   spectator remnants in the cloud.
  These gluons form -- together with their
    partners from the target -- the source
    for  secondaries.
  The parton language is used here  rather loosely,
  to give a rough sketch of the process.
There will be no attempt here to describe
 this process quantitatively by interactions of QCD quanta.
  It will be simply assumed that these
   interactions  result in the longitudinal excitation of
  color strings.
 The concepts of strings in strong interactions actually
 predates QCD. There are many recent attempts
 to  understand the properties of strings  or equivalently
  chromoelectric flux-tubes
  directly from QCD, e.g.\ by invoking the
 dual Higgs mechanism for QCD-monopole condensation
 \cite{SUG95}.

\subsection{Strings  in RQMD}
 \label{secstring}

 In inelastic $hh$ collisions the  fraction  $x^+$ of the
  projectile  lightcone momentum $P^+=E+p_l$
   ($p_l>0$ assumed)
 which goes
   into target string excitation
  is determined from
\begin{equation}
        \label{ppmnew}
   dP \sim \frac{d x^+ }{ x^+ }
  \quad ,
\end{equation}
 with the lower  limit set by the
    target hadron's (small)
    plus-component before collision.
   The target momentum
     (negative lightcone component)
    which goes
   into projectile string excitation is
   determined in the same way.
    This string excitation law  is the same as originally
   suggested by the Lund group and realized in the
    Monte Carlo code FRITIOF  \cite{FRI87}.
    The  invariant masses
    of the target and  projectile
     excitations should not fall below some minimum
     to allow for particle production.
     The minimum value is set to
     $m$+0.3 GeV/$c^2$, with
     $m$  usually taken as the ingoing hadron mass.
      Since the
     ingoing pseudoscalar
      mesons
     which are  Goldstone bosons
      have
      exceptionally light  masses,
      $m$ is set in these cases to 0.7 (0.85) GeV
        for mesons without (with) strange valence quarks.
     The excited states decay either stringlike or
      -- for excited masses below some `smeared-out'
      cut-off value
       (2 GeV/$c^2$ for nucleons) -- as
      resonances.

\begin{figure}[h]

\centerline{\hbox{
\psfig{figure=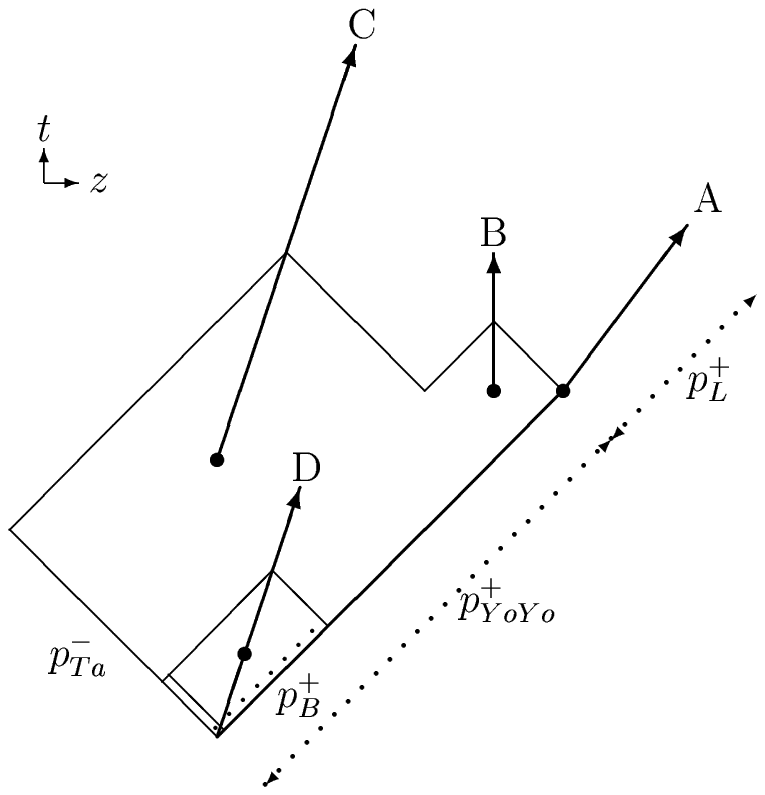,width=17.0cm,height=24cm}}}
  \caption[
           ]
          {
          \label{hyoyodec}
                                 }
 \end{figure}

The string topology
 generated in RQMD for $hh$ interactions
 and displayed  in
 fig.~\ref{hyoyodec}
is determined from the basic assumptions that all
ingoing  valence quarks
 keep  on moving into the same  direction and
no net color flows between  target
 and projectile.
A sea quark pair
gets polarized in course of the interaction
resulting in spatial separation of sea quark  charge and anticharge
(see  fig.~\ref{hyoyodec}).
The interacting
constituent quark  combines
 with the sea
anticharge
into a colorless state.
 In the semi-classical picture
the  sea-quark companion
 starts to  move backward
due to the  momentum flow from the target.
   Confinement
  forces the backward moving quark to
  pull out a
string.
The spectator
quarks
neutralize  color in the same way
as the interacting quark
at the other end of the string.
Of course, if energy is lacking
 the backward and forward
sea-quark pair
might be the same. In this case there will be no
string,
but decay of the created hadron excitation into
two hadrons.
  In the terminology employed here
   a  `color string'
    (restricted  to be a
      longitudinal excitation)
   is  spanned between sea quarks
     which are identified with partons.
    Such a configuration is assumed to fragment
      in the same way
     into hadrons as a $\bar{q}$$q$  configuration
      produced in $e^+$$e^-$ annihilation. In contrast,
    the ingoing valence  quarks
   are assumed to be `dressed' (constituent quarks)
    and better prepared
   for hadronization. An ingoing spectator quark
   fragments
    differently (harder)  than a partonic quark
      with equal momentum
      giving rise to the
   so-called leading particle effect.
In the model quark spectators
 are able to transfer
all their primordial momentum
to the hadron(s)
into which they fragment.
  The corresponding constituent quark structure functions
   are specified in
  \cite{SOR95ZFP,SOR93ZFP}.
The interacting quark keeps only a fraction of its original
momentum,
in contrast to the spectator quark(s).
  The momentum which it keeps is dubbed $p_{B}^+$
      in fig.~\ref{hyoyodec}.
     The difference to the original momentum
   $p_{YoYo}^+$ is used up as the forward momentum
    of the excited string, i.e.\ for particle creation.
 $p_{B}^+$ is determined stochastically
 using the string fragmentation function $f(z)$ \cite{LUND83}
   as the probability measure.
     The  backward momentum  of the string
       $p_{Ta}^-$
          transferred from the target  is determined from the
     equivalent of eq.~(\ref{ppmnew}).
      The decay properties of a color string are completely
      determined by its light cone momentum
      and  the flavor
      at its end.
  The spectator momentum  $p_l^+$
    does not enter at all into the calculation
     of the string evolution or its decay products.
      (see fig.~\ref{hyoyodec}).

   The decay of elementary color strings into hadrons
    is calculated employing the concept of
    `left-right symmetric string fragmentation' developed
    by the Lund group  \cite{LUND83},
     with default JETSET 6.2 default parameters
     for $f(z)$ \cite{SJO86}.
   The formation points of hadrons from string decay are
    calculated as the average of the two break points
    from which the quark constituents are emerging.
    This is schematically displayed in fig.~1
     from which one can also read off the prescription for
     the formation points of hadrons containing
     one of the original valence quarks.
     In particular, the formation length of the leading hadrons
     coincides with the concept of
     so-called `constituent formation length'
      \cite{BIG87}  in which the formation length shrinks to 0
    in the extreme limit  $z$$\rightarrow$1.
  The standard high energy 2-string scenario
   is modified in three cases,
   generation of additional strings from sea excitations
   which can be related to multi-Pomeron-exchange \cite{ABR84},
   projection of low-mass excitations onto resonance states
   and diffractive inelastic interactions \cite{GOU83}.

    Corrections to the 2-string-excitation scheme arise
   at lower energies,
    if the excited masses which are randomly chosen according
    to eq.\ (\ref{ppmnew}) are below the cut-off values set
    for string fragmentation.
    If both outgoing states are below string excitation
    threshold, two hadrons (or resonances) are formed in the out-state
  \begin{equation}
    \label{h1h2}
    h_1 + h_2  \rightarrow h_1^*+h_2^* \quad .
  \end{equation}
    Production of two baryon resonances
    is  the dominating process
    in $pp$ collisions
   at the AGS energies of 10 to 15 GeV.
    Furthermore, even at the highest beam energies
    strings are actually rarely produced in nonannihilating events
   during the rescattering stage of nucleus-nucleus collisions.
   Therefore
      production (and absorption) of resonances
    in 2$\rightarrow$2 processes  effectively replaces the
    2-string excitation
     component and  fills up the inelastic cross section,
     in addition to $s$ channel resonance formation and
    meson exchange processes in the $t$ channel
    (see section \ref{secresc}).

  Let us assume now that  above condition leading
  to a reaction of the type in  (\ref{h1h2}) is met.
     The two produced excitations have to be projected onto
    hadronic states.
    Resonances are propagated explicitly in RQMD
    and may scatter themselves.
   All lightcone momenta which enter into
    eq.\ (\ref{ppmnew}) are calculated as if the
   ingoing hadrons would be groundstate hadrons,
    keeping the CMS energy the same.
   Thus  the class of out-channels
  in  (\ref{h1h2}) is populated with the  same
   probability and  cross section
   (given  the same total  cross section),
   irrespectively whether $h_1$ and $h_2$ are excited states
  or not.
   The need to respect
    detailed balance requires some modifications
    which were neglected in earlier versions of the
    RQMD  model \cite{SOR91}.
    The improved model for  such
    $2 \leftrightarrow 2$
    transitions will be
    described in the following.
  The flavor  of each hadron is kept the same,
   an assumption which can be dropped eventually
  in further refinements of the model to allow for
  flavor-exchange between the collision partners.
  The
   out-channel
  is chosen randomly with proper weights to give
  the following   cross sections
    for $2 \leftrightarrow 2$
  where each term in the sum corresponds to the
  production cross section of
  a particular 2-hadron (resonance) state:
   \begin{eqnarray}
    \label{sigklij}
        \sigma (k,l\rightarrow \mbox{`2'})  &=&
         \sigma_0 (s)
           \sum _{i,j}
                p_{ij}^2
                         \\ \nonumber
                 &  &    (2 S_i+1)  (2 S_j+1)
                \cdot SF(kl,ij)
      \quad .
   \end{eqnarray}
   $i$$-$$l$ label  hadron states,
    $p_{ab} $ denotes the CMS momentum
     and $S_a$ the spin of hadron $a$.
    $\sigma_0 (s)$ is
    determined by the normalization
    condition.
     Depending on the initial choice --
     a nondiffractive or a diffractive
      inelastic interaction --
      the sum runs over all allowed states
     $i$ and $j$ or is restricted by the constraint that
     either hadron $i$ or $j$ is a groundstate hadron.
    $SF(kl,ij)$ denotes a symmetry factor which is unequal to 1 only
     for $k\ne l$ and $i=j$.
    In this case it is 1/2
    to respect the detailed balance
    relation.
   Since eq.~(\ref{sigklij}) is implemented in RQMD
   employing the Monte Carlo technique,
    the transition $kl\rightarrow ii$ is assigned first
    a probability which ignores the symmetry factor.
    Afterwards the transition is accepted with probability 1/2.
    In case of rejection an elastic collision is realized.
    The physics content of
   eq.\ (\ref{sigklij}) is rather simple.
     The out-states
    are chosen according to phasespace and available
    spin degrees of freedom,
    its functional form in accordance with the detailed balance relations.
    The   basic  idea is that statistics governs
     the population of out-states
   if many  states
    are available in binary collisions.

 Extending the RQMD approach
   from $h$$h$
  to $h$$A$ and $AA$
  collisions
 by  allowing for multiple string excitations
 is straightforward.
  Each  nondiffractive
  constituent quark interaction generates
 a new string if the energy is
 sufficient \cite{SOR95ZFP}.
The  cross section of interacting
constituent quarks is given from the additive quark model
$\sigma_{q-q}=1/9\cdot \sigma_{NN}$
which keeps the original projectile
cross section constant while
the projectile is traversing the target.
  Multiple  interactions
  can  be pictured
  as
  continuing `undressing' of a
   constituent quark propagating through the target.

\begin{figure}[h]

\centerline{\hbox{
\psfig{figure=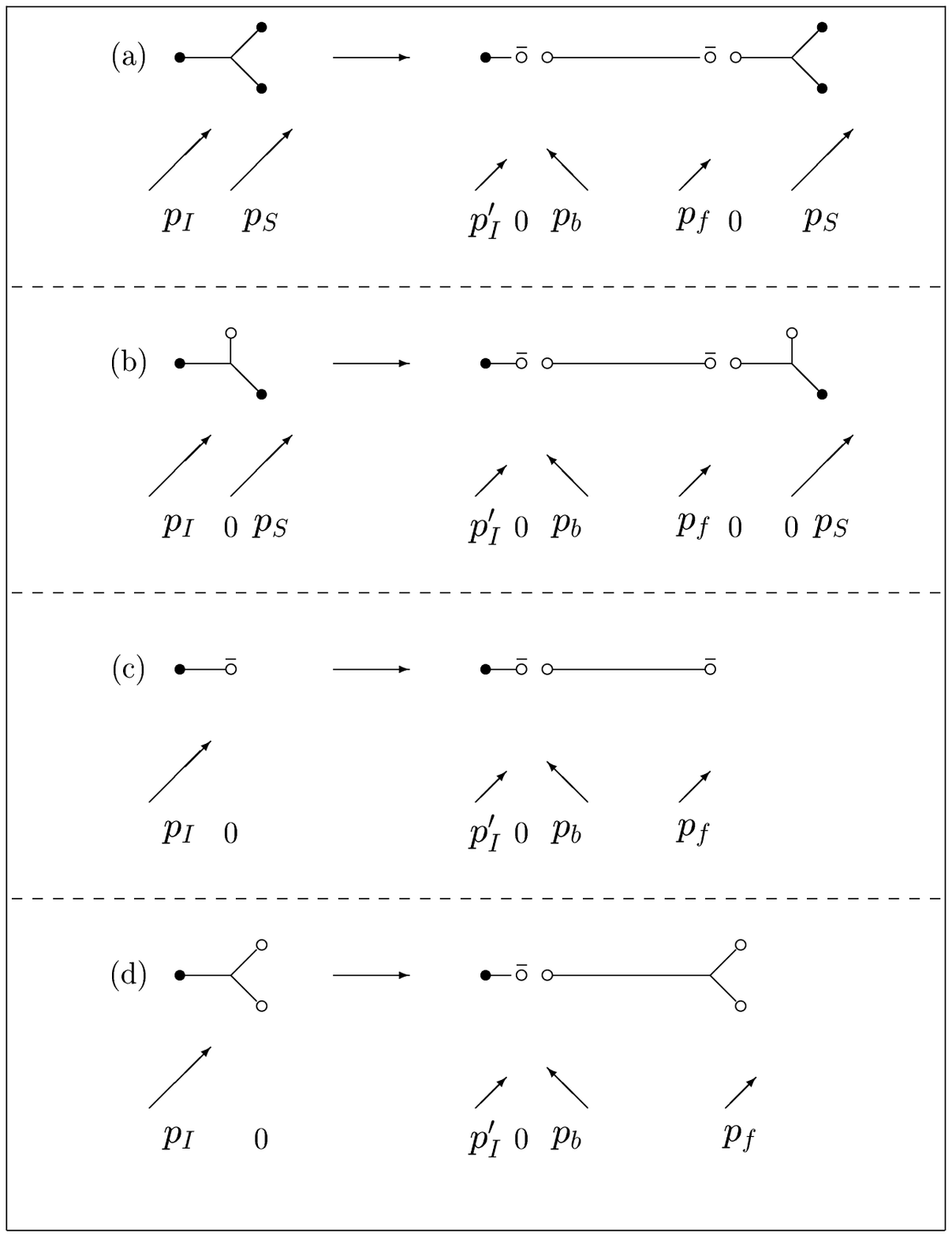,width=17.0cm,height=24cm}}}
  \caption[
          ]
          {
     \label{bstring}
                                          }

 \end{figure}

 It is an important question for the degree of
 baryon stopping and the achievable baryon densities
  in nucleus-nucleus reactions
 how the baryon number is shifted in
  multiple nucleon collisions with a target.
  The minimal-stopping approach  is to
  concentrate all momentum which is not used up
  for target string excitation into a leading diquark
   which fragments like in a $N$$N$ collision.
 Historically, this  was the first
  approach which has been  applied in
   string fragmentation models \cite{AGN87,CAP87}.
 It is well-known by now that
  the minimal-stopping approach  is at variance
   with the experimental data in
  ultrarelativistic $pA$ and $AA$ collisions \cite{MIT94}.
 The shift of baryon number in rapidity is stronger.
  It contains a component which goes with the
   number of additional projectile collisions.
  In the beginning several people have
   parametrized the dependence of
  the nuclear stopping power as a function of the
  number of additional collisions \cite{HKH85}.
  A  model for baryon stopping which contains
  a strong dependence on the number of
  collisions was suggested in \cite{SOR93ZFP} and is
  applied in RQMD. It is in good agreement with
  available  baryon measurements
  of the NA34, NA35 and NA44 group in 200AGeV collisions
  \cite{SOR95ZFP}. In this approach
 the light cone momentum fraction
 of the baryon after fragmentation is determined
 from the number of spectator constituent quarks
 which it keeps in its wave function.
  The topology of string excitations in
  multiple baryon collisions and the distribution
  of the ingoing momentum between the
   active degrees of freedom (strings and constituent quarks)
  are displayed graphically in fig.~\ref{bstring}.
  Each further
  inelastic interaction in a target
   removes an additional constituent quark
  from the
  spectator remnant
  replacing it by a sea quark
   with zero momentum (in this approximation).
  It may even happen that none of
 the original constituent quarks
   ends up in the outgoing baryon. In this case
  a partonic sea diquark moves at the forward end
   of one of the projectile strings, and
  the baryon momentum is determined from
  string fragmentation.

  The ingoing hadrons are complex objects themselves.
  Multiple collisions of a hadron  at high energy
   are not sequential
   collisions of a single object, but
  in each collision a different hadron component
   is involved.
  Simple consideration of time scales tells that
   multiple soft interactions of the same object are highly
   suppressed. A fast particle which
   needs on the order of 1 fm/c
  in its rest system to finish  the first interaction
   has left the target far behind which suppresses
   a subsequent interaction.
  This argument can be given a more rigorous meaning
  by analysing planar diagrams with multiple `ladder'
   exchange \cite{AFS62}.
  Using the same line of argument
  multiple quasi-elastic collisions
   ($2\rightarrow 2$)
  and  inelastic diffractive interactions
   should be suppressed as well at high energies.
  Of course, the basic reason behind the suppression of
  multiple collisions within short time intervals
  is of quantum-mechanical nature. The uncertainty
  relation associates a finite time and distance interval
  to an interaction
   which can be characterized by exchange of energy and
  momentum.
  Recently a first step was done to
 take this  effect into account in RQMD.
  Since the particle propagation in RQMD is realized
   semi-classically, only an approximate solution can be found
  in this framework.
  It was discussed a long time ago by Low and Gottfried that a
   classical space-time concept of propagation can be applied for fast
  particles \cite{LG78}, because rapidity and longitudinal
   position are commuting in this limit.
  Therefore one is allowed to specify $z$, $t$ and $y$ after a
  collision.
   The space-time interval after which
     a quasi-elastic or diffractive collision is considered
   as finished is now defined by
 \begin{equation}
   x^+_{min} =    x^+_{Coll}  +  \Delta x^+
   \quad ,
 \end{equation}
 assuming that the particle is moving in forward (=+) direction.
  (A corresponding relation holds for the collision partner.)
  If the collision cannot be finished before the next collision
  is going to take place, the first collision
  being `too soft'  is discarded.
  The collision point $x_{Coll}$
  is defined from the minimum distance value
   which the two classical hadron trajectories can have in their
  2-body center of mass frame.
   $\Delta x^+$ is calculated using
 \begin{equation}
   \label{sclunc}
    \Delta x^+  \sim 1 / \Delta p^-
   \quad ,
 \end{equation}
    with $\Delta p^-$ the absolute value of the
  lightcone momentum which the hadron
  has picked up from the collision partner.
 (As usual in RQMD the longitudinal component is defined in
  the 2-body CMS by the direction of particle motion before
  collision.)
 Relation (\ref{sclunc}) is motivated by the Heisenberg relation
  $\Delta x^+ \cdot \Delta p^- \ge 1$. The proportionality factor
  is presently set to 1.
  Applying the criterion of
 relation (\ref{sclunc}) to all 2$\rightarrow$2 collisions
  which are  generated by RQMD in the dynamical evolution of nuclear
  collisions suppresses  soft collisions. By their very definition,
   not many particles are produced
  in such collisions. Therefore the resulting effect on the final
  particle yiels is small in A(200AGeV) on A collisions,
   on the few percent level.
  The quantum-mechanical suppression
  of very soft multiple collisions
   in $AA$ interactions at other  beam
   energies is currently studied. The results
   will be presented elsewhere.

\subsection{Color ropes in RQMD}

    \label{secrope}
 The   quarks which
  have been polarized from the sea in course of the initial
 interactions are receding from each other.
 Their masses are usually small  compared
 to their  lightcone momenta and neglected.
 Therefore these quarks are moving on the lightcone and form
  the sources of the
 chromoelectric field filling the region between
  the forward and backward moving charges.
  The field is compressed into tubes
   with the tube cross section kept independent
  from the field strength.
 Thus a collection of tubes
  is spanned in $AA$ collisions.

\begin{figure}
\centerline{\hbox{
\psfig{figure=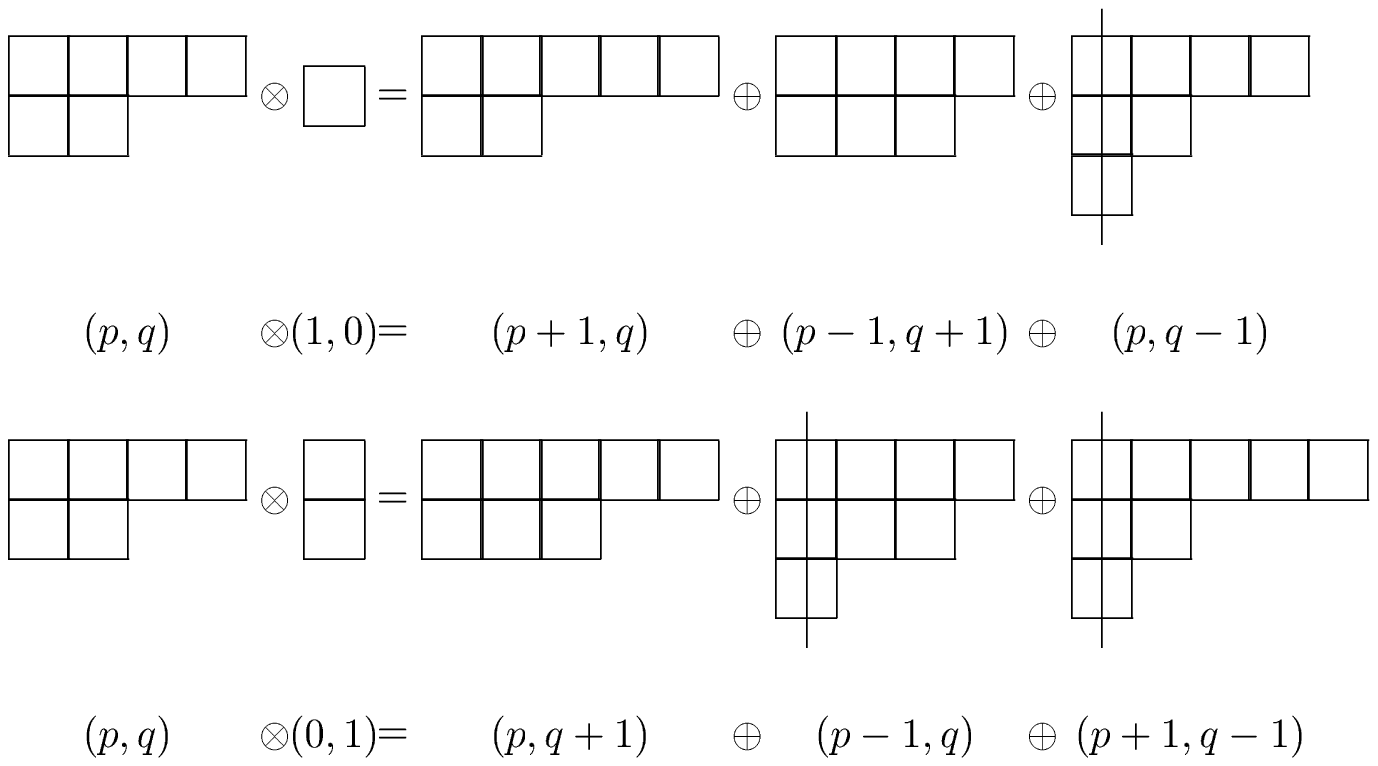,width=17.0cm,height=24cm}}}
  \caption
       [
                                  ]
          {
    \label{su3mltp}
            }
\end{figure}

First, I am going to describe the physics of a
 single chromoelectric flux-tube. The formation of
 tubes with stronger than the elementary string fields
 could  be of relevance for $AA$ collisions but also for
 soft multiparticle production
 in hadron-hadron collisions at collider energies \cite{MPR92}.
 The created SU(3)-valued color fields
   inside each tube
  are added coherently.
The elementary triplet charges
of quarks and antiquarks in one of the
receding rope `condensator' plates
are coupled
stochastically
to  the
total $SU_3$ color charge as the
 source of the rope field:
    \begin{equation}
    \label{Gl1}
      (p,q)  \otimes (1,0) \rightarrow
      (p+1,q)  \oplus  (p-1,q+1)
                     \oplus  (p,q-1)
        \quad  ,
    \end{equation}
   with the  statistical weight
   given from the dimension of each  multiplet
    \begin{displaymath}
     d(p,q) =
        \frac{1}{2}  \cdot  (p+1)  \cdot (q+1)
              \cdot  (p+q+2) \quad .
    \end{displaymath}
 The possible couplings
  of an arbitrary $SU_3$ charge with an elementary
  charge are graphically displayed in fig.~\ref{su3mltp}.

The chromoelectric field of the rope is determined
in  the flux-tube picture
by Gauss' law:
    \begin{equation}
   E^\alpha(p,q) \cdot A(p,q)= g \:
    \mbox{\cal{F}}(p,q)^{\alpha}
  \quad   .
    \end{equation}
 $(p,q)$  characterizes  the
 multiplet  of the QCD charge.
   $\mbox{\cal{F}}(p,q)^{\alpha}, \alpha=1,8$
  are the generators of the SU(3) group in the
  corresponding representation.
 It follows that the energy density
 of a rope field
 is   proportional to
 the eigenvalue of the Casimir operator
  $C(p,q)=
    \mbox{\cal{F}}(p,q)^{\alpha} \cdot
    \mbox{\cal{F}}(p,q)^{\alpha} $.
In  RQMD
 the transverse size of a rope
$A(p,q)$
is being kept
independent on the representation of the source.
 This is
required
in the flux-tube model
to get  scaling
of the rope tension
with the Casimir eigenvalue
which has been observed for both $SU_2$ and $SU_3$
gauge groups \cite{BER82}.
Thus the rope tension,  the energy per unitlength,
also scales
with the Casimir eigenvalue
 \begin{equation}
  \kappa(p,q)=3/4 \cdot C(p,q) \cdot
                   \kappa_{el} \quad.
 \end{equation}
 The  tension
 of  a triplet flux-tube (elementary string)
 $\kappa_{el}$
  can be related to the Regge slope parameter
   $\alpha '$ which gives   0.9 GeV/fm.
 It is well-known that
  the flux-tube picture emerges from
   QCD  with static quarks in  lowest order
  of the strong coupling expansion
  \cite{KOG75}.
 Recently the assumed
  flux-tube poperties
 --
  dominance of the longitudinal electrical field components,
 scaling of the field strength with
  $C(p,q)$  and charge independence of the
 transverse extension of the field --
 have been confirmed by a calculation
of Wilson loops for higher dimensional charges
in 3-dimensional $SU_2$ lattice gauge theory \cite{TRO93}.
Note that these new lattice results
provide  important information concerning the
dynamics underlying confinement.
If confinement is due to a bulk property of the
QCD vacuum (like as the pressure in bag models),
then the transverse size of higher dimensional
flux-tubes is expected to increase.
Consequently, the rope tension would scale
less rapidly,
 as the square root of the Casimir eigenvalue only.
Thus  phenomenological models of confinement
like the bag model are not compatible with the recent
results of lattice simulations.

 Quark-antiquark pairs are created from the chromoelectric
 field
 and screen the original field \cite{CNN79}.
 In a first approximation
 the total pair creation rate and the flavor
 composition
 can be calculated employing
 Schwinger's vaccuum persistence rate for a
 constant electric field \cite{SCH51}.
 This is done by treating the pair creation
as a tunneling process
 in the semi-classical
  WKB approximation (see Appendix).
Employing a local density approximation,
the calculated pair creation rate
determines stochastically the space-time points
 in which the field strength is degraded by created quarks.
 This procedure generalizes the
 Artru-Mennessier scheme
    \cite{ARM74} to the situation
  of fields with variable strength.
 Of course, some modifications of the decay probability
  are expected for very short times
 \cite{HER90} and
 near to the sources of the rope field \cite{MAR89}.
 The pair creation probability per unit time
  and length is obtained by integrating over
  the transverse rope area $\pi$$\cdot$$r_{tube}^2$
  with transverse rope radius set   to 0.8 fm.
   The radius parameter is fixed by the requirement to
  get reasonable production rates  from elementary
  string decays in comparison to
   $e ^+$$e^-$ hadroproduction data.
Strange quark production is easily enhanced
 by increasing the field strength,
 because the mass difference between the light flavors becomes
 irrelevant for sufficiently large fields. In contrast,
 3-diquark creation inside the rope stays rather weak
in the average.
The small probabilities to create a diquark pair
by the rope field have their root in the assumed
2-step process
as outlined in  Ref.~\cite{SOR95ZFP}.

 The collective field is gradually degraded, because the
  initial `macroscopic'
 spatial separation of charge and anticharge cannot be
  maintained.
  The first mechanism to degrade the field strength is quark pair
  production.
Pair creation  points,
 the created  flavor and
 transverse momenta
are  sampled stochastically.
The  local field strength
is calculated self-consistently,
 taking
screening of
 the original field
by already produced quark pairs
into account.
 The rope tension is lowered in the forward lightcone
  of each break point
 from   $\kappa(p,q)$ to $\kappa(p',q')$.
 E.g., $(p',q')$=$(p,q-1)$ if a quark is
  pulled out of the vacuum and
 attracted by the charge $(p,q)$.

\begin{figure}

\centerline{\hbox{
\psfig{figure=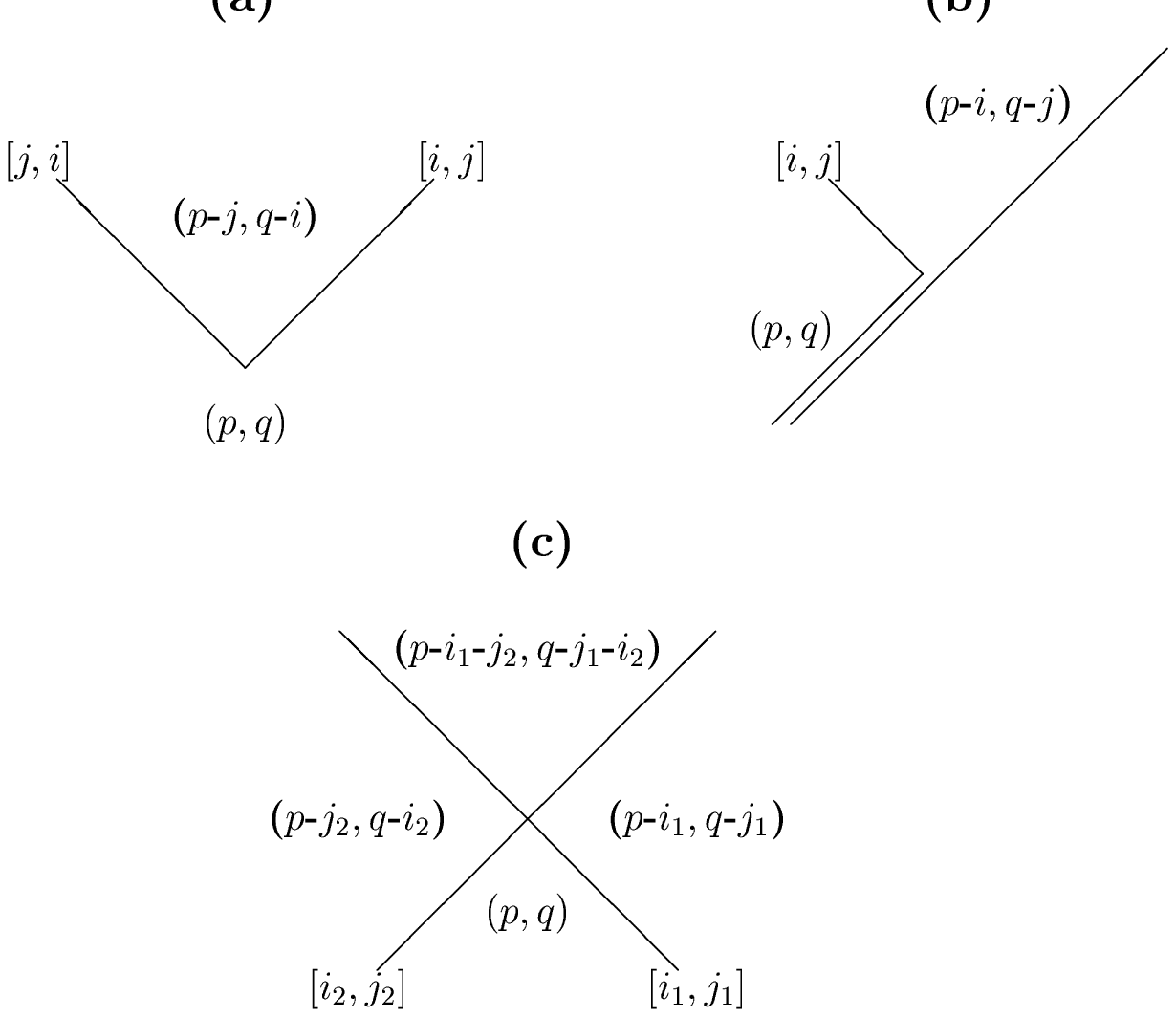,width=17.0cm,height=24cm}}}
  \caption[
       	  ]
         	    {
   \label{fdegra}
               		    }
 \end{figure}

 There are two other processes which degrade the original
 field strength, turning points of quarks in the
  rope end plate and crossing points of two color charges
  inside the rope. All these processes are
  displayed together in fig.~\ref{fdegra}.
  The turning point of a quark is determined
  from the condition that  its original momentum has
   been used up by propagating under the force of
 the rope. Its momentum loss per unit-time
   is given from
 \begin{equation}
   \label{dpdtquark}
   \frac{dp}{dt} = \pm \frac{\kappa(p,q)}{p+q}
 \end{equation}
  with the sign depending on the direction of motion.
  Of course, this is the direct generalization
  of the situation in  string decay
  to account for the finite total momentum.
   Using  eq.~(\ref{dpdtquark})
  independent  fragmentation is recovered
   in the  limit
 that  the fragmentation products of two strings have no overlap
  in rapidity space.
 \footnote{
   The hadron rapidity distribution
   from string fragmentation is approximately constant
   within some limits $y_{min,max}$.
    The limits are related to the
     energy-momentum of the string via the relations
    $y_{min,max}$$\approx$$\ln P^{-,+} $ -$\ln (2m)$,
    with $m$ a typical hadron mass.
   }
   In this case
  the initial quark momenta  show complete mismatch, e.g.\
  one string is specified by ($P^+$,$\delta^-$), the other
  by   ($\delta^+$,$P^-$) with $\delta ^{+,-}$$ \ll$$ P^{+,-}$.
  The region with non-triplet field strength shrinks to zero for
   $\delta ^{+,-}$$ \rightarrow$0
   and $P^{+,-}$$ \rightarrow$$\infty$.
  Similarly, those parts of string  world sheets
   which do  not overlap with other strings
  can make no contribution to build up a region of
  larger  field strength.
   In RQMD these nonoverlapping parts of generated strings
  are split off from the beginning in the rope fragmentation process.
  The reason is the following.
  As displayed in fig.~\ref{su3mltp}
   two quarks (anti-quarks) may coalesce into a diquark
   in the $\overline{3}$ (3)-representation
   during the process of rope charging.
  Guided from the general coalescence picture one might
   expect that such a diquark would break
   if the momentum mismatch is too large between the
  two constituents,  and an
  additional meson would be created.
  I have checked by introducing a `reasonable' parameter
  for diquark break-up
   that practically the same results are achieved from rope fragmentation
  as in the parameter-free default procedure.

\begin{figure}

\centerline{\hbox{
\psfig{figure=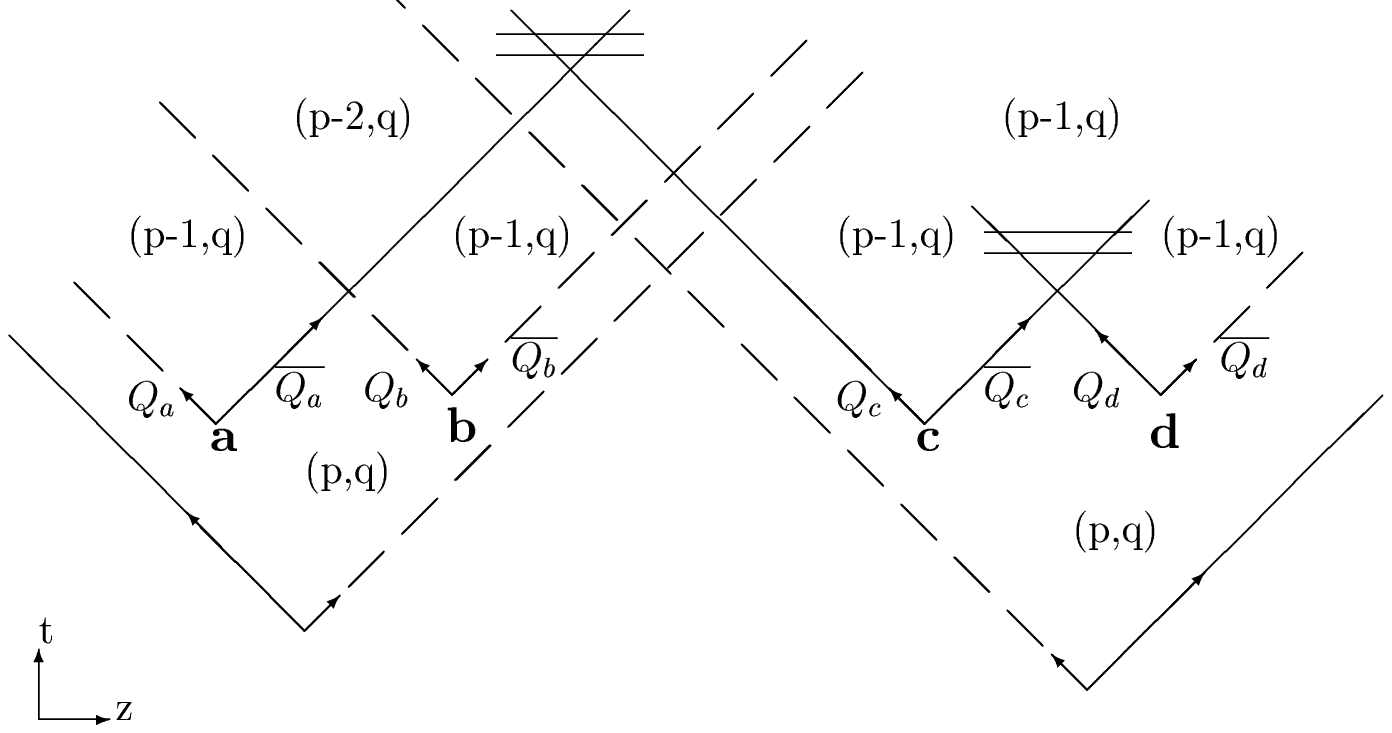,width=17.0cm,height=24cm}}}
  \caption[
       	  ]
         	    {
   \label{ropedeconf}
               		    }
 \end{figure}

 The third process of field degradation, crossing  trajectories
   of two color charges which do not form
  a singlet, has no analogue
  in an elementary string decay.
 It had also not been considered in early studies of
 the materialization time after which the rope field
 is degraded to zero \cite{BCZ86}.
The two crossing charges need  not  form a white
state as long as
there are other color charges available which can
neutralize their charge (see fig.~\ref{ropedeconf}).
 Of course,
 two  triplet or two antitriplet charges always cross
 without being able to form a color singlet.
The probability that quark and antiquark form a white state
 is given in the model as
  (1+$p'\cdot q'$$)^{-1}$, with $p'$ ($q'$) being the
 number of
 3 ($\bar{3}$) charges available to
 neutralize the
 antiquark (quark) charge after
  crossing.
This probability becomes very small for regions of
high field strength.
 Thus color is confined globally in a rope, but not
 locally.
In contrast,
 color charges are neutralized always locally
 in a fragmenting string.

 The quark pairs are produced in a rope with zero longitudinal
 momentum, but afterwards they are accelerated
 in the force field of the outer charges
 (see fig.~\ref{ropedeconf}).
The accelerating force
is given by the difference between
the  rope tensions before and after
pair production $\kappa$-$\kappa'$,
 the same force which is driving the pair creation
 process.
 The classical trajectories are calculated neglecting
  the finite quark masses as it is usually done
  for string fragmentation.
  Therefore all charges are moving with the velocity
  of light all the time. This simplifies the calculation
 considerably and is, in fact, the only lorentzinvariant
 propagation without reference to the global rope rest system.

Finally,  all
original and newly produced quarks
will end up
in a color singlet
with a corresponding partner.
The color singlets are projected onto the basic
hadron multiplets, with the same relative weights
as for string fragmentation.
 The generation of three-quark systems
 and  projection on
 baryon states require some care \cite{SOR95PLB}.
 Independent choice of three
 quark flavors  in the rope endplates tends to  overpopulate
 baryons with quarks of unequal  flavor.
Thus  each chosen configuration is assigned a  proper weight
 to avoid an  unphysical flavor $SU_6$ breaking.
 The positions of the breakpoints are slightly readjusted
 that all  hadron momenta  which are determined from the
 momentum sum of their constituents fulfill the
  mass shell constraints.
  The whole rope fragmentation scheme
  is constructed in a way that total net flavor and
   energy-momentum
  are conserved.
 The hadron formation points are calculated from the quark trajectories.
 A formation point is defined
 defined by
 \begin{equation}
    x^\pm_h =  x^\pm (\mbox{Yo-Yo}) - p^\pm _h/(2\kappa _{el} )
   \quad ,
 \end{equation}
    where  $ x^\pm (\mbox{Yo-Yo})$  is
 the first meeting point of the two quark constituents
 forming a hadron  with momentum  $p^\pm_h$.
  In case of a triplet charges as the source of the rope field
  this prescription reduces to the corresponding definition of
 formation points in string decay  (cf.\ fig.~\ref{hyoyodec}).

It is clear
    from relations (\ref{Gl1})
 that  in the process of rope charge formation
 a triplet  and an antitriplet  charge
  may eventually form a color singlet
 and do not contribute to the total rope
charge (see also  fig.~\ref{su3mltp}).
  \footnote{
    This process was  neglected in the
    first RQMD calculations which included
    rope formation \cite{SOR92}.
    It changes the results from rope fragmentation
    on the order of 5 \%, because the probability
    to form a color singlet by statistical combination
    of 2 randomly chosen (anti-)triplet charges is 1/18.
        }
  In the most extreme case the total charge is
   zero and the region of the rope is field-free!
   The probability that this may happen is rather small,
    however \cite{BIR84}.
  The quarks which form a color singlet in the rope endplates
  are projected onto  hadron  states of the basic
   flavor  $SU_3$  multiplets.
  The momentum  of such a hadron is given from
   the sum of the quark light cone momenta
  (if the hadrons are assumed massless, in the real calculation
   up to a
   small correction).
   The hadron formation point cannot be  determined
   from the rope dynamics to which the quark constituents
   do not contribute
  in this approximation.
 It is usually expected that formation point
   and momentum are related
   in soft production
    processes, approximately
    $x^\pm  \sim p^\pm $, which is used here
    with the scale factor  $\kappa _{el}$.

 The  fragmentation of a single color rope
  as it is implemented in the RQMD model
  has just been described.
  In this picture several flux-tubes are formed
  in very energetic $AA$ reactions which may cover
   the whole transverse area of overlap between the two
   ingoing nuclei.
 How is
   the fusion of strings into several ropes
  realized  in RQMD?
 At first the strings are generated
  independently.
  The hadronization of each string is calculated
   in a Monte Carlo-type fashion,
  and the information is stored.
   This decay is considered as `virtual' until a first
   `would-be' hadron
   of a decayed string is emerging at its formation point.
   All  other strings which have been generated so far
   are now examined whether
   the transverse coordinates of their origin have a smaller
   distance to  this first string than
    the flux-tube radius $r_{tube}$.
   `Transverse' and `longitudinal' refers to the
   coordinates in the rest system of this  string.
   Thus the specification of the
   fusion process is lorentz invariant. Of course,
    the longitudinal direction coincides usually
   approximately with the beam direction.
   Furthermore, it is checked whether the world sheets
   which are swept out by each $Q\overline{Q}$ pair
   during half a period of string motion
    overlap in the $t$-$z$ plane.
    Note that this criterion is trivially satisfied at
    asymptotic  energy, because all string origins
    will be identical in this case.
   If no other string is found which fulfills the criteria
   the  string decay is considered as `real'
    and all hadrons emerging
  from its fragmentation are propagated subsequently.
 All strings which have been accepted
  are collected, and their independent
  decay is rejected. Instead, they fuse into a rope.

  In principle,
   the interactions between  flux-tubes
   can be more complicated than in the model adopted in RQMD, e.g.\
   transverse `communication'
   between different tubes.
   So far such interactions are not taken into account.
   It should be noted, however, that the recent lattice
  results \cite{TRO93} provide some justification for neglecting
   these interactions in a first approximation.
    Trottier  and   Woloshyn have shown that the
     color ropes in lattice gauge theory do not
    expand in transverse dimensions. This result has been
   obtained in the static limit, with an infinite amount
   of time available and with only the nonperturbative vacuum
  outside the tube.

The need to go  beyond independent string fragmentation
 for an understanding of the early stage in ultrarelativistic
  AA collisions is generally accepted nowadays. Of course, the
  appropriate  replacement by more sophisticated approaches
  in line with QCD
  is currently under debate.  It should be added that
the string fusion approach of Refs.\  \cite{AME93,ABFP95}
     based on the dual parton model is  similar to the
   concept which is realized in RQMD.
  A major difference is that in the DPM calculations
   only fusion of at most two strings is considered so far.
   Fused strings  break only as a whole in these DPM Monte Carlos
  while in RQMD stepwise field degradation is introduced.
  The RQMD approach seems better suited in  case that  many
   strings overlap and fuse.
  Nevertheless, possible effects of a different fragmentation
   procedure could be compensated partially modifying the
  fusion strength.
   A direct comparison of results from the two string
  fusion approaches which will be undertaken lateron
   shows rather similar trends for central Pb on Pb collisions
  at CERN-SPS energy. The beam energy may be  too
   low  that  single-particle observables display
    sizable sensitivity to
   details  of string fusion based models.

\section{$hh$ interactions in the rescattering stage}

\label{secresc}
   The interactions in the hadronic resonance gas stage
   are described by binary collisions between hadrons.
   Although preequilibrium processes are present in $AA$ collisions
   \cite{SOR95ZFP}, the interactions are usuallly
  of nonasymptotic type, in the energy region of resonance
    production,  formation and absorption.
    Most of the work in the RQMD framework has been devoted
    to develop a reasonable model for all
    so-called nonexotic reactions.
    In these reactions
    an  $s$ channel resonance can be formed as an intermediate
     state.
   A basic motivation is that the interaction strength
    is much larger here.
     (The energy-momentum entering
    the denominator
      of  the corresponding $T$ matrix element
    can be  near to a pole.)
    Experimental justification comes from large
     differences of cross sections in nonexotic
     versus exotic channels, e.g.\
     $\overline{K}N$ versus $KN$
     or $\pi \pi $ with isospin 0 or
     1 as compared to isospin 2,
     at low and medium energies.
   If   ingoing hadrons  form
    a  state with quantum numbers not allowed
    for a single hadron by the quark model (exotic state),
    the low and medium energy interaction is
    solely given by $t$ channel hadron (Reggeon)
     exchange.
    Only nonexotic reactions do  allow  quark-antiquark
    annihilation in the entrance channel.
    This process is very important for
    the flavor dynamics in $AA$ collisions.
    If the system starts with strangeness below chemical
     equilibrium values, quark-antiquark annihilation
     in nonexotic reactions
     drives the system towards chemical equilibration.
    Therefore this section presents
     a detailed description of  the RQMD model for
      the nonexotic reactions.
   A modeling of these processes is clearly needed,
    because hadronic resonances which are treated
    as quasi-particles may interact themselves.
    For strangeness creation these processes
    are actually much more important than the
    interactions solely between groundstate hadrons.
  \footnote{
   For instance, $\pi N $ interactions are insufficient to explain
   the $\Lambda$ and $ K$ enhancement in S induced
    collisions at 200AGeV,
   in contrast to e.g.\ the conjecture in Ref.\ \cite{CAP95}.
    The strangeness suppression  in $\pi N$ collisions
    at the relevant invariant mass above  1.6 GeV/c
    is  rather similar to $pp$ collisions at 200AGeV.
   }
 It was a very early observation in hadron physics
 that bumps show up in the  energy dependence
  of cross sections which can be identified with the
  excitation of discrete
  Breit-Wigner type resonance states,
   e.g.\ the $\Delta $(1232).
 Hadronic interactions at somewhat larger
  energies are
 dominated by quark exchange and  annihilation
(describable as Reggeon exchange)
  whose energy dependence are given by a negative
 power of $s$.
 At even higher energies
 Pomeron exchange in elastic interactions
   -- exhibiting some kind of universality --
  induces an  approximately energy independent
 total cross section
  for hadron-hadron collisions.
    These  three components
     -- with their characteristic and different
      energy dependences --
     can be  identified with
      three pieces of interaction in RQMD:
      formation of discrete resonance states,
      annihilation into a resonance
      or string  `continuum'
    and   excitation of two (or more) strings
    which was described in the section before.
   The importance
   of each component varies
  for different hadronic reactions.
  Four  classes of interaction
   can be distinguished
   according to
  the number of ingoing (anti-)baryons
  ($B$ denoting baryon, $M$ denoting meson):
  $BB$, $BM$, $MM$ and $\overline{B} B$.
  The charge reversal invariance of strong
  interactions can  be used to generate all other
 interactions
  of antibaryons.
   $MM$ and $MB$ interactions in states with
   nonexotic quantum numbers get some contribution
   from each component. In contrast,
    $B\overline{B}$ annihilation may occur   only
   in the  continuum, because the minimum
   invariant mass is too large.

   A description  of the interactions
   which are
   induced by  quark exchange and annihilation
   has to interpolate smoothly
   from the low energy region of $s$ channel resonance
   formation
  to the high energy interactions which are better
  described in terms of $t$ channel Reggeon exchange.
   Of course, the old discussion about `duality'
   of these two descriptions has
   never lead to  satisfying results \cite{COL71}.
   For the construction of the
   $hh$ interaction in RQMD
     related to quark annihilation and exchange I
   follow a pragmatic approach, adding together a few
    lowlying resonances and a Regge-type parametrization.
  The formation of $s$ channel resonances is calculated
   in the $MB$ and the $MM$ sector
   from  multichannel Breit-Wigner formulae.
  It is assumed for RQMD that
  resonance formation
  determines completely the interactions
  in nonexotic channels
  up to some  CMS energy
   $\sqrt{s_0}$.
    This statement is not
   completely true
   as   will be explained lateron.
   (There are some small corrections
  due to $t$ channel background processes.)
   Above $s_0$  when
    a description in terms of discrete
   resonance levels becomes invalid  --
   the `annihilation'
    cross section
    (for $\pi \pi$, $\pi N$, $\overline{K} N$, etc.)
   is assumed to decrease
   with energy as
    $1/\sqrt{s}$.
  Since the tails of the Breit-Wigner resonances decay  much faster
   with energy, the resulting gap to the total
   annihilation
    cross section
      ($\sigma _{Ann}$)
     is filled by annihilation into a `continuum'
     which serves as the bridge to
     Reggeon exchange,
      at high energy
     supposedly dominant
      over $s$ channel resonance formation.
 \footnote{
   The
    cross section related to Reggeon exchange
   is dubbed here `annihilation' cross section
    for terminological convenience, though
    part of  the physics may really be  quark
     `exchange'. I shall come back to this point later.
   }

   The choice of the exponent -1/2 for  the power of $s$
     in the energy dependence of
      $\sigma _{Ann}$
   is motivated by
    the existence  of energy thresholds in
     $2\rightarrow 3,4,\ldots $ reactions. They
      can be
     described by multiple  Reggeon exchange  \cite{COL71}
     and  effectively increase
    the  maximum value -1 for the power  of $s$
     in  2$\rightarrow $2 Reggeon exchange processes
    as determined  from the  $\varrho $ trajectory
      (and, of course, from experimental data).
    One can test the assumed energy dependence by looking at
   reactions which are tied to Reggeon exchange and
     {\it not } to high energy processes determined by
     Pomeron cuts.
   For instance,  the
     energy behaviour of the inclusive cross section
    $K^- p \longrightarrow
     \Lambda + X$
     -- one of the rare cases in which flavor tagging
     allows for rather good distinction --
      is consistent  with
    $1/\sqrt{s}$  in the relevant energy region
       ($s> 4$GeV$^2$).

  Since the annihilation cross section decreases
  with energy, a gap to the total cross section
  ($\sigma _{tot}$)
  opens up.
  $\sigma _{tot}$  is either given from experimental data
   or calculated using the additive quark model
    \cite{SOR93ZFP}.
   The physics filling the gap is the high energy component
   of $hh$ interactions,
   Pomeron exchange for high energy elastic interactions
   and Pomeron cuts in  inelastic collisions \cite{AGK74}.
  It follows from the considerations above that
  the energy dependence of
   the high energy component
    in reactions with {\it measured} total
    cross sections (e.g.\ $\pi N$, $\overline{K} N$)
    turns out to be  approximately
   proportional to
     (1-$\sqrt{s_0/s}$).
    This functional dependence on energy is used
    to switch on the
    high energy component of the
     total and elastic AQM cross sections
    for reactions with unmeasured cross sections.

  After the general idea
    has been presented
   how the annihilation cross sections
   are  constructed
    as a sum of Breit-Wigner and continuum contributions,
   I am going to discuss
    the various classes of reactions specifically.
    The general scheme is realized  somewhat differently
    in different channels depending on the level of
    experimental  knowledge.

\subsection{
        Formation of  $s$ channel
                              resonances}
 \label{secschres}
Let us turn first to
meson-meson interactions.
Each  of the three vector mesons
 $\rho (770)$, $K^* (892)$ and $\phi (1020)$
 decays into two pseudoscalar mesons.
It is experimentally well-known that
 $\rho (770)$ and $K^* (892)$
 dominate the phaseshifts
 in the $p$ wave for $\pi$$\pi$, $I$=1 and $\pi$$K$, $I$=1/2 scattering
and can be described well by isolated
 Breit-Wigner  resonances above a  small background.
The formation of these 3 and 26 additional
meson
resonances
in the mass region up to 1800 MeV/$c^2$
is taken into account in RQMD.
 Note that not all $SU_3$ flavor
 nonets are complete due to lack of
 experimental information
 about resonance masses and widths. However,
 all
 lowlying multiplets --
the scalar, the two axial vector and the
tensor mesons -- are completely included.
 The groundstate mesons
 and the resonances of these multiplets
 are propagated explicitly in RQMD, while
 the other resonances do appear only in the
 intermediate states.

Godfrey's and Isgur's
quark model calculation \cite{ISG85}
 which is rather successful in explaining
 measured branching ratios of resonance
 decays
has been taken to extract resonance couplings
to unmeasured decay channels.
Partial decay widths
of resonances which can decay only off-shell
into a $\phi $,
have been related
to measured branching ratios  employing
flavor $SU_3$ symmetry and a
correction for phasespace kinematics.
This applies to  the $K^*$ resonances
of the lowlying tensor and
two axial vector nonets.

In RQMD Breit-Wigner type
multi-channel cross sections
for $s$ channel  resonance formation
are usually
 summed up
incoherently.
Exceptions  are  $\pi$$\pi$, $I$=0 and
 $\pi$$ K$, $I$=1/2 reactions
 in  the $s$ wave.  Here  interferences
 are taken into account, because
 very broad resonances  are present in
 these channels
  ($f_0 $(1400) and $K_0^*$(1430))
  which interfere with
  a strong background from attractive interactions
 \cite{MAR76}.
  In addition,
  the  narrow
  $f_0 $(975) state interferes destructively
   in the scalar-isoscalar channel
   with the other  $f_0$ resonance and a background.
  The phases of resonances and background are added
  in the elastic channels
 \begin{equation}
   \label{deltasum}
    \delta _0 = \delta _b + \sum _{R} \delta _R
   \quad ,
 \end{equation}
  which is the appropriate way to get an unitary $S$ matrix
 as long as only one channel is open.
  The background phase $\delta _b$ can be calculated from
   one-meson exchange in the $t$ channel
  employing the $K$ matrix formalism
   (see the subsection \ref{mmtexch}).
   $\delta _R
    = {\rm arctan}\left( {\Gamma_{el}/2 \over m_R-\sqrt{s}}\right)$
   is the phase attributed to each resonance.
 In particular, unitarity is important for
  the $\pi \pi$ interaction
   in the scalar-isoscalar channel. Just below  invariant
  mass of 1 GeV the  phaseshift goes through $180$ degrees
  due to the strong destructive interference effects.
 If more than one channel opens up with increasing energy
  (in $\pi \pi$ interactions the $K \bar{K}$ channel),
   the situation becomes even more involved.
  In order to ensure continuity of the $S$ matrix
  across the thresholds
   eq.~(\ref{deltasum}) is also used above particle production
   threshold.
  The inelasticity parameter $\eta$ in
  the diagonal elements of the $S$ matrix
   $S^J_{ii}=\eta_i \exp \left(2 i \delta _i \right)$
   is calculated from the unitarity condition
 \begin{equation}
    \sum _k S^{J*}_{ki}  S^J_{ki} =1
 \end{equation}
 with the (small) non-diagonal elements given from
   addition of Breit-Wigner-type $T$-matrix
  elements.
   No relative phase between resonances and background
   is introduced
  which would allow the construction of
  a completely unitary
   multi-channel $S$ matrix. This is possible
  in the framework of the Davies-Baranger formalism
 \cite{DAV62}. The results
   in the scheme adopted here and
    in this more complicated unitarization approach
    are rather similar,
  however \cite{BER94}.

  In all meson-baryon interactions
       with  quantum numbers
      allowed  for a baryonic state from
      flavor $SU_3$
   resonances may be  formed.
   So far, however,
     $\Omega ^*$ formation  is neglected
   in RQMD.
    The properties of
   nonstrange baryon resonances
   -- with isospin 1/2 ($N^* $)
    and 3/2 ($\Delta ^* $)  --
  have been
  experimentally well explored in
  $\pi N$ collisions, and of
   hyperon resonances (with isospin 0 and 1)
   in
  $\bar{K} N$ interactions.
   The knowledge about multiply strange baryon resonances
    ($\Xi ^*$ and $\Omega ^*$) is poor, however.
  A generalized Breit-Wigner formula
  for the cross section
   is used to calculate the
        $s$ channel
        resonance
   formation probabilities
  (for $N^*$, $\Delta ^*$ and $\Xi ^*$):
\begin{equation}
   \label{mbbw}
  \sigma _{MB\rightarrow B^*} = b(s) \cdot
    \frac{\pi}{p^2} \cdot
       \sum_{R}
    \frac{(2J_R+1)}{(2S_{1}+1)(2S_{2}+1)} \cdot
                 \frac{\Gamma_{R}(MB)\cdot \Gamma_{R}(\mbox{tot}) }
      {(\sqrt{s}-m_{R})^{2} + \Gamma_{R}(\mbox{tot})^{2}/4}
       \quad.
\end{equation}

 $b(s)$ is a normalization factor
 which  renormalizes
   -- for  $N^*$ and $\Delta ^*$ formation  --
 the Breit-Wigner sum to a given
 absolute cross section.
   (For $\Xi ^*$  formation
    incoherent addition gives $b(s)$=1.)
  The renormalization is done  for  nonstrange baryon resonances,
    because the
  total cross sections in $\pi N$ reactions are known.
 The total resonance formation cross section
   calculated with eq.~(\ref{mbbw})
  in $\pi N$
 is actually the total
  $\pi N$ cross section minus an elastic background
 (up to 5 mb)
  determined by a consistency
  condition that the sum of background and  elastic resonance decay
 equals the measured elastic $\pi N$ cross section.
      $\Gamma_{R}(MB)$ denotes a partial decay width into
      the channel with meson $M$ and baryon $B$.
The decay widths are $\sqrt{s}$-dependent via
the relative momentum in the CMS of the decaying
resonance $p$
\begin{equation}
  \label{gammap}
   \Gamma_{R}(MB) \sim
      m_R/\sqrt{s} \cdot
      p^{(2l+1)}/(1+0.2 \cdot \left(p/p(m_R)\right)^{2l})
    \hspace{2em}.
\end{equation}
  This ensures a correct threshold  and high energy
  behaviour of a particular
 channel.
 The particular
  resonance
 which is formed in a  meson baryon annihilation
  is chosen randomly with a weight given by the corresponding
  term
 in the sum of eq.~(\ref{mbbw}).

 In the
   $N^*$ and the $\Delta ^*$ channel
the sum
   in eq.~(\ref{mbbw})
 runs over all
resonances
with a mass below 2  GeV/$c^{2}$.
The resonance masses and decay parameters
as implemented in RQMD
 \footnote{
  Some of the resonance
  parameters have been changed slightly
  as compared to the values given in
 \cite{SOR93ZFP}, because
   they were refitted
    together with the modified continuum component
   to  $\pi N$ data.
          }
are in accordance with the
 listing of the
Particle Data Group \cite{PDG92}.
  It has been checked that an application to $\pi N$
   collisions reproduces approximately the
   hyperon-kaon production cross sections and
   the total pion yields in inelastic collisions
         	   \cite{SOR93ZFP}.

 In the
   $\Xi^*$  channel
         formation  cross sections
      for 6 discrete
          resonance states
        are  incoherently added
            with
         couplings
        to  meson-baryon states
      given from
        $SU_3$ flavor symmetry.
         (Note that flavor $SU_3$ is broken by
         different $M$ and $B$ masses in each
         octet.)
         The
          $F$ and $D$ parameters
          specifying the strength
          of symmetric and antisymmetric
          coupling
         are taken from the literature
            \cite{SAM74}.

 Resonance formation
  as given in eq.~(\ref{mbbw})
  is not realized
 for hyperon resonance formation
  ($\Lambda ^*$ and $\Sigma ^*$).
  Instead,
    the experimentally measured
   exclusive cross sections for
     charge exchange ($K^- p \leftrightarrow \overline{K}^0 n$),
     elastic interactions and hyperon production
      ($\overline{K} N \leftrightarrow \pi \Lambda$/$\pi \Sigma $)
     have been  tabellized in RQMD.
   The $\Sigma ^*$ resonance parameters are not
     determined very well.
   Furthermore, at low energy
    the  $\overline{K} N $ interaction is
    rather complicated, e.g.\  due
    to the existence of interfering resonances.

  The lifetime of formed resonances is stochastically
   chosen according to an exponential decay law. The average
   lifetime (without collisions) is given from the
   inverse decay width.
   The angular distributions in decays of formed resonances
   are certainly an area in which more work is needed.
   All transport models for $AA$ collisions so far
    had ignored this problem,  which is already present
    in the `prototype' reaction
  $\pi N \rightarrow \Delta (1232) \rightarrow \pi N \ $,
   by  decaying the resonances isotropically.
   While the problem can be easily solved for isolated
   resonances, it becomes more involved for overlapping
   resonances. The different partial waves
   should be summed coherently which would generate a
   diffraction-type peak at small scattering angles.
   As a first step isotropic resonance decay is
   supplemented in RQMD by a second component,
   a Gaussian $p_t$ distribution. This  ensures that
   the average $p_t$  which is generated does not exceed
   the standard  value of  400 MeV
   which is usually assigned to a produced quark pair
   in soft production processes at high energy.

\subsection{
  Transition from $s$ channel resonances
   to Reggeon exchange
                         }

 \label{secresregtr}

   The formation of discrete $s$ channel
   resonances,
   which represents the low energy component of
    quark-antiquark annihilation
  in $hh$ interactions,
   has just been described in some detail.
   Such a description is expected to break down
  in the intermediate
 CMS energy region $s=4-16 \mbox{ GeV}^2$.
  Resonance level
 densities are getting too high and
  resonance widths too large that
 a formulation in terms  of discrete
 resonance  excitations could be meaningful.
 Furthermore,
  due to these difficulties
 the empirical information about discrete
 states
   gets poorer
  with increasing invariant resonance mass.
   On the other side,
    Regge theory is
    spectacularly successful
    in this energy region.  I do not discuss here
    the speculation that an infinite tower of
   $s$ channel resonances could effectively
    generate the same interaction as
    $t$ channel Reggeon
    exchange (see e.g.\ the Veneziano model \cite{VEN68}).

   The main emphasis is put in this subsection
   on a discussion
     of  $B$=1 channels. The
     continuum component in $MB$ reactions is of
    utmost importance for nucleus-nucleus collisions due to
     preequilibrium processes
    \cite{SOR95ZFP}, while it is suppressed kinematically
     in $MM$ collisions.
   For instance,
    here the processes are  above threshold for
    $s\bar{s}$ creation.
  There is one more reason to  model  carefully
    $MB$ interactions in this energy region.
    Strange (anti-)baryons are a very promising signature
    of collectivity in
   ultrarelativistic  nucleus-nucleus reactions.
    After they are produced
    in an $AA$ collision, they have to propagate eventually
    through a cloud of nonstrange mesons \cite{SOR95PLB}.
          The strength of transition rates like
              $  \Lambda \rho
           \leftrightarrow
         \overline{K} N $
         or
              $  \Xi K
           \leftrightarrow
           \overline{K} Y$
           is determined by
          the probability  of
          moving  the $s$ quark
           from the baryon to the meson side
        (and vice versa).

 The total strength and
 the energy dependence of the Reggeon
 exchange component follow from the ansatz
  which was introduced  above:
\begin{eqnarray}
    \label{xsregge}
  \sigma _{Regge} (s)
    & = &
  \left(
    \sigma _{tot} (s)
    - \sigma _{BW} (s)
       - \Delta \! \sigma (s) \right)
            \hspace{3em} \mbox{for }
              s \le s_0
       \\  \nonumber
    & = &
  \left(
    \sigma _{tot} (s)
    -  \sigma _{BW} (s)
       - \Delta \! \sigma (s) \right)
         \cdot
         \sqrt{s_0/s}
            \hspace{3.0em}
              s > s_0
\end{eqnarray}
    for $\pi N$, $\overline{K} N $
   and $\pi \Xi$ interactions.
   Eq.(\ref{xsregge})
     is applied for
    $MB$ states of good flavor quantum numbers
    (isospin, hypercharge).
    Interference effects of  amplitudes
     in different  states are neglected.
  The $\sqrt{s_0} $ parameter is set to
 1.8 GeV in the $N^*$,
 2 GeV in the  $\Delta^*$ and
  $Y^*$,
   and 2.15 GeV in the $\Xi ^*$  channel.
  $\Delta \! \sigma (s) $
   is given from the
     elastic background cross sections
       for $\pi N$
     or   the sum of the tabelized
     cross sections
      $\overline{K} N \longrightarrow
      \overline{K} N$,
      $\pi \Lambda $, $\pi \Sigma $
       for $\overline{K} N$ collisions.
    According to eq.~(\ref{xsregge}) the
    contribution of the
   Breit-Wigner resonances
    in the $N^*$/$\Delta^*$/$\Xi$ channels
   have to be subtracted
   to get the total `Reggeon exchange component'.
   Note that the tails of the
     $N^*$ and $\Delta^*$
    resonances above $s_0$
    are calculated
   from   eq.~(\ref{mbbw})
     with the $b(s)$ values `frozen'
     to the value of $b$ at $s$=$s_0$.

      No experimental data are available
    for the total $\Xi \pi $ cross section.
     Its value above $s_0$ is taken
          from the additive quark model (AQM).
      Lack of experimental information is the reason
       that fewer $\Xi^*$  resonances are included in
      the Breit-Wigner sum of eq.~(\ref{mbbw})
         than  in  other
        channels. This is `corrected'
       in eq.~(\ref{xsregge})
          by multiplying the
          constant AQM value by
           $\sqrt{s_0/s}$
        for $s$  smaller than $s_0$
        down to the $\Xi ^*$(1530) resonance,
       if the Breit-Wigner sum gives a smaller value
        for the cross section than this parametrization.

   There is no Reggeon exchange component
     in the $N^*$ and $\Delta ^*$ channel
      for energies below $s=s_0$.
     (The annihilation cross section is filled up
      by $s$ channel resonance formation according to
    eq.~(\ref{mbbw}).)
     In the $\Lambda ^*$ and $\Sigma ^*$
       channels the Reggeon exchange component
      is present down to lower energies, because
       the difference of the total  cross
       section   to the sum of the tabelized
       cross sections
      becomes
       nonzero, even below invariant mass of 2 GeV/c$^2$.

   The total Regge exchange component
   as given by eq.~(\ref{xsregge})
   can be subdivided
    into single Reggeon exchange
      (2$ \longrightarrow$ 2) and
    into multi-Reggeon exchange
      (2$ \longrightarrow$ 3,\ldots) diagrams.
    Multi-Reggeon exchange
      diagrams
       with $n>2$-body final state
      may become important to account
     for some fraction of  multiparticle production at higher energies.
     However, such processes are  suppressed in the
     hadronic rescattering stage of $AA$ collisions \cite{SOR95ZFP}.
      Their presence would be indicated by an additional
      component
      to the  produced
       particle yields
         nonlinearly increasing with the number of participant
        nucleons
       for which there is also no
      hint from experimental data.
    Before the relative weights for these two classes
    of processes in RQMD are discussed, it is  useful to
     describe how a
       particular 2-body final state
        in a  reaction  of the type
   $M_i B_i \longrightarrow M_o B_o$ is selected
    out of the sample of all possible states.

    The exclusive
     cross sections
     of the Regge exchange  component
      2$ \longrightarrow $2
        have  been given the
       following functional form:
 \begin{eqnarray}
     \label{mbmb}
     \lefteqn{
       \sigma (M_iB_i \rightarrow M_oB_o) =
        \frac{1}{p^2(2S_{Mi}+1)(2S_{Bi}+1)}
                \sum _I
                 a_I(s)
         \cdot
                }
      \\   \nonumber    &   &
            \left\{
           \left| C_{MB,I} \right|^2
    \left(
     1-\exp \! \left(-(p_{MB}/\sigma)^2 \right) \right)
         \left((2S_M+1)(2S_B+1) \right)
              r(R)
       \begin{array}{c}
          \mbox{} \\
          \mbox{}
       \end{array}
          \hspace{-1.0em}
            \right\} _{in}
      \\   \nonumber    &   &
        \hspace{8em}
         \cdot
            \left\{ \ldots
       \begin{array}{c}
          \mbox{} \\
          \mbox{}
       \end{array}
          \hspace{-1.0em}
            \right\} _{out}
         \left(  1 + p p'/p_R^2 \right) ^{-\Delta \alpha _R}
              \quad .
 \end{eqnarray}

    Eq.\ (\ref{mbmb})
    contains   several factors
    -- in their ordering from the left to the right --
    related to
     incoming flux and spin averaging,
      absolute normalization ($a_I(s)$),
      isospin projection
      (Clebsch-Gordon-coefficients squared),
      threshold  phase space factor,
      internal spin degrees of freedom,
      suppression parameter
     $r(R)$ of strangeness
           creation or quark exchange between $M$ and $B$,
        and a suppression for exchange of Reggeon $R$
         as determined by the intercept of
          the corresponding Regge trajectory $\alpha _R$.
   If
     the absolute normalization
    is  ignored for the moment, the relative probability
      of some final state is determined rather simply.
    Up to some statistical factors which express the `matching' of the
     isospins and the available spin degrees of freedom
     there are only three factors which contain dynamics:
     the phase space factor,  the strangeness
      parameter $r(R)$,  and
    the Regge factor.
      Since the Regge factor tends to 1 at low relative
      CMS momenta $p$ and $p'$,
       the  approach assumes  kind of `universal'
       (or average) coupling of a meson-baryon 2-body
       state to the intermediate system with baryonic quantum numbers.
     The phase space factor cuts down on the production probability
      near threshold, but then also goes quickly to 1.
       Its functional form is motivated by the
         available  phase space   for
         Schwinger type
         $p_t$ creation  in flux-tube decays which is
         Gaussian distributed, with width $\sigma$=400MeV.
     The
      parameter $r(R)$ is assigned a
       value different from 1 only to account for
          $s\bar{s}$  suppression
        (0.3 in a
            connected and
          0.09 in a disconnected quark-line diagram)
       and to  suppress exchange of quarks between $M$ and $B$
        in some processes (0.5 for
       exchange of baryon number or strangeness).
        The first parameter is the usual strangeness suppression
         which is characteristic for soft particle production.
        The last  parameter value  reflects the  immobility
         of  the heavier  strange quark and of  the
         leading spectator diquark.
      The model presented here
      is very similar  in its spirit
     to Vandermeulen's statistical approach
     \cite{VAN88}, and Mundigl, Vicente Vacas and
     Weise's doorway model
     \cite{MVW91}, both for meson production in
    low energy $\overline{p}p$ annihilation.
      This is not really a surprise, because the common denominator is
      the overwhelmingly large number of open states
      which precludes any kind of detailed dynamical treatment.

         Some additional
           strangeness suppression
          is `hidden' in the
            Regge factor which is smaller for strange
           Reggeon exchange.
  It is a general observation
  that the strangeness suppression
  in nonexotic $hh$$\longrightarrow $$hh$ interactions  is somewhat
   stronger
   -- in addition to phase space suppression --
   than `asymptotically'
   in soft hadronic physics
   ($P(s)/P(u)$ is $\approx 0.3$ for multiparticle production).
    See e.g.\
   the suppression parameters for
     kaon channels in
   $\overline{p}p$
    annihilation \cite{VAN88,MVW91}
   which are needed to fit the experimental data.
  A stronger $s\overline{s}$ suppression
   is expected from microscopic calculations which
 give  a larger suppression for heavy mass quarks
 if a color flux-tube
    has a small length  or
   breaks very early  \cite{HER90}.
   However, it cannot be ruled out that part of the
   stronger suppression effect
    comes from the neglect of quark exchange
     diagrams.
   Their inclusion -- in reactions without
   strange  quark in the initial state --
    favors nonstrange final states.
    It has been found for $MB$ interactions that
     an average strangeness suppression parameter of
    0.2 can reasonably well describe total strangeness
     creation in $\pi N$ reactions \cite{SOR93ZFP}.
     Therefore not too much freedom is left
   for a pure quark exchange component
     in such reactions.

   The Reggeon exchange suppression factor
   contains the momentum $p_R$ as a scale
    at which
         Reggeon exchange becomes important.  Thus the
         parameter choice
         reflects the interpolation from the
       $s$ channel resonance region to the
          high energy Regge approach.
       $  \Delta \alpha _R $ is defined as
 \begin{equation}
     \Delta \alpha _R :=   2 - 2 \alpha _R - \alpha _0
      \quad .
 \end{equation}
   For each
   $M_i B_i $$\longrightarrow $$M_o B_o$ reaction
   only a single Regge trajectory is taken into account,
   the one with largest intercept.
   The vertices which are used in the RQMD calculations
    are presented in tables \ref{reggea}  and \ref{reggeb},
   together with the parameters for the Reggeon intercepts.

    The  multi-Reggeon  component is represented in the model
      by a $hh$ annihilation into a color string which
       decays into more than two hadrons
      (resonances) according to the standard string fragmentation
     scheme.
      In principle, the probabilility
      whether  a 2-body final state
       ($2\rightarrow 2$ process) or
      an $n$-body final state ($n>2$) is chosen
       in  a Reggeon exchange process
       is determined by
      the normalization factors  $a_I(s)$   in eq.~(\ref{mbmb}).
     However,   $a_I(s)$ is  not specified  explicitly in RQMD.
   Instead, the correct Regge behaviour for 2$\rightarrow$2
    processes at large energies provides an implicit normalization
    of the total  2$ \rightarrow $2 Reggeon-exchange
    cross section.
   Regge theory requires
    all 2$ \rightarrow $2
    cross sections to  behave (approximately) as
   \begin{displaymath}
     \sim   s^{2\alpha _R - 2}
   \end{displaymath}
    at  large  energies.
    With the parameter choice $\alpha _0 $=1
    the cross sections of eq.~(\ref{mbmb})
  show  this behaviour under the assumption that
      $a_I(s)$ approaches a constant value at high energies.

 The correct asymptotic behaviour
  for  2$\rightarrow $2 transitions
      is realized in the following way.
     For invariant masses
    of the colliding
      $M_i B_i $ pair
     below some value $\sqrt{s_1}$=2.8 GeV
      (plus 0.15 GeV for each s quark in the
       $B^*$ state due to flavor symmetry breaking)
     it is assumed that all final states are 2-body.
     Above $s_1$ the 2-body final state is suppressed
      by an additional power 1/2
\begin{equation}
    \label{regge22}
  \sigma _{Regge} (2\rightarrow 2)
        \le     \sigma _{Regge} (s) \cdot
               \sqrt{s_1/s}  \quad ,
\end{equation}
  and the appearing gap to the total
  Reggeon-exchange cross section
   is filled up by choosing an $N$$>$2-body final state.
     The $\le$ sign in eq.~(\ref{regge22})
     is to be understood  in the sense that
  $\sigma _{Regge} $(2$\rightarrow $2) is
     assigned  at first
     a value
       which is equal to the right hand side of
     eq.~(\ref{regge22}).
   The  normalization according to
     eqs.~(\ref{xsregge}) and  (\ref{regge22})
   would give
    asymptotically a $1/s$ dependence for
  $\sigma _{Regge}$(2$\rightarrow$2)
    if the equal sign in  eq.~(\ref{regge22})
     would hold.
  This is actually the correct energy dependence if
     Reggeons of the $\varrho $ trajectory
      with the largest intercept are exchanged
     (the $\alpha _\varrho \simeq 0.5 $).
    On the other side,  cross sections for other transitions
   $M_i B_i \longrightarrow M_o B_o$  may drop faster with energy
   due to a smaller intercept parameter
  $\alpha _R$ of the exchanged Reggeon.
     In order to avoid that
    cross sections characterized by a
    comparably larger  intercept $\alpha _R$
    simply blow up due to the normalization condition
    in  eq.~(\ref{regge22}),
     a fraction of the total
        2$\rightarrow $2  Reggeon-exchange cross section
      goes actually into many-body final states.
     This  will be explained in the following.

  After the decision to choose a 2-body state has been made
  in RQMD, a particular state has to be selected,
   respecting its relative weight according to eq.~(\ref{mbmb}).
   This is realized in a Monte Carlo-type fashion using the
   rejection method.
   After some randomly chosen out-state has passed all other
   acceptance tests, the Reggeon factor
\begin{equation}
  \label{reggefactor}
     p_{Regge} (s)=
      \left(  1 + p p'/p_R^2 \right) ^{-\Delta \alpha _R}
\end{equation}
 is taken  as a probability for  final acceptance
 of the 2-body state.
 Below $s_1$, Monte Carlo type rejection of a 2-body channel
   is followed by choosing another 2-body state until
     at last an out-state passes all acceptance tests successfully.
    Above $s_1$, the
     acceptance probability is again given by $p_{Regge} (s)$.
     However, the probability to examine another 2-body state
     is frozen at the value 1-$p_{Regge} (s_1)$.
      As a third possibility, an
      $N$$>$2-body final state may be chosen,  with probability
  \begin{displaymath}
     \mbox{probability}(N>2)=
               p_{Regge} (s_1)-p_{Regge} (s)
            \hspace{3em} \mbox{for } s> s_1
     \quad .
  \end{displaymath}
      This prescription
       guarantees  the continuity of all cross sections
       across $s_1$
     and  the correct energy dependence
     of all 2$\rightarrow$2 cross sections
      asymptotically.

  Which hadrons are included in the in- and out-states
    of eq.~(\ref{mbmb})?
  All  nonstrange baryon resonances
   with a pole mass  below 2 GeV/$c^2$
   whose existence is
   established by the Particle Data Group can be
    produced.
    All baryon multiplets with these $N^*$ and
   $\Delta ^*$ states have been completed by including
     strange baryon resonances
     either according to the data tables,
      flavor $SU_3$ relations
       or by applying a
     simple formula for resonance mass
    ($ m= m (N^*/\Delta ^*) + n_s\cdot 150 MeV$)
     and width
    ($ \Gamma _R= d_R \Gamma (N^*/\Delta ^*) $
     with $d_R$  equal  0.85 for $\Sigma ^*$, 0.62 for $\Xi ^*$
     and 0.45 for $\Omega ^*$).
    Thus an unwanted
   flavor $SU_3$ breaking
   due to the poorer experimental
   knowledge of resonances in the strange resonance
  sector is avoided.
   In a second step the formed resonance states decay
   subsequently.
   The  branching ratios of those resonances whose
    properties are not taken from the data tables or
    flavor $SU_3$ relations have to be specified.
   They are determined similarly as
   given for the intermediate doorway state
  in eq.~(\ref{mbmb}), however, without Reggeon factor:
 \begin{eqnarray}
     \label{Bstarmb}
     \lefteqn{
       \Gamma (B^* \rightarrow MB) \sim
                }
      \\   \nonumber    &   &
           \left| C_{MB,I} \right|^2
    \left(
     1-\exp \! \left(-(p_{MB}/\sigma)^2 \right) \right)
         \left((2S_M+1)(2S_B+1) \right)
              r(B^*)
              \quad .
 \end{eqnarray}
Here the parameter
 $r(B^*)$ is  related to strangeness again.
   Its value is set to 0.2
  if $s\overline{s}$ is created in the
  decay (1 otherwise).
   Of course, strangeness creation in a
    second  step  is much more improbable than
    in the first stage of interaction given by
    eq.~(\ref{mbmb}).

   With the combined  model
    of $s$ channel resonance formation and
     $t$ channel Reggeon exchange
    one can describe reasonably well on the order of 60
    measured exclusive
    meson-baryon reactions
   ($\pi N$ and $\overline{K} N$),
      including their energy dependence.
      Usually the   results
       agree  with
      exclusive 2 and 3-body
      data better than within a factor of two.
     Inclusive yields, especially of strange hadrons
         (kaons, hyperons, etc.), are described even much
       better (within 10 percent).
   The constructed model does not aim for a `perfect' fit to
    experimental data. Instead, the intention was to
      introduce only a minimum number of parameters
     whose values have some physical meaning like
      strangeness suppression, the scale which separates
      the resonance from the Regge region and so on.
    Therefore the model can be meaningfully extrapolated
      to  ingoing $hh$ combinations with experimentally
     unexplored  properties of their interactions.
     No new parameters have to introduced if the
      $\pi$ or $\overline{K}$ and the nucleon
        are replaced by other meson-baryon combinations,
     in particular  resonances.
      Their low and medium-energy interaction are completely
      specified by the same model, essentially
       given by
     eqs.~(\ref{mbbw}) and (\ref{mbmb}).
       The interactions fulfill the property
       of {\it  detailed balance}.
       Detailed balance follows from
          the time reversal invariance of
      strong interactions.
        A cross section $\sigma _{io}$ for a
   $M_i B_i \longrightarrow M_o B_o$ transition is related
      to the cross section  in the reverse direction by
   \begin{equation}
         \sigma _{oi} =
        \frac{p^2(2S_{Mi}+1)(2S_{Bi}+1)}
             {p'^2(2S_{Mo}+1)(2S_{Bo}+1)}
            \sigma _{io}
          \quad .
   \end{equation}
 Thus the model
   avoids the pitfalls of equating cross sections
  with ingoing resonances to groundstate hadrons
   as used in early RQMD calculations
  \cite{SOR89}
   and in other transport models \cite{WER93,ARC92}.

 After the discussion about the quark annihilation processes
   in the continuum has  focused so far on
   $B$=1 channels, I would like to discuss shortly its role in
   channels with total baryon number zero.
   Such  reactions can be either $\overline{B}B$ annihilations
   or $MM$ interactions.
   The $MM$
    annihilation into a
   continuum  is constructed very similarly to the approach
   outlined above for
    $MB$ interactions. The continuum annihilation
     starts to open up
   at $s_0$=((1.6+$n_s\cdot$0.15) GeV)$^2$, because
    the Breit-Wigner resonance tails decay faster than
    $s^{-1/2}$ which again is the assumed energy dependence
    of the total annihilation cross section.
   Note that the  coupling between 2-body
   in- and out-states is not reggeized yet.
   Instead, the couplings are constants with
     strangeness annihilation and creation suppressed
    by the factor 0.2 as in \cite{SOR93ZFP,MVW91}.
  A parametrization  of the $\overline{p}p$
  annihilation cross section is  used
  which was fitted to experimental data \cite{KOC89}.
   The construction of
  the other $\overline{B}$$B$
  absorption cross sections is described in \cite{SOR95ZFP}.
  In principle, one should put  the same effort into a detailed
  description
    of  the out-states
   as   discussed for $MB$ interactions.
   However,
   the choice of the outgoing states
   in the continuum region is of minor importance
  for the reaction dynamics of heavy ion collisions.
   $MM$ collisions around an invariant mass of
    2 GeV/$c^2$
   are rather rare
   in $A$$A$ collisions \cite{SOR95ZFP}, because too much
    internal kinetic energy is required.
    Clearly
    $\overline{B}B$ collisions are not
   kinematically  suppressed  in this invariant mass region.
  Therefore the final
    $\overline{B}$ yield is very sensitive to the
    annihilation strength. However,
   it matters only whether an antibaryon survives in
    baryonic matter or not.
    The specific choice for the
   particle production  model in
    $\overline{B}B$ annihilations
    has not much influence on the meson dynamics
   itself. Note that
  the typical
    $\overline{B}/M$ ratio in central $AA$ reactions
    at 200 AGeV is on the
    one percent level only.
   So far
  the RQMD procedure
   for the choice of
   the outgoing states
   in  the $B=0$ system
   is  to form a mesonic string
   which decays according to the standard rules of string
  fragmentation.
    In addition, it is enforced that
    $G$ parity is conserved in  nonstrange meson
     transitions by rejecting out-states
    with wrong $G$ parity.

 Particle production
   at energies in the `continuum' region, i.e.\
  beyond the regime of
  identified  resonance decays,
  can be
  naturally viewed as a
 tunneling process of quark pairs
 leading to subsequent breaking of  a
 color flux-tube
   \cite{CNN79}.
  Thus the annihilation processes in the continuum
   are translated in RQMD into the language
   of (Yo-Yo type) string decay.
  This  determines the
   space-time structure, most importantly the
    formation points
    of  outgoing hadronic states.

\subsection{
   Meson exchange processes   in the $t$ channel
                         }
 \label{mmtexch}
 Here I am going to discuss the component in
     RQMD which describes
      $t$ channel exchange driven processes
    in $MM$ interactions with two mesons or resonances
    in the final state.
    Meson-exchange processes
    in $MM$ interactions are generated by
    a flavor $SU_3$ symmetric Lagrangian which couples
    a vector meson to two pseudoscalar mesons.
  \footnote{
    In addition, RQMD includes parametrizations of
    cross sections involving nucleons and the $\Delta $(1232) resonance
    which were calculated using the One-pion-exchange
    model (OPE) \cite{FER61,DIM86},
   e.g.\ $NN \leftrightarrow N\Delta$,
     or OPE-inspired
    parametrizations of some measured flavor creating processes
     like $NN \rightarrow \Delta (1232) Y K $ \cite{RAN80,MAT91}.
    These processes are of not much importance in
     200AGeV collisions, the topic of this paper.
    }

   The empirical knowledge about meson-meson interactions
   is rather poor due to the inherent difficulty to prepare
   a mesonic `target'.
    Only $\pi \pi$ and $\pi K$ interactions are
     well explored. It has been outlined above that
    formation of the $\rho $ and $K^* $ resonance
    are dominating the interactions in these channels.
   The decay of the
   vector meson resonances into two pseudoscalars is
    approximately described by a
  flavor $SU_3$ symmetric interaction term in the Lagrangian
\begin{eqnarray}
\label{pspsv}
 L_{\rm int}& = &
     -({\rm i}/2) G_{V} {\rm Tr}([\hat{P},\partial_\mu \hat{P}]
   \hat{V}^\mu)=
     2G_{V} f_{ijk} P_i \partial_{\mu} P_j V^{\mu}_k
   \quad .
\end{eqnarray}
 $\hat{P}$ and $\hat{V}$ denote the pseudoscalar and vector
 meson matrix \cite{GAS66}.
 Flavor symmetry is broken
 only by the mass differences between mesons within the multiplets.

\begin{figure}[h]

\centerline{\hbox{
\psfig{figure=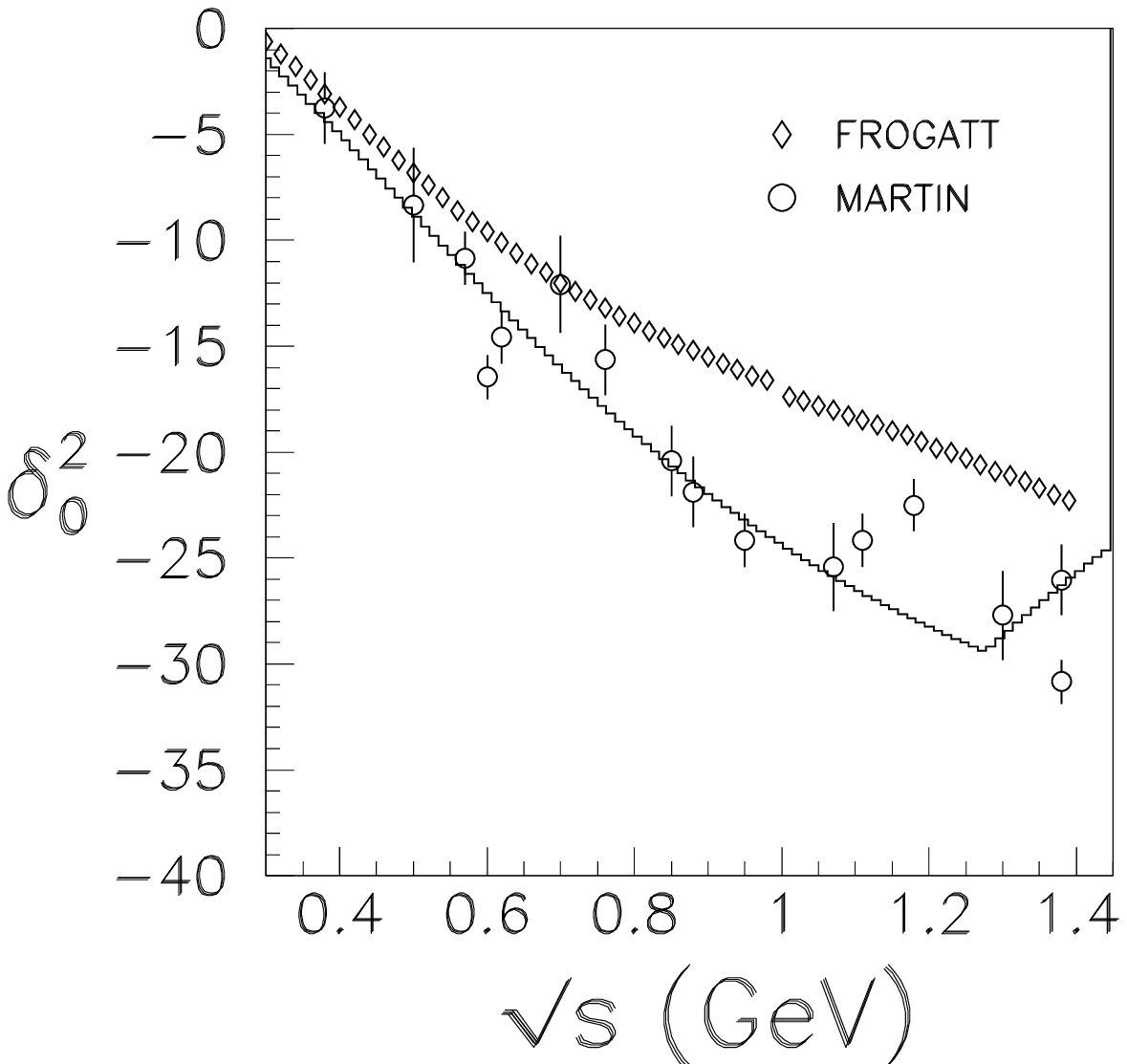,width=8cm,height=8cm}}}

\centerline{\hbox{
\psfig{figure=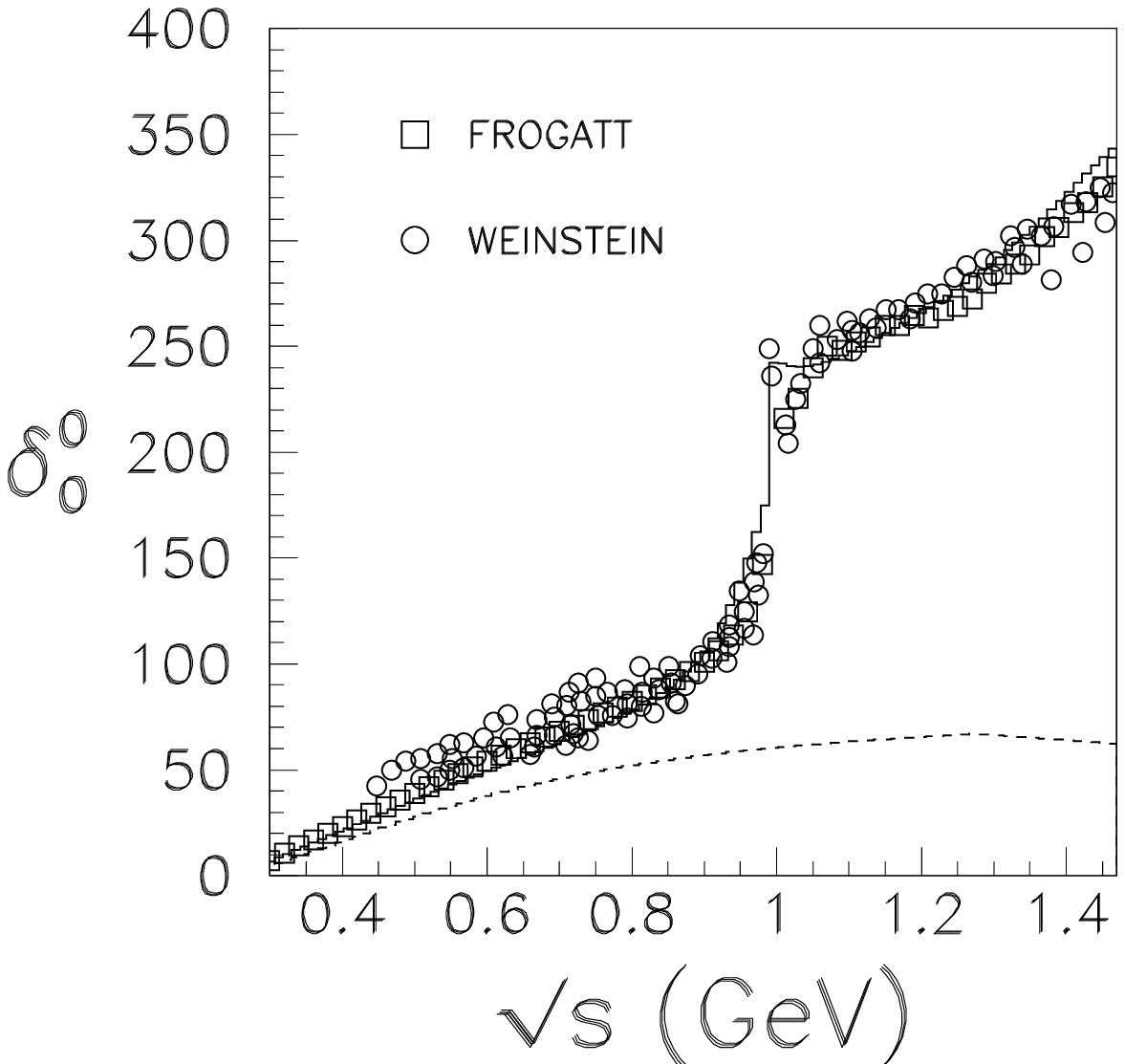,width=8cm,height=8cm}}}
  \caption[
       	  ]
         	    {
 \label{phshpipi}
               		    }
 \end{figure}

In addition to vector meson decay,
 the  $SU_3$ invariant Lagrangian
 in eq.~\ref{pspsv}  generates
  transition rates for  reactions
  $PP\rightarrow PP$, $PV\rightarrow PV$ and
  $PP\leftrightarrow VV$ by exchanging a meson in the
  $t$ channel.
  $G_V$ is fitted to the $\rho $ decay width and set to 3.02.
  All relativistic Born diagrams for flavor changing processes
   and for all important
    exotic channels
   (which have no other interactions at low energy)
   have been calculated and implemented
  in RQMD.
 The interaction vertices are supplemented with monopole form factors
\begin{equation}
     f(t)  =  {\Lambda^2-m^2 \over \Lambda^2-t}
\quad,
\end{equation}
 which express  the finite size of the interacting mesons.
 The  wrong energy dependence in reactions with
  vector meson exchange is `cured' by multiplying
  each PPV-vertex  with
$(s/s_0)^{(\alpha_0-1)/2}$ for $s>s_0$
 ($\sqrt{s_0}$=1GeV+$m_1$+$m_2$).
 $SU_3$ invariant values for the cutoff-parameter $\Lambda $
are chosen,
$\Lambda$= 3.0 $GeV^2$ (1.5 $GeV^2$) for attraction
 (repulsion) in the $s$ wave.
 The calculated interactions
  give remarkably good agreement with the  measured
  phase shifts in
  $\pi \pi $ and
 $\pi K  $ interactions \cite{MAR76},\cite{FRO77}-\cite{AST88}.
 As an example the $s$ wave phaseshift is shown for
  $\pi \pi$ scattering with  I=0 and I=2
 (cf.\ fig.~\ref{phshpipi}).
 Note that the I=2 channel is exotic, while the
  I=0 channel contains  additional contributions
  from $s$ channel resonances, in particular the
  $f_0(975)$ and the $f_0(1400)$.
 A completely unitarizing scheme
  -- eq.~(\ref{deltasum}) below $K\overline{K}$ threshold
   and its generalization, the Davies-Baranger formalism,
   above threshold --
  has been
 applied to calculate the combined effect of $t$ channel background
 and $s$ channel resonances.
 The real-valued matrix elements
   for one-meson-exchange have been identified
   with the $K$ matrix which has been decomposed into
  their components from different partial waves.
   The background $S$ matrix for partial wave $l$ is
  constructed from the $K$ matrix via:
\begin{equation}
\label{kmeq}
    S_{bg}^l=(1+{\rm i}K^l)(1-{\rm i}K^l)^{-1} \quad .
\end{equation}
 While a standard parametrization is used
 for the  $f_0(1400)$ decay width,
 the $K\overline{K} $ molecule picture is employed for
  the $f_0(975)$  \cite{WEI90}.
 Furthermore, the $f_0(975)$ decay width into
 $K\overline{K} $ is analytically continued
 below threshold like in Ref.~\cite{FLA76}
 in order to generate the sharp cusp observable
 in the scalar-isoscalar
 phase shift just below $\sqrt{s}$=1 GeV.

\section{
   Flavor production
    in central Pb(160AGeV) on Pb collisions
                         }

 The two basic building blocks
  of interactions in RQMD -- color ropes and
  hadronic rescattering --  have been presented
  in the preceeding sections. In this section
  the consequences
   of these collective
  interactions  for the flavor dynamics in
   the most central collisions of
   two lead nuclei at a beam energy of 160AGeV
   will be discussed.
  The impact parameters
   for the calculated Pb(160AGeV) on Pb collisions have been
     selected between 0 and 1 fm,
    rather central collisions.
  The calculations
   with the RQMD
   computer code have been done in different modes,
   the default mode and two alternative modes.
    The first alternative mode is defined by switching off
    all  collective interactions, no
    rescattering and no rope formation (`NN mode').
    The  second non-default mode is
   without rescattering but rope formation included.

\subsection{
   Particle multiplicities
                         }
  The results of the different RQMD calculations
   for the final hadron  yields are presented
  in table~\ref{pyield}.
  All members of the basic pseudoscalar meson
   nonet and (anti-)baryon octet
   have been kept stable here with the exception
  of the $\eta '$ which decays already during the
  dynamical evolution generated by RQMD.
   Some  extreme scenarios are a scaling of the totally produced
   hadron yield
   with either the number
   of participants or  the number of binary collisions.
  These scenarios lead to
   an $A^\alpha$ dependence
   of the produced particle multiplicities  with
  $\alpha $ values of 1 and 4/3.
   One would get the $\alpha $-value 4/3 from RQMD
    in the `NN mode'
    by neglecting finite-energy effects.
  In this case the final particle yield
   would simply scale with
   the number of binary collisions, because
    in each collisions new strings can be created and each
   string decay gives
    asymptotically a constant rapidity density.
  The number of created quark pairs in the `NN mode' of RQMD
  is 3061
  which gives
   more than 7.2 created hadrons per participating nucleon.
  This can be compared to the corresponding numbers in
  elementary $pp$ collisions which is 4.9 (at the slightly higher
   beam energy of 200AGeV).
  In terms of an $A^\alpha$ parametrization this
   means an $\alpha$-value of 1.07 by comparing
   Pb+Pb and $pp$ reactions.
  This value is still far away from the `upper limit' 4/3.
    Obviously
   the finite-energy effects --
    finite string masses and
    mutual deceleration of
   projectile and target  --
   change the naive estimate considerably.

 How do the production rates of quark pairs change if
  rope formation is taken into account in
  Pb(160AGeV) on Pb collisions?
  The formation of coherent chromoelectric fields
    which fill the space between the receeding color
   charges sets different initial conditions
   for particle creation than the incoherent
   superposition of elementary strings.
 The effect of ropes on the
    conversion rate of field energy into $q\bar{q}$ pairs,
    the resulting rope lifetime and
   particle multiplicities have been discussed
   already  in the literature  \cite{BIR84,BCZ86}.
   Neglecting partially or completely the screening
   effect of already produced charges and
    employing various approximations, in particular
    boost invariance,
   simple
   relations  for these variables as a function of
   the rope field strength
   have been derived.
   The authors of Ref.\ \cite{BIR84}
   derive a relation valid in  their model
  \begin{equation}
     \label{fkf1}
    F(K)/F(1) \approx  2.12
      \hspace{4em}  K \rightarrow \infty \quad,
  \end{equation}
    with $F$ the meson multiplicity and
   its argument
  the so-called foldness.
    The foldness $K$ is defined by
   the number
    of (anti-)triplet
   charges which screen a rope field completely.
    $K$=$p$+$q$ for a rope whose source is  a color
    $SU_3$ charge in the  ($p$,$q$) representation.
   Eq.~(\ref{fkf1}) implies a strong suppression
   of particle production $\sim 1/N $
    compared to production from $N$ independent strings.
 Note that
   random charging without any constraint
   leads to  the average $K$ value
  \begin{displaymath}
     \langle K \rangle \approx \sqrt {N}
   \quad .
  \end{displaymath}
   The multiplicities from rope fragmentation in RQMD
   cannot be calculated analytically, but show  similarly a
   trend that particle production is strongly dampened
   with increasing rope charge.
    The assumption that each hadronizing string breaks just once
     seems  more appropriate for an estimate of
    the rope effect at 160AGeV on multiplicities than the
    opposite extreme of an infinite number of break points
    which is a consequence of assumed  boost invariance.
    Under this assumption
   the  number of created quark pairs is depleted if ropes are
   formed, because the minimum number of quark pairs to screen
   the original field is smaller ($\sqrt{N}$ versus $N$).
    Thus the total number of quark pairs
     including the quarks of the source charge
    is given simply as 2$N$ from $N$ string decays,
   $\sqrt{N}$+$N$ from rope fragmentation.

   Neglecting finite-energy effects
   grossly overestimates the influence which rope formation
   can have on the whole dynamical evolution of the system.
   After the hadrons which contain
   ingoing constituent quarks have been subtracted from
   the average number of primary hadrons
   in elementary $hh$ collisions at 160GeV is around 2
   per collision partner.
   Subsequent collisions in a nuclear target produce effectively
   even fewer hadrons per collision which come from smaller strings
   or from excited resonance decays.

   The surface and the finite thickness of the
   ingoing nuclei play a  nonnegligible
   role as well for rope formation. The Lorentz contracted lead diameter
   in the C.M.\ frame is approximately 1.4 fm.
   The corresponding passage time for the collision partner
   has therefore a value which is larger than  the
     hadronization scale (1 fm/c).
    Strings or ropes which were formed early in the collision may
    have already hadronized before the projectile has completed its
   passage through the target.
   All these effects tend to suppress the importance of
   rope formation and favor the other
    sources of secondaries, elementary strings and
  resonances.

  The RQMD  approach to
    rope formation and hadronization is superior
   to schematic calculations, also because it
   respects the simple  constraints  arising
   from nuclear geometry and finite beam energy.
  The yield of produced secondaries decreases from 3006 to
   2571  in the calculated central Pb on Pb collisions
  by allowing  string fusion into ropes.
   This 15\%-effect on the total multiplicity from rope
   formation is tiny as compared to the suppression effect
    implied by eq.~(\ref{fkf1}).
  It will be interesting to study the energy dependence of
   the multiplicities in case of rope formation.
  The dampening of particle production due to
    formation of strong chromoelectric fields
   should become much more pronounced
   at collider energies.

   Inclusion of hadronic rescattering leaves the total
   produced particle multiplicities practically unchanged
    (2512 in comparison to
   2571 without rescattering).
   These  values correspond to an
     $\alpha$-value of 1.04
    which is even closer to 1 yet
   than in the `NN mode' of RQMD.
   The reason is that
    particle number conserving processes
     -- in RQMD  2$\rightarrow$ 2 --
   dominate
   the rescattering stage.
   Particle conservation in the expansion stage of
   $AA$ collisions has been conjectured by other
    authors earlier
    based on general grounds
     (kinetic equilibration together with $G$  parity conservation
     which forbids to change the pion
     number by one unit
     in collisions of nonstrange mesons) \cite{GAV91}
   and is confirmed
   by the microscopic transport calculations.
  The small decrease of the particle multiplicity
   due to rescattering is  connected
   with the strangeness enhancement (see below) and the related
  smaller feed-down of the lightest hadronic degree of freedom, the
  pions.
  One should keep in mind that the strange hadrons,
   $\Lambda $, $K_S$, etc.\ are considered stable here.

\subsection{
   Rapidity distributions
                         }
\begin{figure}[h]

\centerline{\hbox{
\psfig{figure=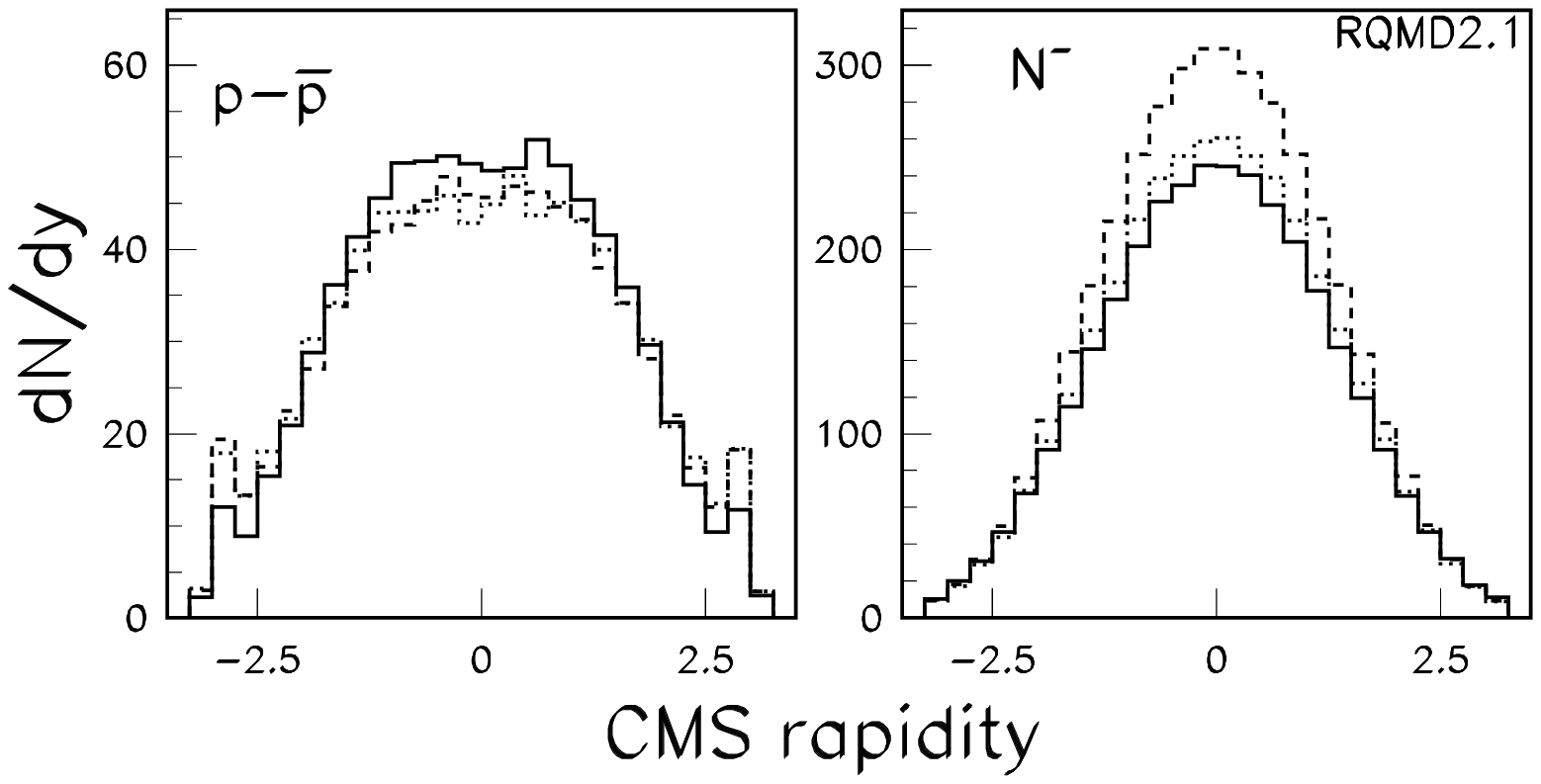,width=18cm,height=18cm}}}

\caption
[
 ]
{
  \label{pbpbpnnegy}
}
\end{figure}

 Here I shall focus on the shapes of the rapidity
  distributions for  different particle species.
 The  rapidity distributions for negatively charged hadrons
   calculated in the different RQMD modes
 are displayed in fig.~\ref{pbpbpnnegy}.
 It is clearly visible that the change in produced particle
 multiplicities is concentrated in the central region.
  The collective effects leave practically no tracks  in the
 fragmentation region of projectile and target ($|y_{CMS}|$$>$2).
  At these rapidities the three rapidity distributions
  -- from the `NN mode', with ropes and with additional
  rescattering fall on top of each other.
  This can be understood from the
   discussion in section \ref{secrope}.
   The  rope fields  influence mainly the
  particle multiplicities  at rapidities
   which are covered by many  strings.
  Also the small absorption effect which rescattering has on the
  particle multiplicities is concentrated around midrapidity.
 On the left hand side of
  fig.~\ref{pbpbpnnegy} the net proton ($p$-$\overline{p}$)
  rapidity distribution is shown for comparison.
   The $p$-$\overline{p}$
    rapidity distributions are remarkably
  similar in the three RQMD modes. Rescattering makes the
   final net baryon distribution only slightly narrower.
   The small effect of hadronic rescattering on the baryon
   stopping in the present  RQMD calculations differs markedly
   from results in the string model QGSM (without string fusion
    but rescattering included)  \cite{AME91}.
  The effect in the calculation presented here is also somewhat smaller
  than in early RQMD studies \cite{AvK91}.
  The difference is mainly caused by the more realistic
   angular distribution in baryon resonance decays
   (see section \ref{secschres} for a discussion).
  Isotropic decay as employed in Ref.~\cite{AvK91} tends to
   shift baryons rather rapidly into the central region.
  One can see from fig.~\ref{pbpbpnnegy}
   that the $p$-$\overline{p}$
    rapidity distribution is
  significantly broader than the produced particle
  distribution.
   However, the baryon stopping mechanism in RQMD is much
   more pronounced than in other approaches
  \cite{CAP87,KAD95}.
  In RQMD it shows no minimum
  at midrapidity but a plateau in the central region.

\begin{figure}[h]

\centerline{\hbox{
\psfig{figure=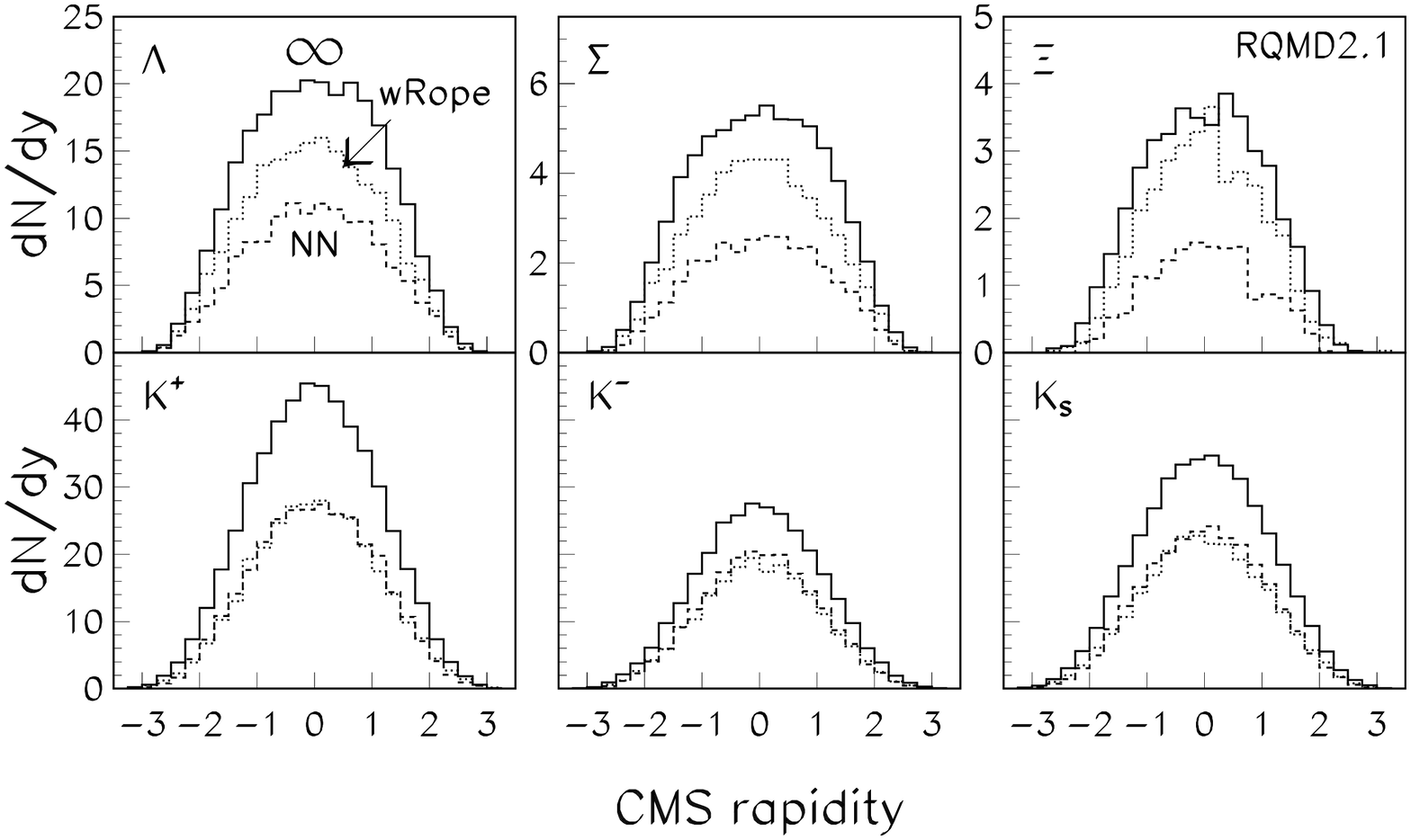,width=18cm,height=12cm}}}

\caption
[
 ]
{
  \label{pbpbstrhad}
}
\end{figure}

  The comparison of the rapidity distributions
   of net protons and negatives
  in fig.~\ref{pbpbpnnegy} demonstrates  that
   the single fireball picture is  inconsistent with
  the calculated rapidity distributions, like in the smaller
  S+A reactions \cite{SOR95ZFP}. The
  primordial nucleon rapidity density would have a width
   between 0.44 and 0.52 units at
   freeze-out temperature between 140 and 200 MeV.
   The heavier a particle,  the narrower is its rapidity distribution
   in a thermal fireball.
   The reverse ordering found from RQMD
    is indicative for the presence of
   longitudinal flow in the system.
   One could imagine different mechanisms driving
   the longitudinal flow,
   internal  pressure (as in Landau scenario)
    or a transparency effect (as in Bjorken scenario).
    RQMD in which the secondaries are initially created by
     the  boost-invariant  decay of a longitudinally stretched tube
   is constructed very similarly to a Bjorken-type
   approach. Some differences arise from finite-energy
    effects which are clearly visible from
   fig.~\ref{pbpbpnnegy}. There is no plateau in the produced particle
    rapidity distribution at all.
    Furthermore,
     the Bjorken ansatz of identifying space-time
    and momentum-space rapidity is
     `softened' in string models resulting in
    a finite width of the local rapidity distribution \cite{SOR93ZFP}.
   The calculated rapidity distribution
   of negatively charged hadrons is narrower in
   central Pb on Pb than in S(200AGeV) on S collisions.
   The width of the distribution is 1.6 units of rapidity versus 1.9
    which makes a difference of 8\% if they
     are normalized to the initial rapidity gap
  between projectile and target.
    The narrowing   arises
   in the model from the stronger attenuation of the
   ingoing nuclear constituents in the heavy system.
  It is interesting to note that
   Landau-type hydrodynamics
    with all matter initially at rest
    would show the opposite effect.
 Fig.~\ref{pbpbstrhad}
  displays the calculated rapidity distributions
  of strange baryons and mesons.
  The distributions of  strange baryons are markedly narrower
  than the  proton density,
   1.7 versus 2.1.
  This cannot be attributed to the
   mass differences
   which results only in small width differences
   in a thermal picture  ($<0.05$).
   Taking longitudinal flow into account
    does not explain this effect either.
  In RQMD it reflects differences in the production dynamics.
  The probability that a baryon carries finally a strange quark is
   correlated with its rapidity. Baryons at central rapidity
   have suffered simply more
  collisions,  initially with nucleons from the other
   nucleus, lateron with secondaries.
  Spatial inhomogeneity
   of the flavor composition is no real surprise,
  even in collision of truely heavy ions.
  On the other side,
   parametrizations of `hydrodynamical flow' under the homogeneity
  assumption have been quite popular for studies of light ion reactions
  \cite{SSH93,PBM95}.
  One can   take the RQMD calculations presented here
  as an indication that results in such simple models have to be
   taken with some grain of salt.

\begin{figure}[h]

\centerline{\hbox{
\psfig{figure=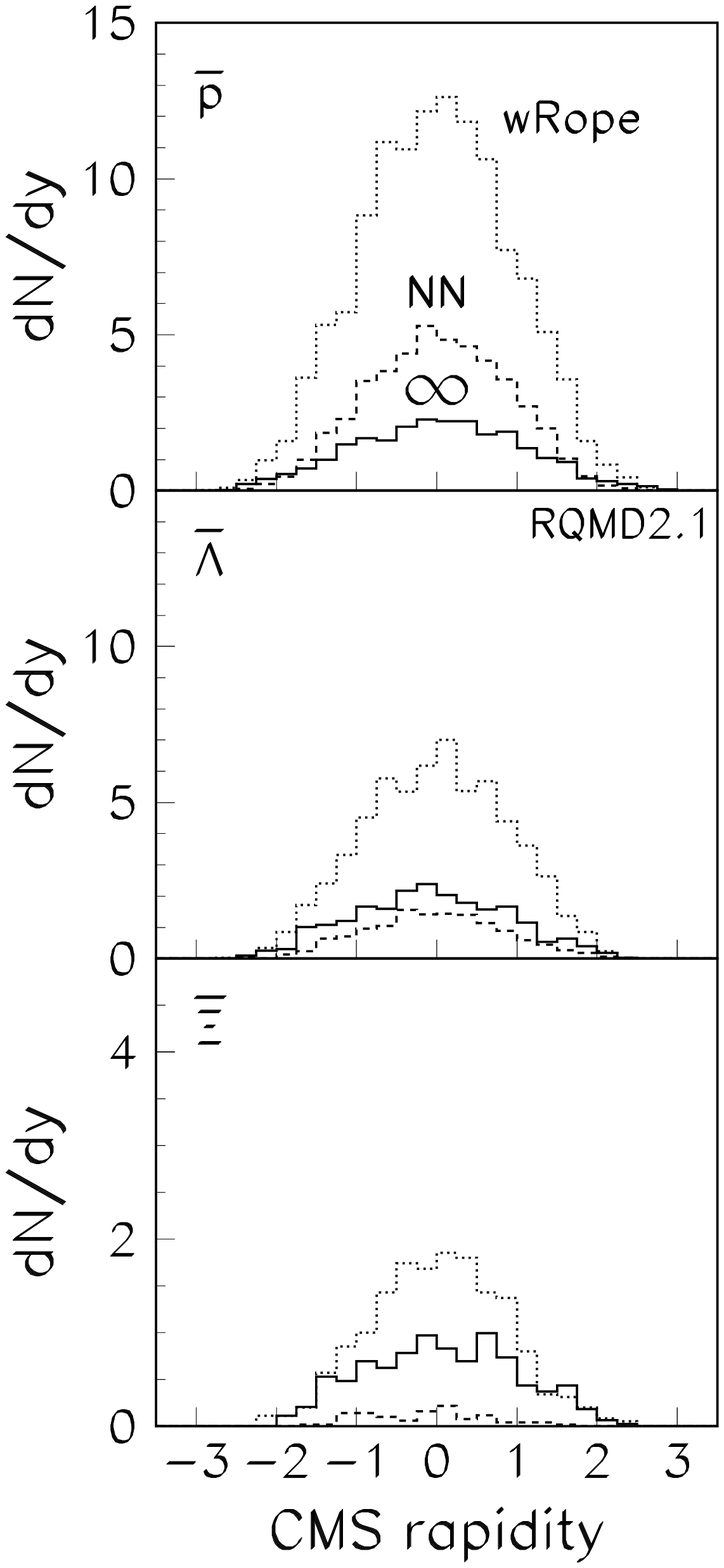,width=9cm,height=18cm}}}

\caption
[
 ]
{
  \label{pbpbbbar}
}
\end{figure}

  The final antikaon and kaon distributions
   have a slightly smaller width (1.3-1-4) than
  the hyperons as a result of the generated longitudinal flow.
  In contrast, the antibaryon rapidity densities have
    much smaller widths (1.1-1.2) than
   mesons and  baryons as can be learnt
 from  the results presented in fig.~\ref{pbpbbbar}.
   Again, this effect is incompatible with a homogeneous
  thermal fireball -- with or without  longitudinal
  flow  -- and
   reflects   different production mechanisms in the model.
  Since antibaryons are mostly produced in
   rope fragmentation (see below),
  they are primarily  formed in
   position and rapidity space where  many
   strings overlap. Baryon pairs emerge only
   from  the regions of highest
   energy density.
   In contrast, due to smaller
   production threshold
   $s\bar{s} $ pairs are created
   also in the diluter
   regions as well.
  It had been already noted  by the experimentalists
   in the presentation of first data for S(200AGeV) on S collisions
  that the  different shapes
  of $\Lambda$ and $\overline{\Lambda}$  dN/dy distributions
    provide evidence for  differences  in the production
    mechanisms \cite{NA3590}.
   The systematics of measured
    strangeness and antibaryon production in S on A collisions
   at beam energy of 200AGeV \cite{NA3594} can be exploited
    to understand better  the dependence on initial
    baryon and energy density. The latter
     variables are related to the final
   baryon and meson rapidity densities.
   It was argued in \cite{SOR95ZFP} that
    the narrow antibaryon distributions in S+A reactions
   \footnote{
     Note that
     the $\overline{p}$ distributions
     for the calculated reactions with O and S projectiles
    in fig.~5 of Ref.~\cite{SOR95ZFP} are scaled by
      a factor of 10 (not 20 as indicated in the figure).
            }
   arise as a convolution of two effects,
    string or equivalently energy density, which is
  the determining  factor
   in the model for the achievable rope field strength, and
   the smoothly increasing
   baryon density (in position and momentum space)
   towards the target rapidity region.
   In contrast, in central Pb on Pb reactions
   the antibaryon rapidity densities get slightly
    broadened  due to  absorption,
    because the  rather flat net baryon density for
    Pb on Pb is centered around midrapidity.

\subsection{
   Strangeness enhancement
                         }
  How large is the fraction  of created strange quarks
   relatively to the light flavors  from
    RQMD?
  The fraction of created
   $s\overline{s}$ pairs
      normalized to the average of
   $u\overline{u}$ and  $d\overline{d}$ pairs
  \begin{displaymath}
   R_s = \frac{2N(s\bar{s})}{N(u\bar{u})+N(d\bar{d})}
  \end{displaymath}
    turns out as  11.3\% in the `NN mode'  which should be
   compared to 10\% in $pp$ collisions at 200GeV, a small
   increase.
 \footnote{
   It is be neglected here that
   $\eta $ and $\eta '$ meson carry some amount of hidden strangeness.
   }
  The rate of produced
  kaons remains practically
  unchanged in the `NN mode' if they are normalized to
   the pion yield.
  For instance,
  the $K^-$/$\pi$ ratio is  6 percent
  (with isospin averaged pion yield)
   while it is 5.6 \% in $pp$ collisions.
   Several competing effects may give rise to small
    deviations from the $pp$ result.
    Because of energy degradation
    the  average string masses are lower in multiple
    collisions of a projectile hadron impinging on a nuclear target.
  This  effect is
    unfavorable
   for strange particle production.
    On the other side, the ingoing flavor
     has to be used up if only one string is excited
    in course of a nucleon collision. If several
    strings are attached to an ingoing interacting constituent
   quark, all of the additional strings have
    color charges from the sea, possibly an $s$-quark,
     at their end points.
   Furthermore, multi-step processes of the type
   $B $$\rightarrow$ $B^*$$\rightarrow$ $B^{**}$
   which are most relevant near particle production threshold
   may still play a small role even at a beam energy of 160 AGeV.

  The strange baryon rate normalized to
  the total baryon number  increases
   considerably already in the
  `NN mode'.  The ratio of $\Lambda $/$B$
   including all feed-down contributions is 0.05
    in $pp$ collisions
   from RQMD which is in accordance with experimental
   data \cite{JAE75}. It increases
   by  nearly a factor of 2 to a value of 0.09 in
   central Pb on Pb reactions.
  This effect can be understood rather simply
   from the multiple collision effect on baryons
  (see the diagrams in fig.~\ref{bstring}).
  While only  1 or 2 of the original
   valence quarks are replaced
  in elementary $N$$N$ collisions,
  the large majority of valence quarks
   are being replaced in the outgoing
   baryon states  by quarks from the sea
   in Pb on Pb collisions at small impact parameters.
   Thus the probability increases with
    the number of collisions which a baryon undergoes that
    it will carry a single or even multiple units of strangeness
    after the interaction.
  A similar effect is observed in the VENUS approach
   to $AA$ collisions
   which includes the mechanism of
    `double-string' formation \cite{AIC93}.

 A strong  strangeness enrichment
   compared to the `NN mode' and the calculated values for
 $pp$ collisions is expected
   by including the collective interactions
  into the calculations.
  The suppression of heavy quark production in
  strong chromoelectric fields is weakened.
  The different barrier penetration factors for virtual quark
  tunneling are approaching the same values.
  Rescattering processes
  tend to enrich strangeness in nuclear collisions,
   because  -- starting under conditions
    in which strangeness is undersaturated --
  nonstrange quark pairs are preferentially
  annihilated in resonance and string formation processes
    and  sometimes replaced by a strange
  quark pair in the decay.
 Associated production  plays the most important
 role for strangeness enhancement in the hadronic resonance gas.
The  difference between
a $\Lambda $ baryon and  nucleon mass
is just 178 MeV, but  twice  this value
 between anti-kaon and pion mass.
 Furthermore, the preequilibrium contribution
  to the  secondary interactions
   is stronger in the
   meson-baryon than the meson-meson sector \cite{SOR95ZFP}.
  This is qualitatively easily understood, because
    the baryon rapidity distribution is broader
   than the meson distribution. This holds also
    locally
     which is  responsible for the hard tail in the
    collision spectrum of baryons.
  Transport calculations
  with the RQMD model
  show
 good agreement with available data
  for reactions with light ion projectiles
  which are sensitive to this production mechanism
  (net-$\Lambda$ rates, $K ^+$/$K^-$ enhancement)
 \cite{SOR92,SOR95ZFP}.

Most of the strangeness is carried in the final
state by (anti-)kaons which makes their yield
normalized to the total produced particle multiplicity
to a signature for the strength of collective effects in
AA collisions.
It has  been discussed already some time ago
based on studies of S induced collisions at 200AGeV
that rope formation leaves the absolute yield of produced
kaons practically  unmodified
 \cite{SOR92}. The situation does not change if collisions
 with heavy projectiles are considered
 (see table~\ref{pyield}).
 The relative enhancement due to the weaker strangeness suppression
 is completely counterbalanced by the depletion of total
  multiplicity.  The results in the DPM based string fusion
  approach SFMC confirm these conclusions and
  are surprisingly close to RQMD, even in terms of absolute numbers.
  The 4$\pi$ yields for central Pb on Pb collisions
   given in Ref.~\cite{ABFP95}
  are (with RQMD values in parantheses):
   negatively charged hadrons without fusion 924.4 (964.3),
   but with fusion included only 806.2  (829.2).
   In contrast, the kaon yields change much less,
    $K^+$  90.1 versus  88.4  (79.2/79.0)  and
    $K^-$  71.5 versus  66.8  (53.0/50.4).

  The $R_s$ value increases in RQMD
  from 11.3 \% to 15 \% if the
   strings can fuse into color ropes and even to
  24\% if the hadrons  interact with each other
 after formation.
 One can learn from these numbers that
  most of the  strangeness enrichment
   is produced  in the hadronic stage
   and not during the quark matter evolution
  in central Pb on Pb collisions.
 Furthermore, the strangeness enrichment
   in comparison to $pp$ interactions
  is even stronger in central collisions
  with Pb than with S projectiles.
 Both aspects are actually connected
 as will be discussed in the following.
  The strangeness enhancement factor
  for quark pairs which are created
   in the rope field is practically projectile-independent
   in comparing central S  and Pb collisions,
   namely 50 \%.
  This is again a consequence of the finite
   energy  and length scales  which tend to
   suppress string excitations
    and longitudinal overlap of strings
    after the first 2-3 collisions of a colliding
   nucleon.
 Therefore the prehadronic processes are not responsible
  for the differences in the final states
   of reactions with S and Pb projectiles.

 There are two reasons why
  the interactions in the hadronic stage are  more effective in
  enriching strangeness
  in reactions with larger  projectile masses.
 The first reason is the increased stopping power which
  shifts relatively more nucleons of the target
   towards the  central rapidity region.
  This allows them to participate more easily in interactions
   with secondaries which are concentrated around midrapidity.
 The prominent role of baryons
  in increasing the strangeness content of the system
  can be directly read off from table~\ref{pyield}.
  The final baryons  contain
  124 $s$ valence quarks, which are 45 \% of the
  $s$ quarks created in total.
  A nuclear {\it hypermatter} state is created in these
  collisions according to the RQMD calculations.
  With respect to
   strangeness production,
  the dynamics of baryons
   is far from being  a perturbation to the whole dynamics
  in $AA$ reactions at the energy of 160AGeV.
\footnote{
  Consideration of baryon degree of freedom is sometimes
  neglected in theoretical studies
  of $AA$ collisions which is based on the small
   $B$/$M$ ratio in the final state at 160-200AGeV.
  The
   important role of baryons for strangeness production which is found
   from RQMD
    has  repercussions
   for the production of
   other particles whose production is inhibited by thresholds,
   for instance
  direct photons at large
   transverse momenta.
  }

 The second reason is related to the time intervals which the system spends
  in these two `phases'.
\footnote{
 A detailed account of the time evolution
  of the created matter in central
  Pb(160AGeV)+Pb collisions will be given elsewhere \cite{SOR95PRCB}.
  }
 After the quark matter and hadronic preequilibrium stages
 which last about  3 (4) fm/c in S (Pb) induced reactions
  have been completed
  the system stays in approximate
  kinetic equilibrium for the rest of the interaction time.
 The temperature drops only slowly
  which is a property of the  resonance gas
  with its many degrees of freedom.
 The hadrons and resonances interact in this stage until freeze-out,
 for  approximately 11 fm/c in central Pb+Pb, but only
  for 3 fm/c in S+A reactions.
 Thus the weight of interactions in
  this third phase is  more pronounced in
 Pb induced reactions than in light ion
  (e.g.\  O or S) collisions, even
  with heavy targets. In the latter case transverse
  expansion  becomes rather effective
  after local kinetic
  equilibrium is reached, because the
    transverse area of the system is  smaller.

\subsection{
   Baryon pair production
                         }
A strong strange antibaryon
 enhancement has been experimentally
observed
for central S collisions on S and heavier targets
\cite{NA3590,NA3594,ABA91}.
Binary interactions in the rescattering stage  tend to conserve
particle numbers. This
excludes them as an important contributor
to the creation of additional antibaryons if the
total interaction times are on the order of 4-6 fm/c only.
Such lifetimes of the interacting system
 in S+A collisions  are
 suggested from interferometry data \cite{NA35HBT94,NA44HBT95}
  consistent with  the RQMD calculations.
The failure of Boltzmann-type
hadron resonance gas dynamics
to account for a sizable enhancement of
strange antibaryons
in  central $AA$ collisions
points towards the importance of
production mechanisms
in denser stages of the reaction,
notably in the prehadronic stage.
 This is supported by a simple argument.
 The  energy needed to
 create a  baryon pair
 is on the order of 2 to 3 GeV.
 Soft production processes
 which are characterized by the scale
 $\Lambda _{QCD}^{-1} \approx 1$fm
 therefore require energy densities of a few
 GeV/fm$^3$ in order to
 overcome the $B\overline{B}$
  suppression in comparison to meson production.

\begin{figure}[h]
\centerline{\hbox{
\psfig{figure=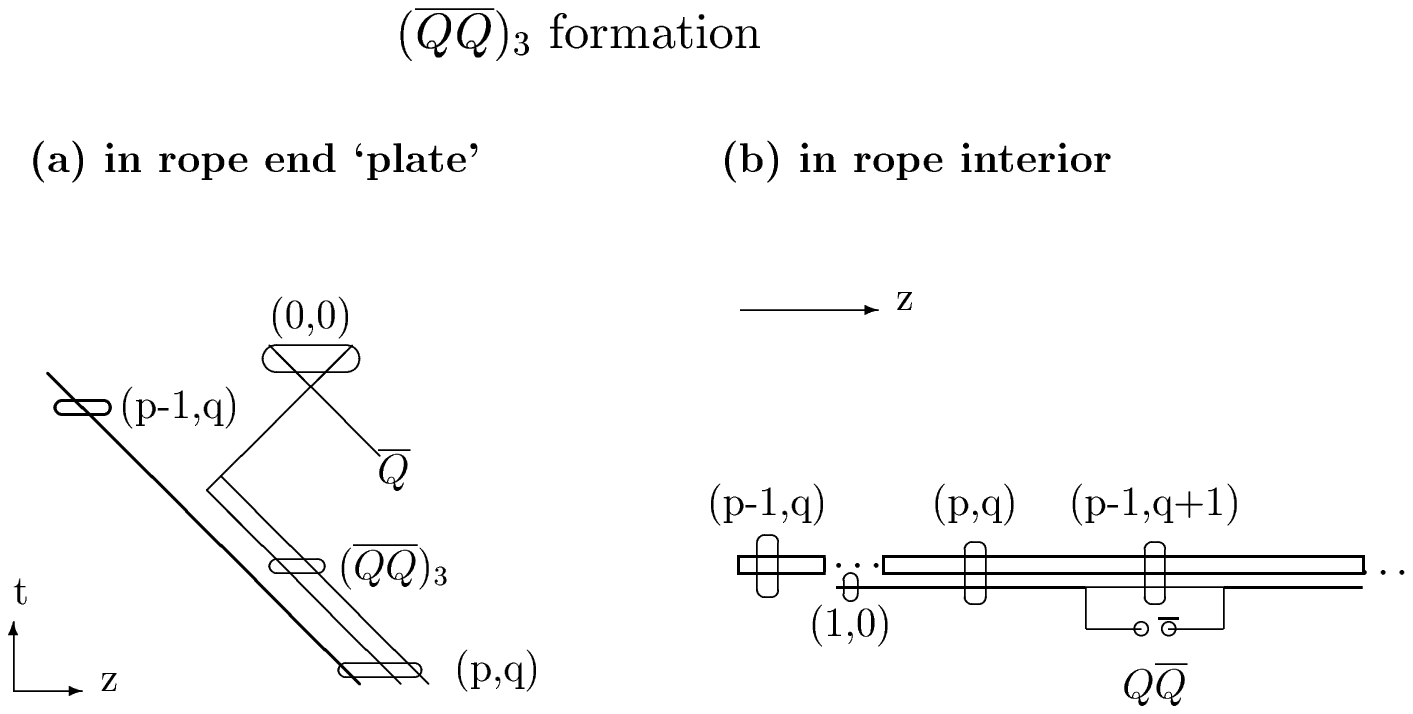,width=17.0cm,height=24cm}}}
\caption
[
 ]
{
 \label{bbbarmecha}
}
\end{figure}

Color ropes are a strong source of additional
baryon pairs in ultrarelativistic $AA$ collisions
\cite{SOR92}.
There are two distinct mechanisms how
a rope may generate
diquark pairs in the triplet representation
 ($3-\overline{QQ}$).
These are
quark coalescence in the rope end plates
and quark pair creation
with color mismatch in screening the field.
The two processes are graphically represented
in fig.~\ref{bbbarmecha}.
In the process of forming the rope charge
elementary triplet charges
of quarks and antiquarks in a
receding rope `condensator' plate
couple
stochastically to the
total $SU_3$ color charge of the rope.
This provides a strong source of
$3-\overline{QQ}$'s
and $\bar{3}$-$QQ$'s
 by quark coalescence.
The statistical
weight that
 two quarks
form
a
$\bar{3}$-$QQ$
 compared to a
$6-QQ$
 is just  1:2.
  The (anti-)diquarks in triplet representation
   will combine with a corresponding anticharge
   to form an (anti-)baryon.

 It turns out that
  in nuclear collisions at 160-200AGeV
  the probability of
   quark coalescence
 in the  charge  at the end of the rope
 is typically much larger
   ($>$90\%)
  than due to
  diquark creation
  inside the field.
  This is very much based on
     the energy gain
    by optimal screening in course of charge creation
    $\sim $(C(p,q)-C(p,q-1))
    and/or
    $\sim $(C(p,q)-C(p-1,q))
    as compared to
   the nonoptimal case
    $\sim $(C(p,q)-C(p-1,q+1)+2/3).
  Note that this result reflects also the
   rather short average length of the flux-tubes
  which are created at these energies.
   Giant flux-tubes with infinite length would be
   completely dominated by creation processes inside.

 Indeed, baryon pair production is strongly increased
 in central Pb on Pb collisions if overlapping
 strings are fusing into ropes.
 The total number of created
 baryon pairs in the `NN mode' turns out
 to be 27.7. Ropes increase this number  to
  84.3.
The SFMC calculations show qualitatively similar trends,
an increase
(of 2$\overline{p}$+$\overline{\Lambda}$)
from  16.7  to 83.9 due to string fusion \cite{ABFP95}.
However, the flavor dependence is different in the two approaches.
In RQMD the increase is strongest for strange antibaryons
which is reflected for instance in the increase of the
$\overline{\Lambda}$/$\overline{p}$  ratio.
The trend is the opposite in the SFMC results.
The RQMD result arises from the `constructive interference' of
 weaker strangeness suppression and
 additional diquark production mechanism if ropes are formed.

  Subsequent antibaryon absorption
  in baryonic matter
  brings  the final antibaryon yield from RQMD down again to
   21.8 in the final state.
  This means that the strong  antibaryon enhancement
   (a factor of 3)
  from the preequilibrium
   quark matter stage is more than `eaten up' by
  the subsequent absorption in baryon-rich matter.
  Only a quarter of the initially produced antibaryons
   survives the interactions in the hadronic gas
   environment
  until freeze-out.
   Like in the case of strangeness enhancement a
  comparison of the yields in the different RQMD modes
   of operation demonstrates that the hadronic stage
   has much more impact on the flavor composition
   in central Pb+Pb than in S induced reactions \cite{SOR95ZFP}.
  The strength of the
   antibaryon absorption shows some flavor dependence.
  While the number of $\overline{p}$
   is even a factor of 2 below the results in the
   `NN mode',
   the final yields of strange antibaryons are still
  considerably enhanced.
  For instance, the   $\overline{\Xi}$ yield
    increases by a factor of 7.7
   compared to the result in the `NN mode',
    even a factor of 13.3 if antibaryons
   would not be absorbed in the hadronic stage.
   Since both antibaryon and strangeness production
    are suppressed in elementary hadronic interactions,
   the rope formation
   effect is particularly strong for hadrons which carry
  these flavors in combination.
  Furthermore, the flavor dependence of antibaryon absorption
   reflects that  $s\overline{s}$ pairs are less frequent
   in the system and their
     annihilation probability is reduced, which mirrors
   the suppressed production probability
  (cf.\ section  \ref{secresregtr}).

 Is there some way to disentangle the antibaryon enhancement effect
 due to color rope formation or some other collective effect
  like QGP formation  \cite{LRT94}
  in the early stage of the reaction
 from strong antibaryon absorption in the later hadronic stage?
 At least, it can be expected that the rather extreme model
  for antibaryon interactions adopted presently in RQMD
  which assumes a free
   $\overline{p}p$ annihilation cross section
 can be tested rather well in heavy ion collisions.
  Since the  annihilation cross section is strongly energy-dependent,
  preferentially low-momentum antibaryons are annihilated.
 This effect is even more pronounced in heavy ion collisions
 at lower beam energies \cite{SPI95}.
  The use of the free $\overline{p}p$ annihilation cross section
  is certainly debatable  \cite{ARC93}, in particular
  at low relative momenta.
  Since this cross section
  corresponds at low energy
 to particle interaction distances clearly
 larger than  1 fm and thus exceeds the
  inter-particle distance  in the dense stage,
   medium effects
  are expected to be very important.
 The real part of the antibaryon self-energy in baryon-rich
 matter may get substantial values as well, because
  vector meson exchange leads to attraction in addition
   to attraction already from scalar  exchange.
  It will have some influence on the shape of the particle spectra,
 although the magnitude of this effect is unclear
  \cite{SPI95}.

\section{
   Summary and Conclusions
                }

  String fusion into color ropes and
   hadronic rescattering  --
  collective interactions in the
   preequilibrium quark matter and
   hadronic resonance gas stage of ultrarelativistic
   nucleus-nucleus collisions --
   have been modeled in RQMD.
  The system  created in central Pb(160AGeV) on Pb reactions
  is characterized by strong longitudinal  flow.
  The nucleon momentum distribution shows the strongest
     elongation along beam direction,
  but with a maximum still at midrapidity.
 The broadness
   arises from  partial transparency
   (corona effect).
  The final antibaryon source is
   clearly more concentrated around midrapidity than
  their antiparticles,  kaons and pions.
   Antibaryons are produced  only
   in the region of highest
   energy density, while
   mesons and even strangeness
    are created
   also in the diluter
   regions as well.
   If these RQMD predictions are experimentally confirmed,
  strangeness and antibaryon enhancement
  in ultrarelativistic heavy ion collisions
   cannot be described by only one source with
  homogeneous flavor composition.
  The  strangeness suppression factor defined by
   the ratio of created quark pairs
    $s\bar{s}$/($u\bar{u}$+$d\bar{d}$)
   is strongly enhanced by  a factor of 2.4
   in comparison to $pp$ results.
  Color rope formation  increases
   the initially produced  yield of antibaryons
    to 3 times the value in the `NN mode', and
    even stronger
     if they carry strangeness.
  Only approximately one quarter of the produced antibaryons
  survives because of
  subsequent strong
  absorption in baryon-rich matter.
  The differences in the final particle composition for
  Pb on Pb collisions to   S induced reactions
   are attributed to the
    hadronic resonance gas stage which is
   baryon-richer and lasts longer in the heavy system.

\section*{ Acknowledgements }
I  acknowledge  gratefully the help of
  M.\ Leltchouk, J.\ Nystrand, G.\ Heintzelman, M.\ Zou, and J.\ Nagle
   in eliminating some bugs in the computer code  RQMD2.1
   and getting it executable on various computers.

 \section*{Appendix: Quark pair creation in a rope}
   The path integral for the tunneling of a virtual charge pair
   is given
   in the WKB approximation  as
     \cite{YND83}:
   \begin{equation}
    \label{Gl3}
      P(p_t)= \left| e^{-2\left|S_{cl}\right|}  \right| ^2  ,
   \end{equation}
 where
   $S_{cl}
      =   \int_{0}^{z_{fin}}  p_l(z)  \: dz   $
    is the  action for the  path
    across the classically forbidden region.  Therefore
   longitudinal momentum  and  action
    $S_{cl}$ are purely imaginary in this
   region, initially:
   \begin{equation}
      p_l^2  +  p_t^2  + m^2  = 0  \quad .
   \end{equation}
    The force acting on a created charge is given from
     the amount by which the  field energy per unit-length
     is lowered due to  screening of the original
    source \cite{GLE83}:
   \begin{equation}
     \label{tunforce}
       F_{e}= (\kappa - \kappa ' )
    \quad  ,
   \end{equation}
   with  $\kappa$ the rope tension before and
    $\kappa '$ the tension after
   pair creation.
  After each of the created charges
     has moved over a distance $z$
    the energy balance reads:
   \begin{equation}
     \label{Gl2}
      2 F_{e} \cdot z =
          2 \sqrt{p_l^2(z) + p_t^2 + m^2}
    \quad  ,
   \end{equation}
  for a charge with constant mass
   $\left|S_{cl}\right|=
    \pi m_t^2/4 (\kappa -\kappa ')$.
   However, it is assumed here that the mass  varies linearly
   with distance which is motivated by the expectation
    that quark masses are `current' on short distance
    scales and `constituent' masses with respect to the
    nonperturbative confining force.
 \begin{eqnarray}
   \label{minterpol}
     m(z) &=&   m_0 + \beta \cdot z ,
                          \\   \nonumber
     z &\leq & \Delta m  /\beta
     \quad ,
  \end{eqnarray}
  with $m_0$ the current
  quark mass ($m_u=m_d=10$  MeV,
 $m_s=160$ MeV), $\Delta m$=350 MeV the difference to the
  constituent mass,
 and $\beta $=0.355 {\sl GeV}/fm  the `speed' of
  quark dressing.
  The parameters are fitted to give
   0.1 for suppression of diquarks in comparison
   to quark production and 0.29 for strange quark suppression
   in elementary flux-tube decays which are the values
  favored by experimental data.

 The absolute value of the action $S_{cl}$
 for the tunneling  of  a quark
  with linear interpolation between current and
   constituent  mass is given by
 \begin{eqnarray}
 \label{sclanal}
       \left|S_{cl}\right| &  = &  I_1  +  I_2    ,
                                        \\[1em]     \nonumber
        I_1  & =  &
          \frac{M^2}{2A} \cdot
            \left\{
              \left(
               \frac{Az_1}{M}-\frac{\beta}{F_{e}}\right) \cdot
             \sqrt{
                  1 - \left(\frac{Az_1}{M}-
                            \frac{\beta}{F_{e}}\right)^2
                                                       }  +
                         \right.        \\     \nonumber
          &    &
                \hspace{3em}
                 \frac{\beta}{F_{e}}\cdot
              \sqrt{
                  1 -  \left( \frac{\beta}{F_{e}}\right)^2
                                                       }  +
                                        \\     \nonumber
          &    &
               \hspace{3em}
              \left.
                \arcsin{\left(\frac{Az_1}{M}-
                        \frac{\beta}{F_{e}}\right)}   +
                \arcsin{\left(\frac{\beta}{F_{e}}
                                                     \right)}
                                                          \right\} ,
                                        \\[1em]     \nonumber
        I_2  & =  &
              \frac{z_2}{2}  \cdot (m_{0t}+\Delta m)
                 \cdot   \Theta \left(m_{0t} - \Delta m \cdot
                  \frac{F_{e}-\beta}{\beta}   \right)
             \cdot
                                        \\     \nonumber
          &    &
               \hspace{3em}
              \left(
                  -  \frac{\Delta m  }{z_2 \cdot \beta}   \cdot
                      \sqrt{ 1-   \left(
                      \frac{\Delta m  }{z_2 \cdot \beta}  \right)^2
                                       }
                      + \frac{\pi}{2}  -
                       \arcsin{\left(
                          \frac{\Delta m  }{z_2 \cdot \beta}
                                      \right)}
                                                \right)
                                                       ,
 \end{eqnarray}
  where the following abbreviations have been used:
 \begin{eqnarray*}
   \lefteqn{
       A = \sqrt{F_{e}^2 - \beta ^2} ,
          \hspace{2.1em}
       B = \frac{m_{0t}\cdot \beta}{A^2}  ,
          \hspace{2.1em}
       M =  m_{0t} \cdot \frac{F_{e}}{A}  ,
                                    }
                                      \\[-1em]  \nonumber
           &  &    \mbox{\hspace{27em}}
 \end{eqnarray*}
 \begin{eqnarray*}
   \lefteqn{
      z_1 = M \!i  n \! \left(
              \frac{\Delta M}{\beta},
                  \frac{m_{0t}}{F_{e} - \beta }
                                \right) ,
                       \hspace{2.1em}
        z_2 =  \frac{m_{0t}+\Delta m}{F_{e}}.
                                    }
                                      \\  \nonumber
           &  &    \mbox{\hspace{27em}}
\end{eqnarray*}

\begin{figure}

\centerline{\hbox{
\psfig{figure=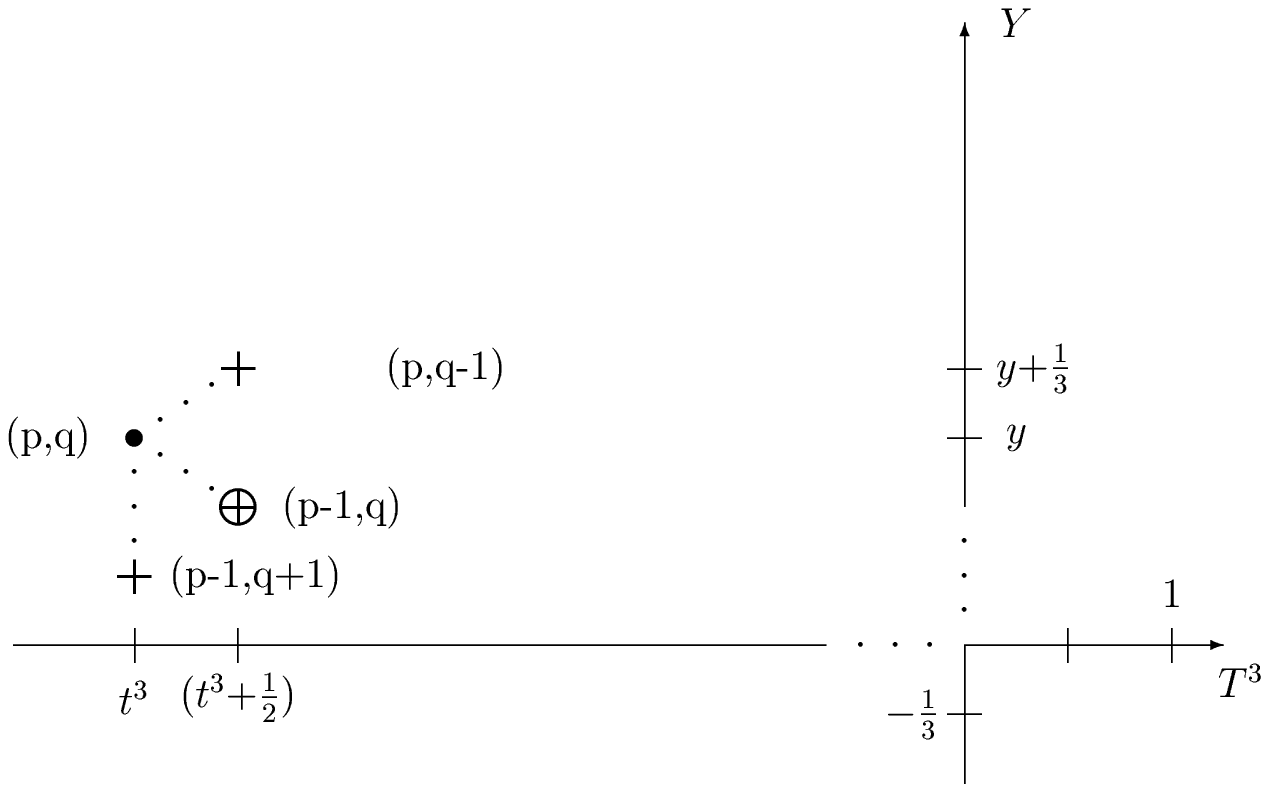,width=17.0cm,height=24cm}}}
  \caption
       [
                                  ]
     {
       \label{su3yt3}
                                    }
 \end{figure}

 Casher, Neuberger and Nussinov (CNN) have rederived the exact
 Schwinger result for the vacuum persistence probability
   in the WKB approximation   \cite{CNN79},
 with the additional benefit of  getting out
 a transverse momentum distribution
 for the created charges. Following their derivation
 the pair production probability
 per unit-time and unit-volume
 in a uniform Abelian field is given by
 \begin{eqnarray}
   \label{Gl4}
  \frac{d^4 p}{d^4x}  &=&
      \frac{\gamma}{4\pi^{3}}
 \cdot F_{e}
          \cdot \sum_{\mbox{\footnotesize  flavors}}
           \:
     \sum_{n=1}^{\infty}
         \int d^2 p_t \frac{P(p_t)^n}{n} .
 \end{eqnarray}
 $\gamma $ denotes the
  degeneracy factor from the
  color degrees of freedom.
 In total there are $2\times 3$=6
  different (anti-)quark color states.
  However, only three of them
  can lower the field strength
  by screening a given rope charge (see fig.~\ref{su3yt3}).

 A charge vector in the $SU_3$ multiplet ($p$,$q$)
   -- with $p$$\ge$$q$ for definiteness --
  could combine with a created (anti-)quark charge
 to  a $SU_3$ charge state in the multiplets
 ($p$,$q$-1), ($p$-1,$q$)
  or
   ($p$-1,$q$+1) to lower the field strength.
  One can see from fig.~\ref{su3yt3} that the sum  of
  original $SU_3$ charge and a color-`up' quark
   do  not have only a nonzero component in the ($p$-1,$q$),
   but also in the
   ($p$-1,$q$+1) multiplet.
  However, the latter possibility is discarded, because
   it is energetically unfavorable.
  Therefore the factor $\gamma $ in the pair creation rate
  eq.~(\ref{Gl4}) representing the color degree of freedom
   is set to 1 in the three cases given above
   ($q$$\ne$0 assumed).
  Screening of a color charge with  $q=0$ can result in
   one state with  charge ($p$-1,$q$)
   or in two states with charge ($p$-1,$q$+1)
   ($\gamma$=2 in this case).
  The  two configurations
  with screened charges in the ($p$,$q$-1), ($p$-1,$q$) multiplets
  will give
  3-$\bar{3}$ color singlet states together with
   the residual rope charge (optimal screening).
 The third
 configuration ($p$-1,$q$+1) consists of
   a $\overline{3}$-diquark
 on one side and an anti-diquark with opposite color charge
 on the other side (color mismatch) and  possibly some
  other  charge.

 The color of a
 diquark  created by  screening
  ($p$,$q$)$\rightarrow$($p$-1,$q$+1)
 will get
 neutralized  in a next step
  by an additionally produced quark pair.
 This is the generalization of the 2-step process
 for baryon pair production in an elementary flux-tube
  \cite{CNN79} to
 the case of stronger chromoelectric fields.
  The model  of diquarks adopted here resembles very much
   the picture which one gets from
    strong coupling QCD \cite{CKP83}.
 A baryon state is represented
 in configuration space
   as a system of three
  quarks, each of them connected to a junction, which
  couples them to a color singlet with the help of the
  Levi-Civita  tensor $\epsilon$ (Y-shaped string).
  An effective diquark carrying $\overline{3}$-color charge
  would consist of two quarks
   and the junction which should be treated
  as an additional
   dynamical degree of
   freedom in principle.
 In a diquark-creating tunneling process  a
   piece of elementary color flux  is replaced by
   the $\epsilon$-junction and two `legs', each having
   half of the elementary string tension (from energy conservation).
  Thus the  force
     creating a diquark pair
   in an elementary flux-tube is weaker by a factor
    1/2  than the one  which produces
    an optimally screening quark. This explains naturally
   the dynamical suppression of baryon pairs as compared
  to meson production in elementary flux-tube decays.

  The produced quark pair acquires some transverse momentum
   in   the tunneling process whose distribution can
   be calculated from eq.~(\ref{Gl4}).
  However, one should not expect that a tube is
   straight and parallel to the distance vector between
   source and sink of the electric field
  on a small distance scale since it is
   subject to `roughening'.
  Zero-point oscillations of a tube's normal modes
   may provide another source of transverse momentum.
  Kokoski and Isgur discuss the effect of roughening for
  the breaking of a flux-tube
  in the strong-coupling limit \cite{KOI87}.
  On the other side, the effect should become irrelevant
  in the classical limit of infinite electrical field strength.
  Here I follow  the pragmatic approach to add
   two uncorrelated
  components for a produced particle's transverse momentum,
    one from tunneling and another one
  from unresolved transverse excitations which give
  for the absolute value:
 \begin{equation}
  p_{tr}^2 = p_{0}^2 +  p_{tunn}^2
   \quad.
 \end{equation}
 The $\vec{p_0}$ component is Gaussian distributed and fixed
 by the requirement that the total created transverse momentum
 in elementary flux-tube breaking is (approximately) 400 MeV.

\newpage
{\noindent  \LARGE   Table Captions:}
\vspace{1.0cm}

{\noindent \large Table 1: }

{\noindent
  Reggeon exchange in meson-baryon collisions:
   planar diagrams.
   The Reggeon
   which is exchanged in the various meson-baryon interactions
     describable by a planar quark-line diagram
    can be read off from the table. Values for
    intercept $\alpha _R$,  strangeness
    and quark exchange  suppression
    parameter $r(R)$ and momentum parameter $p_R$
    are given here which enter into  eq.~(\ref{mbmb}).
   $m$ denotes here a  meson with strangeness $S$=0, $k$
   with $S$=1, irrespectively  which multiplet it belongs to.
   (In contrast, a $K$ denotes a kaon and  $K^*$
   a $K^*$(892).)
   The multiplet of which an exchanged meson is a member is indicated by a
    $V$ (vector) or $PS$ (pseudoscalar meson nonet).
   Alternatively, the parameters are specified as some
  `average' of the $V$ and $PS$ exchange parameters.
   Note that in the uppermost diagram the
    flow of isospin quantum numbers as specified and
   $G$-parity determine whether a $V$ or a $PS$ meson is exchanged.
}
\vspace{0.5cm}

{\noindent \large Table 2: }

{\noindent
  Reggeon exchange in meson-baryon collisions:
   exchange diagrams.
  Contrary to the Reggeon exchange diagrams in Table 1
   the reactions tabelized here include a quark exchange
   between ingoing meson and baryon.
  The notation is the same as in Table 1.
}
\vspace{0.5cm}

{\noindent \large Table 3: }

{\noindent
 The produced hadronic state in Pb(160AGeV) on Pb collisions
  with impact parameters b$<$1 fm, calculated in
 three different operation modes of RQMD 2.1: so-called NN mode
  with ropes and rescattering switched off
  in the right column, rope fragmenation included in the
 middle column and ropes and hadronic rescattering both included
  in left column (default mode).
 All members of the pseudoscalar meson nonet and the baryon octet
 have been kept stable, except the $\eta '$(958) which is decaying
 already during the dynamical evoulution generated by RQMD.
}

\newpage

{\noindent  \LARGE   Figure Captions:}
\vspace{1.0cm}

{\noindent \large Figure 1: }

{\noindent
          Decaying hadronic string in position space
           (scaled by $\kappa _{el}$).
           Four hadrons $A$-$D$ are formed,
           their formation points  indicated by
            dots. Thick lines give the trajectories along which
            the hadrons, respectively in the beginning the
            original constituent quarks,  are propagated.
             $p_L^+$, $p_{YoYo}^+$, and $p_B^+$  are
              the forward light cone momenta for the
               leading (spectator) quark, for string excitation
               and for the interacting  (backward) quark.
               $p^-_{Ta} $ is the backward light cone momentum
              which has been transferred from the target.
}
\vspace{0.5cm}

{\noindent \large Figure 2: }

{\noindent
     Schematic diagrams for string excitations in
     multiple baryon collisions:
      The primary string  excitation is shown at the top
       (a).
      Either the  interacting quark
       collides two or  more times
      (c) or the spectator diquark of interaction
      (a)
       interacts  (b).
        Iteration of  diquark interaction
       is depicted in diagram (d). Here all original
        valence quarks are
        completely
        stripped  off from the baryon which is emerging from
       the fragmentation process.
        Constituent quarks
        are symbolized by full dots
        (interacting quarks on  left, spectators on right side),
        sea quark pairs by open circles.
         The ingoing
         light cone momenta
         -- forward
          $p_I$ ($I$ for interaction) and
          $p_S$ ($S$ for spectator)
          and backward $p_b$
           (provided by the  target) --
          are distributed onto
          outgoing constituents
          and   string excitation
          ($p_f$-$p_b$)
          as
          indicated
          just below each diagram.
}
\vspace{0.5cm}

{\noindent \large Figure 3: }

{\noindent
Possible SU(3) multiplets
 which can be
 built by a combination of ($p$,$q$) states and
  elementary triplet (antitriplet) states.
 $p$ is the number of columns
 with one line, $q$ the number with two lines --
 (p=2, q=2 in  this example).
  Thus $p$ is the number of `quark-like' charges,
  $q$ of `antiquark-like' charges.
  There is no restriction
  in  {\it forming} the total rope charge.
   It is energetically allowed to create
   a charge in rope decay  only
    if it screens the original rope charge, i.e.\
    the resulting
    charge is lowered
 (right above, middle below and
  -- $p$$>$$q$ assumed -- middle above).
    Thus  only (at most) three
    out of six  quark or antiquark states in color
    space can be created in the rope field.
}
\vspace{0.5cm}

{\noindent \large Figure 4: }

{\noindent
 Schematic picture of
  quark pair creation
 in a rope:
  quark trajectories are displayed in the $t$-$z$-plane.
  The field strength is characterized by the
   ($p$,$q$) values of the source
   (which is acting here from  the right side).
  It is indicated in the figure
 how  charge creation and
 crossing of quark trajectories decrease the
  field strength inside the rope.
  Quark and antiquark may form a color singlet
  which splits from the rope
 (symbolized here by two horizontal lines).
  Two examples of hadron formation which are
  displayed here demonstrate that color is
   not necessarily
  locally confined in a rope field.
   Neighboring quark
    $Q_{d}$
  and antiquark
$\overline{Q_{c}}$
   form
   a color singlet. Such a
    topology would be always enforced in
  elementary strings  (($p$,$q$)=(1,0))
    by energy conservation. In contrast,
$\overline{Q_{a}}$
 and   $Q_{c}$
 travel quite some distance before they
 combine into a color singlet.
}
\vspace{0.5cm}

{\noindent \large Figure 5: }

{\noindent
  Three mechanisms of chromoelectric field degradation:
  quark pair creation (a),
   turning point of a quark in the end plate source (b),
    and crossing point of two quark trajectories (c).
   The field strength is characterized by a
    pair of numbers  like $(p,q)$
   which characterizes the charge acting from the right side.
   (The corresponding anticharge
    which is the sink  of the flux on the left side
    is a member of the
     $(q,p)$ multiplet.)
    Square brackets like $[i,j]$  characterize   the
    charge moving on a particular trajectory.
    While $p$ and $q$ can get assigned arbitrarily large integer values,
    a charge which is denoted by $[i,j]$  in the diagrams
    belongs either to a
    triplet ($[1,0]$) or an antitriplet ($[0,1]$).
}
\vspace{0.5cm}

{\noindent \large Figure 6: }

{\noindent
 Phase shifts $\delta^0_0$ and   $\delta^2_0$
 in $\pi$$\pi$ scattering:
  Comparison between calculation and experimental data.
 The data are taken from Refs.\ \cite{FRO77,WEI90}.
}
\vspace{0.5cm}

{\noindent \large Figure 7: }

{\noindent
 Rapidity distributions
 of negatively charged hadrons and net protons
  in Pb(160AGeV) on Pb collisions
  with impact parameters b$<$1 fm, calculated in
 three different operation modes of RQMD 2.1:
  ropes and rescattering switched off (dashed line),
   rope fragmenation included (dotted line)
 and ropes and hadronic rescattering both included
  which is the default mode (straight line).
 The rapidity is calculated in the equal-speed-system
 of projectile and target.
 The negatively charged hadrons include feed-down from weak decays
 (except from $K_S$ and (anti-)$\Lambda $), the net protons
  from all weakly and strongly unstable baryons.
}
\vspace{0.5cm}

{\noindent \large Figure 8: }

{\noindent
 Rapidity distributions
 of strange baryons and kaons
  in central Pb(160AGeV) on Pb collisions.
  The different histograms are related to the
  different RQMD operation modes as explained in the
  caption to fig.~7.
 Note that the $\Lambda $ distribution does not contain
 any feed-down from $\Xi$ decay or  $\Sigma^0$-decay.
  The $\Sigma $-rapidity distribution is calculated by
 averaging over all isospin states. The amount of splitting
  in the yields of the three states is small, however (see Table 3).
 The $\Xi$ baryon distribution in the figure contains the sum of
 both charge states. Again, both states are populated approximately
 with equal strength.
}
\vspace{0.5cm}

{\noindent \large Figure 9: }

{\noindent
 Rapidity distributions of anti-baryons
 ($\overline{p}$, $\overline{\Lambda}$, $\overline{\Xi}$)
  in central Pb(160AGeV) on Pb collisions.
  The different histograms are related to the
  different RQMD modes of operation as explained in the
  caption to fig.~7.
 The $\overline{\Xi}$  distribution  contains the sum of
 both charge states.
  No feed-down from weak and electromagnetic decays is included.
}
\vspace{0.5cm}

{\noindent \large Figure 10: }

{\noindent
 The two different production mechanisms
 of baryon pairs from rope fragmentation are
 depicted here schematically.
 An antidiquark which is part
 of the original
   source charge  of the color flux
 field
  may combine
   with an antiquark
  created by the  field itself
 (left side).
  It is assumed here that the
  original charges making up the
  total rope charge move along
  the light cone.
  Having lost its momentum
   by pulling out the chromoelectric field
  an end plate
  charge turns its direction
  and  gets accelerated again.
  After combining with a corresponding
  anticharge into a colorless state
  it will split from the rope without
  further interaction.
  This process is visualized here in
  the $t$-$z$ plane (with $z$ the direction of
 the electric field).
 A flux tube may
  {\it create} as well a  diquark-antidiquark pair
  in the color triplet configuration
  (right side).
  The scheme employed here is usually
 called two-step (or sometimes
  `popcorn') production mechanism.
  It was first suggested in Ref. \cite{CNN79}
   for the case of
   baryon production from
   elementary flux tubes.
 The strength of the
   flux is indicated
  in the figure
  by
   a pair of numbers like ($p$,$q$)
 which defines the representation
  of the charge source.
  Note that the notation
       ($\overline{QQ}$$)_3$
  to highlight the quark content
   of the color charge means
    a (1,0) (or triplet) charge.
}
\vspace{0.5cm}

{\noindent \large Figure 11: }

{\noindent
        The three configurations
         in  color hypercharge and isospin space
        for screening of
        a charge in the
        multiplet ($p$,$q$)
        -- here the state
        $|$(p,q),$t$=(p+q)/2,$t^3$=-$t$,$y$=(p-q)/3$\rangle$
        --
         by
        a created  (anti-)quark charge.
        With $p>q$ which is assumed here there is one screening
         antiquark color
         (a cross enclosed by a circle)
         and  two  quark colors
          (each represented by a cross).
}

\newpage

 \setlength{\unitlength}{1.7pt}
  \begin{table}[htbp]
  \begin{tabular}{|c|c||c|c|c|c|}
   \hline
     &   &   &   &  & \\
  \begin{tabular}{c}
   Diagram   \\
    (planar)
  \end{tabular}
        &  Subclasses/Examples  & Reggeon  &
   2-2$\alpha_R$  &  $r(R)$     &  $p_R$ (GeV)
                  \\
     &   &   &   &  & \\
   \hline
   $
      \begin{array}{c}
        \mbox{
  \begin{picture}(40,40)(0,0)
    \put(10,5){\line(0,1){30}}
    \put(10,20){\line(1,0){20}}
    \put(30,5){\line(0,1){30}}
    \put(10,0){\makebox(0,0){$m$}}
    \put(10,40){\makebox(0,0){$m$}}
    \put(20,25){\makebox(0,0){$m$}}
    \put(30,0){\makebox(0,0){$B$}}
  \end{picture}
            }
      \end{array}
   $
    &
   $
     \left.
      \begin{array}{c}
        \mbox{
        Isospin + G-parity
              }  \\
        \mbox{
  \begin{picture}(40,40)(0,0)
    \put(10,10){\line(0,1){20}}
    \put(10,20){\line(1,0){20}}
    \put(10,5){\makebox(0,0){$0$}}
    \put(10,35){\makebox(0,0){$0$}}
    \put(20,25){\makebox(0,0){$0$}}
  \end{picture}
              }   \\
        \mbox{
  \begin{picture}(40,40)(0,0)
    \put(10,10){\line(0,1){20}}
    \put(10,20){\line(1,0){20}}
    \put(10,5){\makebox(0,0){$I$}}
    \put(10,35){\makebox(0,0){$0/I$}}
    \put(20,25){\makebox(0,0){$I$}}
  \end{picture}
              }
      \end{array}
     \right\}
   $
    &
  \begin{tabular}{c}
    $V$   \\
       or \\
    $P$$S$
  \end{tabular}
    &
  \begin{tabular}{c}
       1   \\
        \\
       2
  \end{tabular}
    &
  \begin{tabular}{c}
      $\mbox{}$
          \\
      1  \\
      $\mbox{}$
  \end{tabular}
     &
     0.8
    \\
   \hline
   $
      \begin{array}{c}
        \mbox{
  \begin{picture}(40,40)(0,0)
    \put(10,5){\line(0,1){30}}
    \put(10,20){\line(1,0){20}}
    \put(30,5){\line(0,1){30}}
    \put(10,0){\makebox(0,0){$\overline{k}$}}
    \put(10,40){\makebox(0,0){$\overline{k}$}}
    \put(20,25){\makebox(0,0){$m$}}
    \put(30,0){\makebox(0,0){$B$}}
  \end{picture}
              }
      \end{array}
   $
    &
   $
      \begin{array}{c}
        \mbox{
  \begin{picture}(40,40)(0,0)
    \put(10,10){\line(0,1){20}}
    \put(10,20){\line(1,0){20}}
    \put(10,5){\makebox(0,0){$\overline{K}$}}
    \put(10,35){\makebox(0,0){$\overline{K}$}}
    \put(20,25){\makebox(0,0){$\varrho$}}
  \end{picture}
              }   \\
        \mbox{
  \begin{picture}(40,40)(0,0)
    \put(10,10){\line(0,1){20}}
    \put(10,20){\line(1,0){20}}
    \put(10,5){\makebox(0,0){$\overline{K}$}}
    \put(10,35){\makebox(0,0){$\overline{K}^*$}}
    \put(20,25){\makebox(0,0){$\pi$}}
  \end{picture}
              }   \\
        \mbox{
  \begin{picture}(40,10)(0,0)
    \put(10,5){\makebox(0,0){\ldots}}
  \end{picture}
              }
      \end{array}
   $
    &
   $
      \begin{array}{c}
        \mbox{
  \begin{picture}(10,40)(0,0)
    \put(5,20){\makebox(0,0){$V$}}
  \end{picture}
              }   \\
        \mbox{
  \begin{picture}(10,40)(0,0)
    \put(5,20){\makebox(0,0){$P$$S$}}
  \end{picture}
              }   \\
        \mbox{
  \begin{picture}(10,10)(0,0)
    \put(5,5){\makebox(0,0){$\overline{PS+V}$}}
  \end{picture}
              }
      \end{array}
   $
    &
   $
      \begin{array}{c}
        \mbox{
  \begin{picture}(10,40)(0,0)
    \put(5,20){\makebox(0,0){1}}
  \end{picture}
              }   \\
        \mbox{
  \begin{picture}(10,40)(0,0)
    \put(5,20){\makebox(0,0){2}}
  \end{picture}
              }   \\
        \mbox{
  \begin{picture}(10,10)(0,0)
    \put(5,5){\makebox(0,0){1.5}}
  \end{picture}
              }
      \end{array}
   $
    &
     1
     &
     0.8

    \\
   \hline
   $
      \begin{array}{c}
        \mbox{
  \begin{picture}(40,40)(0,0)
    \put(10,5){\line(0,1){30}}
    \put(10,20){\line(1,0){20}}
    \put(30,5){\line(0,1){30}}
    \put(10,0){\makebox(0,0){$m$}}
    \put(10,40){\makebox(0,0){$k$}}
    \put(20,25){\makebox(0,0){$\overline{k}$}}
    \put(30,0){\makebox(0,0){$B$}}
  \end{picture}
              }
      \end{array}
   $
    &
   $
      \begin{array}{c}
        \mbox{
  \begin{picture}(40,40)(0,0)
    \put(10,10){\line(0,1){20}}
    \put(10,20){\line(1,0){20}}
    \put(10,5){\makebox(0,0){$\pi $}}
    \put(10,35){\makebox(0,0){$K$}}
    \put(20,25){\makebox(0,0){$\overline{K}^*$}}
  \end{picture}
              }   \\
        \mbox{
  \begin{picture}(40,40)(0,0)
    \put(10,10){\line(0,1){20}}
    \put(10,20){\line(1,0){20}}
    \put(10,5){\makebox(0,0){$\pi $($\varrho $)}}
    \put(10,35){\makebox(0,0){$K^*$($K$)}}
    \put(20,25)
     {\makebox(0,0){$\overline{K}$}}
  \end{picture}
              }   \\
        \mbox{
  \begin{picture}(40,10)(0,0)
    \put(10,5){\makebox(0,0){\ldots}}
  \end{picture}
              }
      \end{array}
   $
    &
   $
      \begin{array}{c}
        \mbox{
  \begin{picture}(10,40)(0,0)
    \put(5,20){\makebox(0,0){$V$}}
  \end{picture}
              }   \\
        \mbox{
  \begin{picture}(10,40)(0,0)
    \put(5,20){\makebox(0,0){$P$$S$}}
  \end{picture}
              }   \\
        \mbox{
  \begin{picture}(10,10)(0,0)
    \put(5,5){\makebox(0,0){$\overline{PS+V}$}}
  \end{picture}
              }
      \end{array}
   $
    &
   $
      \begin{array}{c}
        \mbox{
  \begin{picture}(10,40)(0,0)
    \put(5,20){\makebox(0,0){1.6}}
  \end{picture}
              }   \\
        \mbox{
  \begin{picture}(10,40)(0,0)
    \put(5,20){\makebox(0,0){2.4}}
  \end{picture}
              }   \\
        \mbox{
  \begin{picture}(10,10)(0,0)
    \put(5,5){\makebox(0,0){2.0}}
  \end{picture}
              }
      \end{array}
   $
    &
      0.3
     &
      0.8
    \\  \hline
  \end{tabular}
  \caption[
          ]
            {
               \label{reggea}
                }
  \end{table}

\newpage

 \setlength{\unitlength}{1.7pt}
  \begin{table}[htbp]
  \begin{tabular}{|c|c||c|c|c|c|}
   \hline
     &   &   &   &  & \\[0.2ex]
  \begin{tabular}{c}
   Diagram   \\
    (exchange)
  \end{tabular}
        &  Subclasses/Examples  & Reggeon  &
   2-2$\alpha_R$  &  $r(R)$     &  $p_R$ (GeV)
                  \\[0.5ex]
     &   &   &   &  & \\
   \hline
     &   &   &   &  & \\[-2.8ex]
   $
      \begin{array}{c}
        \mbox{
  \begin{picture}(40,40)(0,0)
    \put(10,5){\line(0,1){30}}
    \put(10,20){\line(1,0){20}}
    \put(30,5){\line(0,1){30}}
    \put(10,0){\makebox(0,0){$\overline{k}$}}
    \put(10,40){\makebox(0,0){$m$}}
    \put(20,25){\makebox(0,0){$\overline{k}$}}
    \put(30,0){\makebox(0,0){$B$}}
  \end{picture}
              }
      \end{array}
   $
    &
   $
      \begin{array}{c}
        \mbox{
  \begin{picture}(40,40)(0,0)
    \put(10,10){\line(0,1){20}}
    \put(30,10){\line(0,1){20}}
    \put(10,20){\line(1,0){20}}
    \put(30,35){\makebox(0,0){$\Lambda ^{(*)}$}}
  \end{picture}
              }   \\
        \mbox{
  \begin{picture}(40,40)(0,0)
    \put(10,10){\line(0,1){20}}
    \put(30,10){\line(0,1){20}}
    \put(10,20){\line(1,0){20}}
    \put(30,35){\makebox(0,0){$\Sigma  ^{(*)}$}}
  \end{picture}
              }   \\
        \mbox{
  \begin{picture}(40,40)(0,0)
    \put(10,10){\line(0,1){20}}
    \put(30,10){\line(0,1){20}}
    \put(10,20){\line(1,0){20}}
    \put(30,35){\makebox(0,0){else}}
  \end{picture}
              }
      \end{array}
   $
    &
   $
      \begin{array}{c}
        \mbox{
  \begin{picture}(10,40)(0,0)
    \put(5,20){\makebox(0,0){$P$$S$}}
  \end{picture}
              }   \\
        \mbox{
  \begin{picture}(10,40)(0,0)
    \put(5,20){\makebox(0,0){$V$}}
  \end{picture}
              }   \\
        \mbox{
  \begin{picture}(10,40)(0,0)
    \put(5,20){\makebox(0,0){$\overline{PS+V}$}}
  \end{picture}
              }
      \end{array}
   $
    &
   $
      \begin{array}{c}
        \mbox{
  \begin{picture}(10,40)(0,0)
    \put(5,20){\makebox(0,0){2.4}}
  \end{picture}
              }   \\
        \mbox{
  \begin{picture}(10,40)(0,0)
    \put(5,20){\makebox(0,0){1.6}}
  \end{picture}
              }   \\
        \mbox{
  \begin{picture}(10,40)(0,0)
    \put(5,20){\makebox(0,0){2.0}}
  \end{picture}
              }
      \end{array}
   $
    &
     0.5
     &
     0.8
    \\[0.1ex]
   \hline
  \begin{picture}(40,40)(0,20)
    \put(10,5){\line(0,1){30}}
    \put(10,20){\line(1,0){20}}
    \put(30,5){\line(0,1){30}}
    \put(20,25){\makebox(0,0){$B$}}
  \end{picture}
    &
   $
      \begin{array}{c}
        \mbox{
  \begin{picture}(40,40)(0,0)
    \put(10,10){\line(0,1){20}}
    \put(10,20){\line(1,0){20}}
    \put(30,10){\line(0,1){20}}
    \put(10,5){\makebox(0,0){$\pi ^-$}}
    \put(10,35){\makebox(0,0){$\Delta ^-$}}
    \put(30,5){\makebox(0,0){$p$}}
    \put(30,35){\makebox(0,0){$\pi ^+$}}
    \put(20,25){\makebox(0,0){$n$}}
  \end{picture}
              }   \\
        \mbox{
  \begin{picture}(40,40)(0,0)
    \put(10,10){\line(0,1){20}}
    \put(10,20){\line(1,0){20}}
    \put(30,10){\line(0,1){20}}
    \put(10,5){\makebox(0,0){$K^-$}}
    \put(10,35){\makebox(0,0){$\Xi ^-$}}
    \put(30,5){\makebox(0,0){$p$}}
    \put(30,35){\makebox(0,0){$K^+$}}
    \put(20,25){\makebox(0,0){$\Lambda$}}
  \end{picture}
              }
      \end{array}
   $
    &
   $
      \begin{array}{c}
        \mbox{
  \begin{picture}(10,40)(0,0)
    \put(5,20){\makebox(0,0){$N$}}
  \end{picture}
              }   \\
        \mbox{
  \begin{picture}(10,40)(0,0)
    \put(5,20){\makebox(0,0){$Y$/$\Xi$}}
  \end{picture}
              }
      \end{array}
   $
    &
   $
      \begin{array}{c}
        \mbox{
  \begin{picture}(10,40)(0,0)
    \put(5,20){\makebox(0,0){2.8}}
  \end{picture}
              }   \\
        \mbox{
  \begin{picture}(10,40)(0,0)
    \put(5,20){\makebox(0,0){3.8}}
  \end{picture}
              }
      \end{array}
   $
    &
   $
      \begin{array}{c}
        \mbox{
  \begin{picture}(10,40)(0,0)
    \put(5,20){\makebox(0,0){0.5}}
  \end{picture}
              }   \\
        \mbox{
  \begin{picture}(10,40)(0,0)
    \put(5,20){\makebox(0,0){0.3}}
  \end{picture}
              }
      \end{array}
   $
     &
     1.4

    \\
   \hline
     &   &   &   &  &  \\[-2.5ex]
   $
      \begin{array}{c}
        \mbox{
  \begin{picture}(40,50)(0,-5)
    \put(10,5){\line(0,1){30}}
    \put(10,20){\line(1,0){20}}
    \put(30,5){\line(0,1){30}}
    \put(10,0){\makebox(0,0){$m$}}
    \put(10,40){\makebox(0,0){($s\overline{s}$)}}
    \put(20,25){\makebox(0,0){$m$}}
    \put(30,0){\makebox(0,0){$B$}}
  \end{picture}
              }
      \end{array}
   $
    &
   $
      \begin{array}{c}
        \mbox{
  \begin{picture}(40,50)(0,-5)
    \put(10,5){\line(0,1){30}}
    \put(10,20){\line(1,0){20}}
    \put(30,5){\line(0,1){30}}
    \put(10,2){\makebox(0,0){$\pi ^-$}}
    \put(10,40){\makebox(0,0){$\phi$}}
    \put(30,0){\makebox(0,0){$p$}}
    \put(30,40){\makebox(0,0){$n$}}
  \end{picture}
              }
      \end{array}
   $
    &
    &
      3.8
    &
      0.09
     &
      1.4
    \\  \hline
  \end{tabular}
  \caption[
          ]
            {
               \label{reggeb}
                }
  \end{table}

\newpage

  \begin{table}[htbp]
  \begin{tabular}{|c|rrr|}
   \hline
     &      Ropes  +  \hspace{0.3em} &
    Ropes +   \hspace{0.4em} &  no Ropes +  \hspace{0.4em}
                                                 \\
     &     Rescattering   &   no Rescattering
                                 &   no  Rescattering
                                                 \\
   \hline
           &        &         &                  \\
$p$       &       157.8     &    199.7   &   184.5
                                                 \\
$n$       &       161.4     &    217.6   &   204.9
                                                 \\
$\Lambda$ &        48.7     &     35.3   &    26.0
                                                 \\
$\Sigma^+$&        17.7     &     12.9   &     7.6
                                                 \\
$\Sigma^0$&        17.8     &     13.1   &     8.0
                                                 \\
$\Sigma^-$&        17.9     &     13.3   &     8.3
                                                 \\
$\Xi^0$   &         5.4     &      4.2   &     2.1
                                                 \\
$\Xi^-$   &         5.4     &      4.2   &     2.0
                                                 \\
$\overline{p}$      &         5.6     &     27.9   &    11.3
                                                 \\
$\overline{n}$      &         5.6     &     27.9   &    11.4
                                                           \\
$\overline{\Lambda}$     &         3.8     &     10.7   &     2.3
                                                           \\
$\overline{\Sigma^+}$    &         1.5     &      4.6   &     0.8
                                                           \\
$\overline{\Sigma^0}$    &         1.5     &      4.6   &     0.8
                                                           \\
$\overline{\Sigma^-}$    &         1.5     &      4.6   &     0.8
                                                 \\
$\overline{\Xi^0}$    &         1.1     &      2.0   &     0.1
                                                 \\
$\overline{\Xi^-}$  &         1.2     &      2.0   &     0.2
                                                 \\
$\pi^+$   &       642.7     &    692.9   &   856.2
                                                 \\
$\pi^0$   &       678.7     &    724.9   &   884.2
                                                 \\
$\pi^-$   &       680.3     &    728.8   &   888.9
                                                 \\
$K^+$     &       130.3     &     79.0   &    79.2
                                                 \\
$K^-$     &        75.4     &     50.4   &    53.0
                                                 \\
$K_s$+$K_l$ &     200.6     &    127.0   &   132.1
                                                 \\
$\eta $   &        81.7     &     84.7   &    86.7

         \\  \hline
  \end{tabular}
  \caption[
          ]
            {
 \label{pyield}
                }
  \end{table}

\clearpage

\newpage

\newpage

\newpage

\newpage

\newpage

\newpage

\newpage

\newpage

\begin{thebibliography}{999}

\bibitem{QM95}
{\bf Quark Matter '95}, Proceedings of the $11^{th}$
international conference on Quark Matter, Monterey, U.S.A., 1995.
A.M.\ Poskanzer, J.W.\ Harris, and L.S.\ Schroeder, eds.:
          {\em Nucl.\ Phys.} {\bf A590} (1995).

\bibitem{QM93}
{\bf Quark Matter '93}, Proceedings of the $10^{th}$
international conference on Quark Matter, Borlange, Sweden, 1993.
E.\ Stenlund, H-A. Gustafsson,
A.\ Oskarsson, and I.\ Otterlund, eds.:
          {\em Nucl.\ Phys.} {\bf A566} (1994).

\bibitem{SOR89}
 H.\ Sorge, H.\ St\"ocker, and W.\ Greiner:
     {\em Ann.\ Phys.\ (N.Y.)} {\bf 192} (1989)   266;
     {\em  Nucl.\ Phys.} {\bf A498} (1989)  567c.

\bibitem{BIR84}
T.S.\ Biro, H.B.\ Nielsen, and J.\ Knoll:
 {\em Nucl.\ Phys.}  {\bf B245} (1984) 449.

\bibitem{BDK84}
G.\ Bertsch, S.\ DasGupta, and H.\ Kruse:
 {\em Phys.\ Rev.}  {\bf C 29} (1984) 673.

\bibitem{AIC87}
 J.\ Aichelin, A.\ Rosenhauer, G.\ Peilert, H.\ St\"ocker,
   and  W.\ Greiner:
 {\em Phys.\ Rev.\ Lett.}  {\bf 58} (1987) 1926.

\bibitem{KOP75}
 J.\ Koplik, A.H.\ Mueller:
   {\em  Phys.\ Rev.} {\bf D 12} (1975) 3638.

\bibitem{BIA76}
 A.\ Bialas, M.\ Bleszynski, and W.\ Czyz:
     {\em   Nucl.\ Phys.} {\bf B111} (1976) 461;
   A.\ Bialas, W.\ Czyz, and L.\ Lesniak:
      {\em   Phys.\ Rev.} {\bf D 25} (1982) 2328;
      {\em   Z.\ f.\ Phys.} {\bf C13} (1982) 147.

\bibitem{NAG94}
      J.\ Nagle et al.:
             {\em Phys.\ Rev.\ Lett.} {\bf 73} (1994) 1219;
              2417.

\bibitem{GON95}
    M.\ Gonin, O.\ Hansen, B.\ Moskowitz, F.\ Videbaek,
        H.\ Sorge, and R.\ Mattiello:
           {\em  Phys.\  Rev.} {\bf C 51} (1995) 310.

\bibitem{SOR92}
     H.\ Sorge, M.\ Berenguer,
             H.\ St\"ocker, and W.\ Greiner:
            {\em Phys.\ Lett.} {\bf B289} (1992) 6.

\bibitem{BER94}
   M.\ Berenguer, H.\ Sorge,  and W.\ Greiner:
       {\em Phys. Lett.} {\bf B332} (1994) 15.

\bibitem{SOR95ZFP}
  H.\ Sorge:
     {\em Z.\ f.\ Phys.} {\bf C67} (1995) 479.

\bibitem{SOR95PLB}
  H.\ Sorge:
   {\em Phys.\  Lett.}  {\bf B344 } (1995) 35.

\bibitem{SOR95PRCB}
  H.\ Sorge:
  to be published.

\bibitem{KOC86}
 P.\ Koch, J.\ Rafelski, and B.\ M\"uller:
 {\em Phys.\ Rep.} {\bf 142} (1986)  167.

\bibitem{SUG95}
  H.\ Suganuma, S.\ Sasaki, and H.\ Toki:
   {\em  Nucl.\ Phys.} {\bf B435}  (1995) 207.

\bibitem{FRI87}
  B.\ Nilsson-Almqvist, and E.\ Stenlund:
     {\em  Comp.\ Phys.\ Comm.} {\bf 43} (1987) 387.

\bibitem{LUND83}
B.\ Andersson, G.\ Gustafson, G.\ Ingelman,
   and  T.\ Sj\"ostrand:
    {\em  Phys.\ Rep.} {\bf 97}  (1983)  31.

\bibitem{WER93}
 K.\ Werner:
    {\em  Phys.\ Rep.} {\bf 232}  (1993)  87.

\bibitem{DPM94}
 A.\ Capella, U.\ Sukhatme, C.-I.\ Tan, and J.\ Tran Thanh Van:
    {\em  Phys.\ Rep.} {\bf 236}  (1994)  346.

\bibitem{AME93}
 N.\ Amelin, M.\ Braun, and C.\ Pajares:
            {\em Phys.\ Lett.} {\bf B306} (1993) 312.

\bibitem{ABFP95}
 N.\ Armesto, M.A.\ Braun, E.G.\ Ferreiro, and C.\ Pajares:
            {\em Phys.\ Lett.} {\bf B344} (1995) 301.

\bibitem{AIC93}
 J.\ Aichelin, K.\ Werner:
            {\em Phys.\ Lett.} {\bf B300} (1993) 158.

\bibitem{CAP95}
 A.\ Capella:
   LPTHE 94-113 (preprint).

\bibitem{SOR91}
 H.\ Sorge, R.\ Mattiello,
 H.\ St\"ocker, and W.\ Greiner:
 {\em Phys. Lett.} {\bf B271} (1991) 37.

\bibitem{HOF95}
    M.\ Hofmann,  R.\ Mattiello,
    H.\ Sorge, H.\ St\"ocker, and W.\ Greiner:
      {\em  Phys.\ Rev.} {\bf C 51}  (1995) 2095.

\bibitem{ARC92}
 T.\ Schlagel, S.\ Kahana, and Y.\ Pang:
 {\em Phys.\ Rev.\ Lett.} {\bf 69} (1992) 3290;
   {\em  Nucl.\ Phys.} {\bf A566}  (1994) 465c.

\bibitem{SCH91}
 J.\ Schaffner, I.N.\ Mishustin,
 L.M.\ Satarov, H.\ St\"ocker, and W.\ Greiner:
 {\em Z.\ f.\ Phys.} {\bf A341} (1991) 47.

\bibitem{KO92}
    C.\ M.\ Ko, M.\ Asakawa, and P.\ Levai:
      {\em  Phys.\ Rev.} {\bf C 46}  (1992) 1072.

\bibitem{WOL93}
 G.\ Wolf:
     {\em Prog.\ Part.\ Nucl.\ Phys.} {\bf 30} (1993) 273.

\bibitem{HJA94}
C.\ Hartnack, J.\ Jaenicke, and J.\ Aichelin:
   {\em  Nucl.\ Phys.} {\bf A580}  (1994) 643.

\bibitem{SOR93ZFP}
H.\ Sorge, L.\ Winckelmann, H.\ St\"ocker, and W.\ Greiner:
     {\em Z.\ f.\ Phys.} {\bf C59} (1993) 85.

\bibitem{AGK74}
 V.A.\ Abramovskii, V.N.\ Gribov, and O.V.\ Kancheli:
 {\em  Sov.\ J.\ Nucl.\ Phys.}  {\bf  18}  1974)   308.

\bibitem{SOR95AGK}
  H.\ Sorge:
  to be published.

\bibitem{BIG87}
   A.\ Bialas, M.\ Gyulassy:
    {\em  Nucl.\ Phys.} {\bf B291} (1987) 793.

\bibitem{SJO86}
 T.\ Sj\"ostrand:
     {\em  Comp.\ Phys.\ Comm.} {\bf 39} (1986) 347.

\bibitem{ABR84}
 P.\ Aurenche, F.\ W.\ Bopp,  and J.\ Ranft:
 {\em Z.\ f.\ Phys.} {\bf C23} (1984) 67.

\bibitem{GOU83}
  K.\ Goulianos:
   {\em Phys. Rep.} {\bf 101} (1983)  169.

\bibitem{AGN87}
  B.\ Andersson, G.\ Gustafson, and B.\ Nilsson-Almqvist:
    {\em  Nucl.\ Phys.} {\bf B281} (1987) 289.

\bibitem{CAP87}
 A.\ Capella, J.A.\ Casado, C.\ Pajares, A.\ V.\ Ramello,
  and
  J.\ Tran Thanh Van:
    {\em Z.\ f.\ Phys.} {\bf C33} (1987) 541.

\bibitem{MIT94}
J.T.\ Mitchell et al. (NA35 Collab.):
 {\em Nucl.\ Phys.}  {\bf A566} (1994) 415c.

\bibitem{HKH85}
  R.\ C.\ Hwa:
     {\em Phys.\ Rev.\ Lett.} {\bf 52} (1984) 492;
  A.\ Klar, J.\ H\"ufner:
     {\em Phys.\ Rev.} {\bf D 31} (1985) 491;
  L.\ P.\ Csernai, J.\ I.\ Kapusta:
     {\em Phys.\ Rev.} {\bf D 31} (1985) 2795.

\bibitem{AFS62}
 D.\ Amati, A.\ Stanghellini, and S.\ Fubini:
 {\em Nuov.\ Cim.}  {\bf 26} (1962)   6.

\bibitem{LG78}
 F.E.\ Low, K.\ Gottfried:
       {\em Phys.\ Rev.} {\bf D 17} (1978)  2487.

\bibitem{MPR92}
 C.\ Merino, C.\ Pajares, and J.\ Ranft:
            {\em Phys.\ Lett.} {\bf B276} (1992) 168.

\bibitem{BER82}
             C.\ Bernard:
            {\em Phys.\ Lett.} {\bf B108} (1982) 431,
           {\em Nucl.\ Phys.} {\bf B219} (1983) 341;
             J.\ Ambj{\o}rn, P.\ Olesen, and C.\ Peterson:
           {\em Nucl.\ Phys.} {\bf B240} (1984) 189;
             C.\ Michael:
           {\em Nucl.\ Phys.} {\bf B259} (1985) 58;
             L.A.\ Griffiths, C.\ Michael, and P.\ Rakow:
            {\em Phys.\ Lett.} {\bf B150} (1985) 196;
             N.A.\ Campbell, I.H.\ Jorysz,  and  C.\ Michael:
            {\em Phys.\ Lett.} {\bf B167} (1986) 91.

\bibitem{KOG75}
  J.\ Kogut, L.\ Susskind:
     {\em Phys.\ Rev.} {\bf D 11} (1975) 395.

\bibitem{TRO93}
H.D.\ Trottier, R.M.\ Woloshyn:
                {\em Phys.\ Rev.} {\bf D 48} (1993)  2290.

\bibitem{SCH51}
J.\ Schwinger:
  {\em Phys. Rev.} {\bf 82} (1951) 664;
E.\ Brezin, C.\ Itzykson:
                {\em Phys.\ Rev.} {\bf D 2} (1970)  1191.

\bibitem{CNN79}
 A.\ Casher, H.\ Neuberger, and S.\ Nussinov:
                {\em Phys.\ Rev.} {\bf D 20} (1979)  179.

\bibitem{ARM74}
 X.\ Artru, G.\ Mennessier:
   {\em Nucl.\ Phys.} {\bf B70} (1974) 93.

\bibitem{HER90}
 M.\ Herrmann,  J.\ Knoll:
 {\em Phys. Lett.} {\bf B234} (1990) 437.

\bibitem{MAR89}
 C.\ Martin, D.\ Vautherin:
   {\em Phys. Rev.} {\bf D 40}  (1989) 1667.

\bibitem{BCZ86}
  A.\ Bialas, W.\ Czyz:
   {\em  Nucl.\ Phys.} {\bf B267} (1986)  242.

\bibitem{COL71}
 P.D.B.\ Collins:
  {\em  Phys.\ Rep.} {\bf 4}  (1971)  103.

\bibitem{ISG85}
 S.\ Godfrey,  N.\ Isgur:
      {\em Phys.\ Rev.} {\bf D 32} (1985)  1357.

\bibitem{MAR76}
 B. R.\ Martin, D.\ Morgan, and G.\ Shaw:
     {\it Pion--Pion Interactions in Particle Physics.}
             Academic Press, London (1976).

\bibitem{DAV62}
K.T.R.\ Davies,  M.\ Baranger:
 {\em Ann. Phys.\ (N.Y.)} {\bf 19} (1962)  383.

\bibitem{BER94}
 M.\ Berenguer:
 Ph.\ D.\ Thesis, Univ.\ Frankfurt (1994), unpublished.

\bibitem{PDG92}
Particle Data Group, K. Hikasa et al.:
  {\em Phys.\ Rev.} {\bf D 45} (1992).

\bibitem{SAM74}
 N.P.\ Samios, M.\ Goldberg, and B.T.\ Meadows:
         {\em Rev.\ Mod.\ Phys.} {\bf 46} (1974)  49.

\bibitem{VEN68}
 G.\ Veneziano:
 {\em Nuov.\ Cim.} {\bf 57A}  (1968) 190.

\bibitem{VAN88}
 J.\ Vandermeulen:
    {\em Z.\ f.\ Phys.} {\bf C37} (1988) 563.

\bibitem{MVW91}
S.\ Mundigl, M.\ Vicente Vacas, and W.\ Weise:
 {\em Nucl.\ Phys.}  {\bf A523} (1991) 499.

\bibitem{KOC89}
 P.\ Koch,  C.\  Dover:
      {\em Phys.\ Rev.} {\bf C 40} (1989)  145.

\bibitem{FER61}
   E.\  Ferrari, F.\ Selleri:
     {\em Phys.\ Rev.\ Lett.} {\bf 7} (1961)  387.

\bibitem{DIM86}
   V.\ Dimitriev, O.\ Sushkov, and C.\ Gaarde:
      {\em  Nucl.\ Phys.} {\bf A459} (1986)  503.

\bibitem{RAN80}
 J.\ Randrup, C.M.\ Ko:
   {\em  Nucl.\ Phys.} {\bf A343}  (1980) 519.

\bibitem{MAT91}
 R.\ Mattiello:
  Diploma Thesis, Univ.\ Frankfurt (1991), unpublished.

\bibitem{GAS66}
    S.\ Gasierowicz:
     {\it Elementary Particle Physics.}
        John Wiley \& Sons -- New York (1966).

\bibitem{FRO77}
 C.D.\ Frogatt,  J.L.\ Petersen:
    {\em Nucl.\ Phys.} {\bf B129} (1977)  89.

\bibitem{EST78}
 P.\ Estabrooks et al.:
  {\em Nucl.\ Phys.} {\bf B133} (1978)  490.

\bibitem{AST88}
  D.\ Aston  et al.:
   {\em Nucl.\ Phys.} {\bf B296} (1988)  493.

\bibitem{WEI90}
  J.\ Weinstein,  N.\ Isgur:
        {\em Phys.\ Rev.} {\bf D 41} (1990)  2236.

\bibitem{FLA76}
  S.M. Flatt\'e:
      {\em Phys.\ Lett.} {\bf 63B} (1976)  224.

\bibitem{GAV91}
 S.\ Gavin, P.\ Ruuskanen:
 {\em Phys. Lett.} {\bf B262} (1991) 326.

\bibitem{AME91}
 N.S.\ Amelin, E.F.\ Staubo, L.P.\ Csernai,
 V.D.\ Toneev, K.K.\  Gudima, and D.\ Strottman:
     {\em Phys.\ Lett.} {\bf B261} (1991) 352.

\bibitem{KAD95}
 K.\ Kadija, N.\ Schmitz, and P.\ Seyboth:
 MPI-PhE/95-07 (preprint), subm.\ to
    {\em Z.\ f.\ Phys.} {\bf C}.

\bibitem{AvK91}
      A.\ v.\ Keitz, L.\ Winckelmann, A.\ Jahns,
      H.\ Sorge, H.\ St\"ocker, and W.\ Greiner:
            {\em Phys.\ Lett.} {\bf B263} (1991) 353.

\bibitem{SSH93}
  E.\ Schnedermann, J.\ Sollfrank, and U.\ Heinz:
 {\em Phys. Rev.} {\bf C 48} (1993) 2462.

\bibitem{PBM95}
 P.\ Braun-Munzinger et al.:
            {\em Phys.\ Lett.} {\bf B344} (1995) 43.

\bibitem{NA3590}
J.\ Bartke et al. (NA35 Collab.):
    {\em Z.\ f.\ Phys.} {\bf C48} (1990) 191.

\bibitem{NA3594}
J.\ B\"achler et al. (NA35 Collab.):
    {\em Z.\ f.\ Phys.} {\bf C61} (1994) 551.

\bibitem{JAE75}
  K.\ Jaeger et al.:
 {\em Phys. Rev.} {\bf D 11} (1975) 1756.

\bibitem{ABA91}
 S.\ Abatzis  et al. (WA85 Collab.):
 {\em Phys. Lett.} {\bf B244} (1990) 130;
 {\em Phys. Lett.} {\bf B259} (1990) 508;
 {\em Phys. Lett.} {\bf B270} (1991) 123.

\bibitem{NA35HBT94}
 G.\ Roland et al. (NA35 Collab.):
  {\em Nucl.\ Phys.} {\bf A566} (1994)  527c.

\bibitem{NA44HBT95}
H.\ Beker et al. (NA44 Collab.):
     {\em Phys.\ Rev.\ Lett.} {\bf 74} (1995) 3340.

\bibitem{LRT94}
J.\ Letessier, J.\ Rafelski, and A.\ Tounsi:
 {\em Phys. Lett.} {\bf B323} (1994) 393.

\bibitem{ARC93}
 S.H.\ Kahana, Y.\ Pang, T.\ Schlagel, and C.B.\ Dover:
      {\em Phys.\ Rev.} {\bf C 47} (1993) R1356.

\bibitem{SPI95}
C.\ Spieles, M.\ Bleicher, A.\ Jahns, R.\ Mattiello, H.\ Sorge,
   H.\ St\"ocker, and W.\ Greiner:
  preprint UFTP 385-95, nucl-th/9506008,
  subm.\ to   {\em Phys.\ Rev.} {\bf C} (1995).

\bibitem{YND83}
F.J.\ Yndurain:
  {\it Quantum Chromodynamics}, Springer,
  New York-Berlin-Heidelberg-Tokyo  (1983).

\bibitem{CKP83}
 J.\ Carlson, J.\ Kogut, and V.R.\ Pandharipande:
       {\em Phys.\ Rev.} {\bf D 27} (1983)  233.

\bibitem{GLE83}
  N.K.\ Glendenning, T.\ Matsui:
 {\em Phys. Rev.} {\bf D 28} (1983) 2.

\bibitem{KOI87}
 R.\ Kokoski, N.\ Isgur:
    {\em Phys.\ Rev.} {\bf D 35} (1987)  907.



\end{thebibliography}
\end{document}